%% file: Mult.tex
\documentclass[final,letter]{article}
\pdfoutput=1

\usepackage{anysize}
\usepackage{amsmath, amssymb, mathtools, physics}
\usepackage{algorithm, algpseudocode}
\usepackage{graphicx, float, calc}
\usepackage{array,multirow}
\usepackage{twoopt}
\usepackage{xcolor}
\usepackage{etoolbox}
\usepackage[bottom]{footmisc}
\usepackage[absolute]{textpos}
\usepackage{lmodern}

\marginsize{1in}{1in}{1in}{1in}

\input{definitions}

\algnewcommand{\CommentLine}[1]{\State \(\#\) #1}
\algnewcommand{\CommentLinex}[1]{\Statex \(\#\) #1}
\algrenewcomment[1]{{\hfill\makebox[.5\linewidth][l]{\# #1}}}

\title{HIGH PERFORMANCE QUANTUM MODULAR MULTIPLIERS}
\author{%
Rich Rines\footnotemark[2]
\thanks{Massachusetts Institute of Technology Lincoln Laboratory, Lexington, Massachusetts 02420, USA}
and
Isaac Chuang\thanks{Massachusetts Institute of Technology, Cambridge, Massachusetts 02139, USA}
}

\begin{document}

\maketitle

\begin{abstract}
We present a novel set of reversible modular multipliers applicable to quantum computing, derived from three classical techniques: 1) traditional integer division, 2) Montgomery residue arithmetic~\cite{Montgomery1985}, and 3) Barrett reduction~\cite{Barrett1987}.  Each multiplier computes an exact result for all binary input values, while maintaining the asymptotic resource complexity of a single (non-modular) integer multiplier. We additionally conduct an empirical resource analysis of our designs in order to determine the total gate count and circuit depth of each fully constructed circuit, with inputs as large as 2048 bits.  Our comparative analysis considers both circuit implementations which allow for arbitrary (controlled) rotation gates, as well as those restricted to a typical fault-tolerant gate set.
\end{abstract}

\begin{textblock*}{5.4in}(1.25in,10.9in)
\centering
\fontsize{4.15pt}{5pt}\selectfont
Distribution Statement: A. Approved for public release - distribution is unlimited

\vspace{4pt}\hfill
\begin{minipage}[c]{\textwidth}
\fontsize{4.15pt}{5pt}\selectfont
This material is based upon work supported by the Assistant Secretary of Defense for Research and Engineering under Air Force Contract No. FA8721-05-C-0002 and/or FA8702-15-D-0001. Any opinions, findings, conclusions or recommendations expressed in this material are those of the author(s) and do not necessarily reflect the views of the Assistant Secretary of Defense for Research and Engineering.
\end{minipage}
\end{textblock*}

\input{intro}
\newpage{}
\input{mod-mult}
\newpage{}
\input{division}
\newpage{}
\input{montgomery}
\newpage{}
\input{barrett}
\newpage{}
\input{fourier}
\newpage{}
\input{resources}
\newpage{}
\input{conclusion}

\section{Acknowledgments}
The authors would particularly like to acknowledge Kevin Obenland at MIT Lincoln Laboratory, whose invaluable discussions, insight, and expertise in both the design of high-performance reversible arithmetic and the efficient computational analysis of reversible circuits was critical to the success of this work. We also graciously acknowledge the support of the NSF iQuISE IGERT and the Intelligence Advanced Research Projects Activity (IARPA).

\newpage{}
\bibliographystyle{unsrt}
\bibliography{mmult}

\newpage{}
\appendix
\input{adders}

\end{document}

%% file: definitions.tex
\newlength{\heightperqubit}
\newcommand\includecircuit[2][4]{%
  \includegraphics[
      width=\textwidth,%
      height=#1\heightperqubit,%
      keepaspectratio]%
    {fig-#2}}

\def\clap#1{\hbox to 0pt{\hss#1\hss}}

\newcommand{\inputtikz}[1]{\includegraphics{tikz-#1}}
\newcommand{\includeplot}[1]{\includegraphics{plot-#1}}

\newcommand{\mathdisp}[2]{\mathchoice{#1}{#2}{#2}{#2}}

\newcommand{\makeBig} [1]{\vphantom{\Big(\Big)}\smash{#1}}
\newcommand{\makebigg}[1]{\vphantom{\bigg(\bigg)}\smash{#1}}

\let\defket\ket

\renewcommand{\ket}[2][]{\ensuremath{\mathdisp%
                         {\ifstrempty{#1}{\defket{#2}}{\defket{#2}_{#1}}}%
                         {\ifstrempty{#1}{\smash{\defket*{#2}}}{\smash{\defket*{#2}}_{#1}}}%
                       }}
\newcommand{\fket}[2][]{\ensuremath{\mathdisp%
                          {\ifstrempty{#1}{\defket{#2}^{\Phi}}{\defket{#2}_{#1}^{\Phi}}}%
                          {\ifstrempty{#1}{\smash{\defket*{#2}}^{\Phi}}{\smash{\defket*{#2}}_{#1}^{\Phi}}}%
                        }}

\newcommand{\fKet}[2][]{\ifstrempty{#1}{\defket{\makeBig{#2}}^{\!\Phi}}{\defket{\makeBig{#2}}_{\!#1}^{\!\Phi}}}
\newcommand{\fkett}[2][]{\ifstrempty{#1}{\defket{\makebigg{#2}}^{\!\!\Phi}}{\defket{\makebigg{#2}}_{\!\!#1}^{\!\!\Phi}}}

\newcommand{\GATE}[1]{\ifmmode{\text{\textsc{#1}}}\else{\textsc{#1}}\fi}
\def\X/{\GATE{X}}
\def\Ry/{\ensuremath{\GATE{R}_{y}}}
\def\CRy/{\ensuremath{\GATE{CR}_{y}}}
\def\Z/{\GATE{Z}}
\def\H/{\GATE{H}}
\def\T/{\GATE{T}}
\def\S/{\GATE{S}}
\def\NOT/{\GATE{NOT}}
\def\QFT/{\GATE{QFT}}
\def\QFTd/{\ensuremath{\GATE{QFT}^\dagger}}
\def\CNOT/{\GATE{CNOT}}
\def\SWAP/{\GATE{swap}}
\def\Toffoli/{\GATE{Toffoli}}
\def\CX/{\CNOT/}
\def\CCX/{\GATE{Toffoli}}
\def\Fredkin/{\GATE{Fredkin}}

\def\multiplication/{\GATE{Multiplication}}
\def\uncomputation/{\GATE{Uncomputation}}
\def\reduction/{\GATE{Reduction}}

\newcommand\qhat{\smash{\ensuremath{\tilde{q}}}}
\newcommand\xyapp{\smash{\ensuremath{\widetilde{Xy}}}}

\newcommand\defeq   {{{}\triangleq{}}}
\newcommand\flr	[1]	{\ensuremath{\lfloor{#1} \rfloor}}
\newcommand\ceil[1]	{\ensuremath{\lceil	{#1} \rceil}}
\newcommand\ord [1] {\ensuremath{\mathcal{O}(#1)}}
\newcommand\clog[2][]{\ensuremath{\ceil{\log_{#1}#2}}}

\def\bdiv{%
  \nonscript\mskip-\medmuskip\mkern5mu%
  \mathbin{\text{div}}\penalty900\mkern5mu%
  \nonscript\mskip-\medmuskip}

\newcommand{\lra}[1][\hspace{1em}]{\ensuremath{{}\xrightarrow{#1}{}}{}}
\newcommand{\lraover}[2][\hspace{1em}]{\ensuremath{{}\xrightarrow[{\hphantom{{#1}}}]{#2}}{}}

\newcommand{\cc}[1]{} 

\newcounter{refitemiter}
\newcounter{refitemcntr}
\newcommand{\refiter}{}
\newcommand{\refparser}[5][\ref]{%
\setcounter{refitemcntr}{0}%
\setcounter{refitemiter}{0}%
\expandafter\renewcommand{\refiter}[1]{%
  \stepcounter{refitemiter}%
  \ifthenelse{\value{refitemcntr}=\value{refitemiter}}%
    {and~#1{#2##1}}%
    {\ifthenelse{\value{refitemcntr}=2}%
      {#1{#2##1} }%
      {#1{#2##1}, }%
    }%
}%
  \forcsvlist{\stepcounter{refitemcntr}\cc}{#5}%
  \ifthenelse{\value{refitemcntr}=1}%
    {#3~\expandafter#1{#2#5}}%
    {#4 \forcsvlist{\refiter}{#5}}%
}

\newcommand\fig [1]   {Figure~\ref{fig:#1}}
\newcommand\alg [1]{Algorithm~\ref{alg:#1}}
\newcommand\tab [1]    {Table~\ref{tab:#1}}
\newcommand\sect[1]  {Section~\ref{sec:#1}}
\newcommand\apx [1] {Appendix~\ref{sec:#1}}
\newcommand\algline [2]{line~\ref{alg-line:#1:#2}}
\newcommand\step[1]{\refparser{alg-line:}{step}{steps}{#1}}
\newcommand\alglines[3]{lines~\ref{alg-line:#1:#2}-\ref{alg-line:#1:#3}}
\newcommand\eq[1]{\relax\ifmmode{\label{eq:#1}}\else{Equation~\eqref{eq:#1}}\fi}
\newcommand\eqn[1]{\relax\ifmmode{\label{eqn:#1}}\else{Equation~\eqref{eqn:#1}}\fi}

\newcommand{\MESTtext}{\ensuremath{\text{\textsc{MonEst}}}}
\newcommand{\MPROtext}{\ensuremath{\text{\textsc{MonPro}}}}
\newcommand{\BPROtext}{\ensuremath{\text{\textsc{BarPro}}}}
\newcommand{\REDCtext}{\ensuremath{\text{\textsc{REDC}}}}

\newcommand{\DIVtext} {\ensuremath{\text{\textsc{DIV}}}}

\newcommand{\MAC}{\ensuremath{\text{\textsc{MAC}}}}
\newcommand{\MUL}{\ensuremath{\text{\textsc{MUL}}}}

\newcommand{\U}{u}
\newcommand{\UU}{\ensuremath{\smash{\tilde{u}}}}

\newcommand{\MonFunc}	[4]{\ensuremath{%
  \ifstrempty{#4}{#1}{{#1}^{#4}}%
  \ifstrempty{#2}{}{({#2}\ifstrempty{#3}{}{\mid{}{#3}})}%
}}

\newcommand{\REDC}  [3][]{\MonFunc{\REDCtext}{#2}{#3}{#1}}
\newcommand{\MonPro}[3][]{\MonFunc{\MPROtext}{#2}{#3}{#1}}
\newcommand{\MonEst}[3][]{\MonFunc{\MESTtext}{#2}{#3}{#1}}
\newcommand{\BarPro}[3][]{\MonFunc{\BPROtext}{#2}{#3}{#1}}

\newcommand{\DIV}   [2][]{\MonFunc{\DIVtext}{#2}{}{#1}}

\newcommand{\QQ}{\ensuremath{\text{\textsc{Q-}}}}
\newcommand{\QF}{\ensuremath{\Phi\text{-}}}



%% file: intro.tex

\section{Introduction}
\label{sec:intro}

\subsection{Efficient Modular Multiplication Implementations}

Circuits implementing reversible modular arithmetic operations are of significant contemporary interest due to their prevalence in important
quantum algorithms~\cite{Shor1994}. Resource-efficient implementation of these operations is critical to their eventual implementation on a quantum computer. 
In this paper, we describe three novel techniques which asymptotically reduce the resources required for exact, reversible modular multiplication to those necessary for non-modular integer multiplication.  Existing proposals for efficient reversible modular multipliers largely fall in one of two categories: (1) the composition of reversible modular adders, each comprising the equivalent of three reversible non-modular adders; or (2) approximated division, in which reversible modular reduction is simplified by allowing the return of an incorrect output for some subset of input values~\cite{Zalka1998, Kutin2006}.  Compared to those in the first category, our circuits achieve a factor-of-three reduction in asymptotic gate count and circuit depth while requiring only $\ord{\log n}$ additional qubits.  They perform comparably to those in the second category, but without employing arithmetical approximations--regardless of the impact of such approximations on the overall fidelity of quantum algorithms, our constructions demonstrate that accuracy need not be sacrificed in order to obtain an efficient modular multiplication circuit.

Classically, large-integer modular multiplication is a critical component of cryptographic functions such as the RSA public-key encryption system~\cite{RSA1978}.
The standard strategy for modular multiplication is to first calculate the integer product of the input values, and then reduce the result via division. However, due to the inefficiency of integer division on most processors, systems implementing extensive modular arithmetic typically employ specialized methods to eliminate the division requirement of modular reduction.  In particular, Montgomery residue arithmetic~\cite{Montgomery1985} and Barrett reduction~\cite{Barrett1987} are commonly used techniques in applications requiring many reductions under the same modulus.  In the Montgomery method, the problem is converted to an ``$N$-residue'' representation in which modulo-$N$ reduction instead requires calculations under an easier modulus (e.g. a power of two).
Barrett reduction takes advantage of the constant modulus by pre-computing a fixed-point reduction factor once, so that individual reductions can be computed using only multiplication and bit-shift operations.

In this work, we describe three novel reversible modular multipliers, employing (1) a new implementation of the standard division-based reduction procedure, as well efficient reversible adaptations of (2) classical Montgomery multiplication and (3) Barrett reduction.
Our designs principally comprise common abstract circuit primitives, such as reversible adders and subtracters, enabling their implementation within various architectural and arithmetical models.  In particular, the Montgomery and Barrett multipliers introduced are uniquely amenable to arithmetic in the quantum Fourier transform basis, sidestepping the bottlenecks plaguing many previous implementations utilizing this representation~\cite{Draper00,Beauregard2003}.
We discuss the relative trade-offs of various implementation methods for each modular multiplication strategy, and perform a detailed resource analysis of each.

\subsection{Prior Modular Multiplication Implementations}
\label{sec:intro:prior-art}
 
The first quantum circuits for modular multiplication were developed as part of a typical arithmetical hierarchy~\cite{Beckman1996, Vedral1996, Fowler2004},
\begin{equation*}
  (\text{Integer Adder}) \lra (\text{Modular Adder}) \lra
  (\text{Modular Multiplier}),
\end{equation*}
in which a quantum modular multiplier is constructed from a sequence of modular adders, which in turn consist of multiple integer addition steps.  The complexity of modular multiplication is then driven by the complexity of reversible implementations of integer addition.

The first quantum integer addition circuits were based on the classical ripple-carry adder, which has circuit depth that is linear in the bit-width of its addends~\cite{Vedral1996, Cuccaro2004}.
The circuit depth and total size of modular multiplication using the ripple-carry adder is $\ord{n^2}$ for $n$-bit inputs. 
Logarithmic-depth addition circuits were subsequently proposed~\cite{Draper2006} that reduce the depth of modular multiplication to $\ord{n\log_2n}$, with the total number of
gates remaining $\ord{n^2}$. 

Quantum adders that operate in the quantum Fourier transform basis~\cite{Draper00} have also been used to implement modular addition, multiplication, and exponentiation. 
By moving integer addition to the Fourier basis, 
Beauregard~\cite{Beauregard2003} presented a modular multiplication circuit that has depth \ord{n^2}
and total circuit count $\ord{n^3}$, but requires fewer qubits than previously described circuits. 
However, Fourier-basis arithmetic employs arbitrarily angled rotation gates, which may require
more resources to implement on typical architectures than the reversible AND (i.e. \Toffoli/) gates required by the other adders.  For this reason, it is difficult to do a direct comparison of Fourier-basis and binary circuits.

Using Fourier-basis addition, Kutin has devised a unique and efficient procedure for approximate quantum modular multiplication with linear algorithmic depth and a linear number of 
qubits~\cite{Kutin2006}. This circuit uses an approximate multiplication technique introduced by Zalka~\cite{Zalka1998}, which requires uncertain assumptions 
regarding the distribution of systematic errors. The argument is made that these errors do not lead to an appreciable level of error.

An exact, linear-width quantum modular multiplication procedure was proposed by Pavlidis and Gizopoulos~\cite{Pavlidis2014}.  
By applying Fourier-basis arithmetic to construct exact reversible constructions of a large array of arithmetic operations, they introduce a novel quantum Granlund-Montgomery 
division procedure enabling exact quantum modular multiplication with linear depth and linear width.  
Still, the leading terms in this construction remain prohibitive, requiring $9n$ qubits, $800n^2$ total quantum gates, and a parallelized depth of $1000n$ rotation gates, 
with almost 90\% of these costs devoted to the division procedure.

As in classical systems~\cite{Wallace1964}, a trade-off exists between algorithmic depth and width.  Allowing a super-linear supply of ancillary qubits,
various procedures for sub-linear time quantum multiplication have been introduced.  Gossett's carry-save multiplier achieves logarithmic depth but requires $\ord{n^2}$ qubits~\cite{Gossett1998}.  
Similarly, modular exponentiation can be performed by arranging the component multipliers in a logarithmic-depth binary-tree structure at the cost of a quadratic circuit width~\cite{VanMeter2005}.  
Combining these models with constant-depth teleportation-based fan-out, Pham and Svore have introduced a 2D nearest-neighbor carry-save architecture allowing modular exponentiation 
in $\ord{\log^2(n)}$ depth with $\ord{n^4}$ qubits~\cite{Pham2013}.  

Classically, multipliers exploiting the fast Fourier transform (FFT) and convolution theorem have long dominated in asymptotic depth~\cite{Karatsuba1962, Schonhage1971, Furer2009}.  
Accordingly, the fastest known quantum multipliers are direct applications of these constructions~\cite{Zalka1998, Kowada2006}.  However, a signature of both classical and 
quantum convolution-based multipliers is a prohibitive initial cost--for example, Zalka's quantum Sch\"onhage-Strassen multiplier~\cite{Zalka1998} has an 
asymptotic depth of $\sim2^{16}n^{0.2}$.

We summarize these results in \tab{intro:comparison}.  Because of the difficulty of making side-by-side comparisons of Fourier-basis and binary circuits, we have separated the two and characterized each in terms of their principle operation (\Toffoli/ gates in the case of binary arithmetic, total controlled-rotations in the Fourier case.)  The characteristics for binary modular multipliers utilizing modular adders are determined from the particular adders employed, assuming three adds per modular addition and $2n$ modular adds per modular multiplication.  All other circuits are presented as reported in the reference.  We will discuss the various trade-offs that exist, and where our proposed circuits fall among these, in \sect{resources}.

\begin{table}[H]
  \centering
  \caption{\label{tab:intro:comparison}%
      Resource comparison of in-place quantum modular multipliers. Only the leading order term is shown for each count.
      }
  \vspace{6pt}
  \begin{tabular}{r l l l l}
  \multicolumn{5}{c}{\textbf{Binary Arithmetic:} } \\[2pt]
  {Proposal}    & {Architecture}          & {Qubits}   & {Gates${}^\dagger$}& {Depth${}^\dagger$}  \\
  \hline\\[-9pt]
    {${}^\star$Cuccaro et. al.}~\cite{Cuccaro2004}
      & Modular Addition (Ripple-Carry)   & $3n$       & $12n^2$         & $12n^2$ \\
    {${}^\star$Draper et. al.}~\cite{Draper2006}
      & Modular Addition (Prefix)         & $5n$       & $60n^2$         & $24n\log_2n$ \\
    {Zalka}~\cite{Zalka1998}
      & Sch\"onhage-Strassen (FFT)        & $24...96n$ & $2^{16}n$       & $2^{16}n^{0.2}$ \\
    {Pham-Svore}~\cite{Pham2013}
      & Carry-Save (nearest-neighbor)     & $16n^2$    & $384n^2\log_2n$ & $56\log_2n$ \\
    \multirow{6}{*}{\textbf{This work} $\left.\begin{array}{r}\\\\\\\\\\\end{array}\right\{$}
      & \textbf{Exact Division (Prefix)}        & $\boldsymbol{5n}$ & $\boldsymbol{20n^2}$ & $\boldsymbol{8n\log_2n}$ \\
      & \textbf{Montgomery Reduction (Prefix)}  & $\boldsymbol{5n}$ & $\boldsymbol{20n^2}$ & $\boldsymbol{8n\log_2n}$ \\
      & \textbf{Barrett Reduction (Prefix)}     & $\boldsymbol{5n}$ & $\boldsymbol{20n^2}$ & $\boldsymbol{8n\log_2n}$ \\
      & \textbf{Exact Division (Ripple)}        & $\boldsymbol{3n}$ & $\boldsymbol{4n^2}$ & $\boldsymbol{4n^2}$ \\
      & \textbf{Montgomery Reduction (Ripple)}  & $\boldsymbol{3n}$ & $\boldsymbol{4n^2}$ & $\boldsymbol{4n^2}$ \\
      & \textbf{Barrett Reduction (Ripple)}     & $\boldsymbol{3n}$ & $\boldsymbol{4n^2}$ & $\boldsymbol{4n^2}$ \\
  \hline\\[-9pt]
  \multicolumn{5}{l}{${}^\star$Reference proposes an adder only.  We assume $3$ adders per modular add, $2n$ modular adds per multiply.}\\
  \multicolumn{5}{l}{${}^\dagger$Total gate counts and depths provided in \Toffoli/ gates.}\\[12pt]
  \end{tabular}
\end{table}
\begin{table}[H]
  \centering
  \caption{\label{tab:intro:fourier-comparison}%
      Resource comparison of Fourier-basis in-place quantum modular multipliers.
      Gate counts are reported in two ways: (1) the total number of rotations assuming infinite-precision control, and (2) the number of gates after removing those with exponentially-small rotation angles~\cite{Barenco1996}. Only the leading order term is shown for each count.
      }
  \vspace{6pt}
  \begin{tabular}{r l l l l l}
  \multicolumn{5}{c}{\textbf{Fourier-basis arithmetic:} } \\[2pt]
  {Proposal}    & {Architecture}       & {Qubits}          & {Gates (1)${}^\ddagger$}   & {Gates (2)${}^\ddagger$}  & {Depth${}^\ddagger$}  \\
  \hline\\[-9pt]                                                                        
    {Beauregard}~\cite{Beauregard2003}                                                  
      & Modular Addition               & $2n$              & $4n^2$                     & $\ord{n^3\log n}$     & $8n^2$ \\
    \multirow{2}{*}{{Kutin}~\cite{Kutin2006} $\Big\{$}     
      & Approximate Division           & $3n$              & $3n^2$                     & $2n^2$                & $6n$  \\
      & Approximate (nearest-neighbor) & $3n$              & $5n^2$                     & $4n^2$                & $11n$ \\
    {Pavlidis}~\cite{Pavlidis2014}                                           
      & Granlund-Montgomery Division   & $9n$              & $800n^2$                   & $\sim250^2$           & $1000n$ \\
    \multirow{3}{*}{\textbf{This work} \ \ \ $\Bigg\{$}                                 
      & \textbf{Exact Division}        & $\boldsymbol{2n}$ & $\boldsymbol{2n^2\log_2n}$ & $\boldsymbol{2n^2}$   & $\boldsymbol{8n\log_2n}$ \\
      & \textbf{Montgomery Reduction}  & $\boldsymbol{2n}$ & $\boldsymbol{5n^2}$        & $\boldsymbol{2n^2}$   & $\boldsymbol{14n}$ \\
      & \textbf{Barrett Reduction}     & $\boldsymbol{2n}$ & $\boldsymbol{5n^2}$        & $\boldsymbol{2n^2}$   & $\boldsymbol{14n}$ \\
  \hline\\[-9pt]
  \multicolumn{5}{l}{${}^\ddagger$Total gate counts and depths in arbitrarily-angled controlled-rotations.}
  \end{tabular}
\end{table}

\subsection{Outline of the Paper}

The remainder of the paper is organized into 6 sections and one appendix. 
In the next section (\sect{mod-mult}) we provide background required by all the circuit constructions described in this paper. We also use these constructions to describe the most common implementation of modular multiplication, building up the multiplier as a sequence of modular adders.
In \sect{div} we describe the first of our modular multiplication circuits; 
a circuit that uses a new implementation of the standard division technique. In \sect{montgomery} we describe a circuit that uses the Montgomery
reduction technique and in \sect{barrett} we describe the implementation of a modular multiplier based on Barrett reduction.
In \sect{fourier} we describe details of the implementation of our schemes that are specific to Fourier basis arithmetic.
In \sect{resources} we present a resource analysis of each of the new implementations that compares the resources required by our circuits to those required
by the base implementation constructed from modular adders. We perform the resource analysis for circuits utilizing a variety of adders including: carry-ripple, carry look-ahead and Fourier basis.
Finally in the appendix we describe details related to the integer adders and multiplier circuits used in the constructions, including several new implementations.

%% file: mod-mult.tex

\section{Quantum Modular Multiplication}
\label{sec:mod-mult}

In this section we describe the basic methods and circuits that are used to construct quantum modular 
multipliers. We also describe the standard circuit that is most often used for quantum modular multiplication,
which performs the modular multiplication using modular adders. This circuit will provide the baseline of comparison 
for our new methods.

Most quantum algorithms require ``quantum-classical'' modular multipliers, in which one multiplier input is a fixed classical parameter.  
The circuits introduced in this paper are therefore described in this form.  We will also describe the modifications
required to implement full ``quantum-quantum'' modular multipliers. These circuits would be useful, for example,
in the quantum circuit for elliptic curves~\cite{Proos:2003}.

In the quantum-classical case, our goal is described by the quantum operation,
\begin{equation}
  \eq{mod-mult:mod-mult}
  \ket[n]{y}  \lra  \ket[n]{Xy\bmod N},
\end{equation}
where $X$ is a classically-determined multiplicand, $n\defeq\clog[2]{N}$ is the number of bits required to hold the modulus $N$, and we have used subscripts to indicate the size of the requisite quantum registers (ignoring any intermediate ancilla qubits).  In the quantum-quantum case, $x$ is held in an additional quantum register, which is preserved through the operation:
\begin{equation}
  \ket[n]{x}\ket[n]{y}  \lra  \ket[n]{xy\bmod N}\ket[n]{y}.
\end{equation}
In general, we will reserve uppercase variable names for constant classical parameters.

\subsection{Modular Multiplication}

A non-modular quantum multiplier can be constructed using the techniques of grade-school arithmetic: we compute and accumulate a sequence of partial products.  In binary arithmetic, each partial product is computed by multiplying a single bit of the multiplier $y$ by the shifted multiplicand $2^kX$, so that the product $yX$ is computed by accumulating $\sum_k{y_k(2^kX)}$ for each bit $y_k$ of $y$.  A binary quantum multiplier therefore consists of a sequence of quantum additions, controlled by the bits of a quantum input \ket{y}.

A modular multiplier can be constructed by combining the steps of non-modular multiplication and integer division operators, where the modular product is returned as a remainder from the latter.  In general, an information-preserving divider will compute both quotient and remainder terms.  For reversible modular modular reduction, we therefore must take additional steps to uncompute the quotient.
Applying Bennett's technique for constructing reversible circuits from classical operations~\cite{Bennett1973}, in which we pseudo-copy the result into an ancilla register and reverse the entire computation, we implement,
\begin{alignat}{2}
  \ket{ 0}\ket{y}\ket{0} \lra 
    &&\ket{Xy}\ket{y}&\ket{0}  
  \nonumber\\ \lra
    &&\ket{q}\ket{Xy-qN}\ket{y}&\ket{0} 
  \nonumber\\ \lra
    &&\ket{q}\ket{Xy-qN}\ket{y}&\ket{Xy-qN} 
  \nonumber\\ \lra
    &&\ket{Xy}\ket{y}&\ket{Xy-qN}
  \nonumber\\ \lra
    &&\ket{ 0}\ket{y}&\ket{Xy-qN},
\end{alignat}
where $\ket{Xy-qN}=\ket{Xy\bmod N}$.  Unfortunately, the above computation is not very resource efficient.  In addition to requiring additional ancilla registers to store results during the uncomputation, both the multiplication and division components of the modular multiplier must be implemented twice: once to compute the result, and again in reverse to uncompute and clear the ancilla space.  

\begin{figure}
\centering
\inputtikz{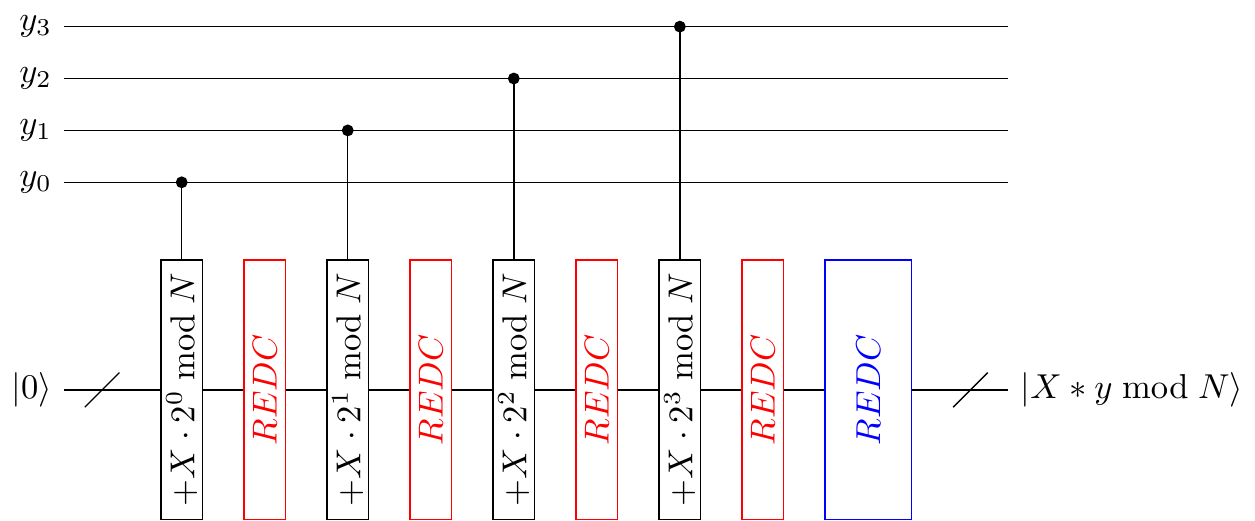}
\caption{A 4-bit quantum modular multiplier that multiplies a quantum register by a classical constant. 
We either reduce modulo $N$ after every addition (red blocks), or once for the entire multiplier (blue block).
In both cases the multiplier consists of $n$ conditional additions of $n$-bit constants.
}
\label{fig:quantum:mult:mult_class}
\end{figure}

Due to the complexity of reversible division-based modular reduction, early modular multipliers were instead constructed with reductions embedded within the multiplication, one after each addition of a partial product~\cite{Beckman1996, Vedral1996, Beauregard2003, VanMeter2005}.  If we classically reduce each partial product modulo $N$ prior to its addition, then at most one value of $N$ needs to be subtracted with each step.  We can therefore equivalently implement reversible circuits for modular addition, so that no garbage is left after any step of the multiplier.  The disadvantage of this
approach is that each reversible modular addition requires multiple integer adders, increasing the overhead of the modular multiplier linearly by this factor.  In \fig{quantum:mult:mult_class} we illustrate the two methods for modular multiplication: using modular adders, or using an
integer multiplication followed by a division step.

The circuits introduced in this paper derive from the first, division-style reduction strategy.   We are able to construct modular multipliers with a single integer multiplication step and a division step whose size and depth is logarithmic in the size of the inputs. 
While quantum circuits for modular multiplication have previously been developed employing this reduction style~\cite{Zalka1998,Kutin2006,Pavlidis2014}, 
the efficient reversible implementation of the modular reduction stage in this construction has remained a bottleneck.  
In the case of~\cite{Zalka1998,Kutin2006}, approximations are used that deterministically introduce errors for some inputs, which are argued to have an insignificant impact on
the error rate of the circuit. In~\cite{Pavlidis2014}, a precise division-style reduction circuit is introduced. However, the overhead of their circuit
is similar to the compute, copy, uncompute approach described above.

\subsection{In-place Modular Multiplication}
\label{sec:mod-mult:in-place}

The operation shown in \eq{mod-mult:mod-mult} is an example of an \emph{in-place} operation,
where one of the input registers is overwritten by the output.
To construct an in-place modular multiplier we can use the standard method that utilizes the reverse of
the inverse function. For example, the out-of-place quantum-classical modular multiplier performs the
transformation: $\ket{y}\ket{0} \lra{} \ket{y}\ket{X*y\bmod N}$. The inverse function with
the modular product as input performs the transformation: $\ket{X*y\bmod N}\ket{0} \lra{} \ket{X*y\bmod N}\ket{y}$.
Both these functions produce the same output and therefore we can chain them as:
\begin{equation}
\label{eqn:quantum:mult:inplace}
\ket{y}\ket{0} \lra[(*X)_{fwd}] \ket{y}\ket{Xy\bmod N}
\lra[\SWAP/] \ket{Xy\bmod N}\ket{y}
\lra[(*X^{-1})_{rev}] \ket{Xy\bmod N}\ket{0}.
\end{equation}
For the quantum-classical multiplier the inverse of the modular multiplication by $X$ is a modular
multiplication by $X^{-1}\bmod N$. Since $X$ is classical this inverse can be calculated offline using
the extended Euclidean algorithm. The in-place modular multiplier is then just two
out-of-place multipliers.

\subsection{Controlled Modular Multiplication}
\label{sec:mod-mult:control}

For most applications of the modular multiplier we will need to condition its operation on an
external control qubit. There are three main ways that this can be done. The first way is to add the
external control to all gates in the circuit. This adds significant overhead to the circuit and 
require every gate to share a common control qubit, either sequentializing the circuit or 
forcing a pseudo-copy of the control to a set of ancillas to allow gates to execute in parallel. 

The second technique is
to incorporate the global control into the individual controls already required by each adder in the multiplier. 
For each addition we can use a \Toffoli/ gate to combine the external control 
with its local control in the input register (i.e., bit $y_i$ from \fig{quantum:mult:mult_class}) onto an ancilla bit, and then use this bit to control the adder. 

A related and third way to control arbitrary quantum operations is to shift the input state into a subspace of the overall computational Hilbert space that is
returned to its initial state after the operation.  
This methodology generally requires an additional qubit register serving as an auxiliary ``quantum cache''~\cite{Zhou2011}; however, in the case of in-place quantum multiplication, 
we can bypass this requirement at the cost of $\ord{n}$ additional \Fredkin/~\cite{Fredkin1982} (controlled-\SWAP/) gates.  
For the out-of-place multiplier, if the quantum input register \ket{y} is zero, we expect the output of the multiplication by it to be zero regardless of $X$. 
The multiplications by $X$ and $X^{-1}$ will therefore have the same output.
Further, because we implement multiplication using controlled additions into an ancilla register, if \ket{y}=\ket{0} the value in the accumulation register is
left unchanged.  We can therefore swap the input into the accumulation register, so that we compute,
\begin{equation}
  \ket{y}\ket{0}  \lra[\SWAP/]
  \ket{0}\ket{y}  \lra[*X]
  \ket{0}\ket{y+0*X} = 
  \ket{0}\ket{y}.
\end{equation} 
The fact that the additions are done modulo $N$
does not matter because, for $y<N$, no reductions are ever required. 
If we then lift the subsequent $\SWAP/$ step of \eqn{quantum:mult:inplace}, 
the two out-of-place multiplications will cancel one another and \ket{y} will be returned at the end of the operation:
\begin{equation}
  \label{eqn:quantum:mult:swap-control}
  \ket{y}\ket{0} \lra[\SWAP/]
  \ket{0}\ket{y} \lra[(*X)_{fwd}] 
  \ket{0}\ket{y} \lra[(no\ \SWAP/)] 
  \ket{0}\ket{y} \lra[(*X^{-1})_{rev}] 
  \ket{0}\ket{y} \lra[\SWAP/]
  \ket{y}\ket{0}.
\end{equation}
In total, we require three sets of \Fredkin/ gates for this implementation of the controlled multiplier: two negatively-controlled sets to swap the input register into and 
out of the ``cache'' register, and an additional positively-controlled set to control the $\SWAP/$ at the center of \eqn{quantum:mult:inplace}.  
Each can be implemented with one \Toffoli/ gate and two \CNOT/ gates per bit.


\subsection{Modular Addition}
\label{sec:mod-mult:mod-add}

\begin{figure}
\centering
\inputtikz{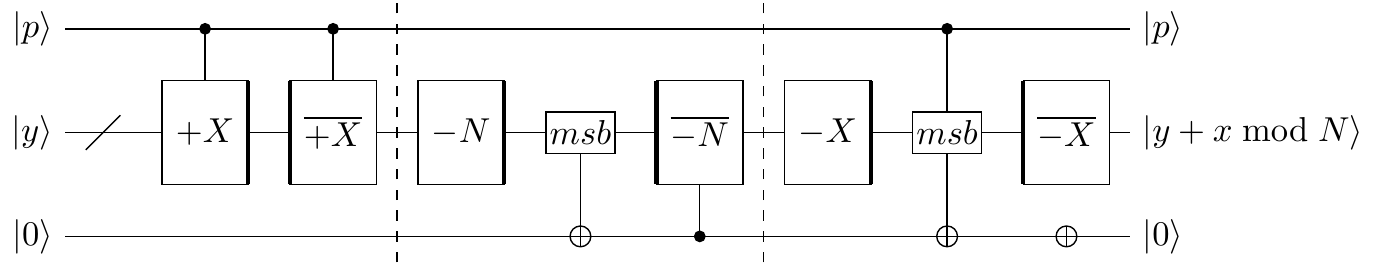}
\caption{A controlled Modular quantum-classical adder constructed from 3 integer adders. 
Adders with thick lines on the right side
indicate forward out-of-place additions and adders with thick lines on the left side are the corresponding
reverse functions. A pair of these two comprise a full in-place adder. The $msb$ function extracts the
most significant bit from the register. 
}
\label{fig:quantum:mult:mod_adder}
\end{figure}

There is an established method for constructing a reversible modular adder out of 
reversible integer adders. The method that we describe is a slight modification to those described 
in~\cite{Beckman1996}\cite{Fowler2004}. This method requires $3$ in-place integer adders for an
in-place modular quantum-classical or quantum-quantum addition.
A circuit for this adder is shown in \fig{quantum:mult:mod_adder}. 
The circuit is for a quantum-classical modular adder. To make this a full quantum-quantum adder
we would replace the additions involving $X$ with full quantum integer additions. We have broken each
integer adder into forward and reverse sections. An in-place adder requires both sections; therefore
the modular adder is comparable in complexity to three in-place adders. Note, this adder is controlled
by a single external qubit $\ket{p}$ as would be required if we use it as part of a multiplier.

Both input values to the modular adder are assumed to be reduced, i.e. $<N$.
The modular adder works by first adding $X$ in-place to the second quantum register containing $y$.
We then subtract $N$ out-of-place. If $X+y > N$, indicating that a reduction is necessary, then
the result of this subtraction will be positive, otherwise it is negative. For two's complement 
addition a negative result will have the most-significant-bit (msb) set and this can be used
as a control to the reverse segment of the subtraction to either overwrite the input or reverse
the action of the forward subtraction. However, we must copy the msb in order to control the reverse
subtraction and this produces a bit of garbage. Other than this bit of garbage, we have the result of
the modular addition in the $y$ register. To clear the garbage bit, we subtract $X$ from the result and
note that if we performed the reduction in the second step we now have $X+y-N-X=y-N<0$ in the register
and if we did not perform the reduction we have $X+y-X=y>0$ in the register. Therefore 
we can use the sign of this subtraction to clear the garbage bit. 
We then uncompute the subtraction of $X$ to complete the modular addition.

\subsection{Quantum-Quantum Modular Multiplication}
\label{sec:mod-mult:mod-shift}
\label{sec:mod-mult:quantum-quantum}

\begin{figure}
\centering
\inputtikz{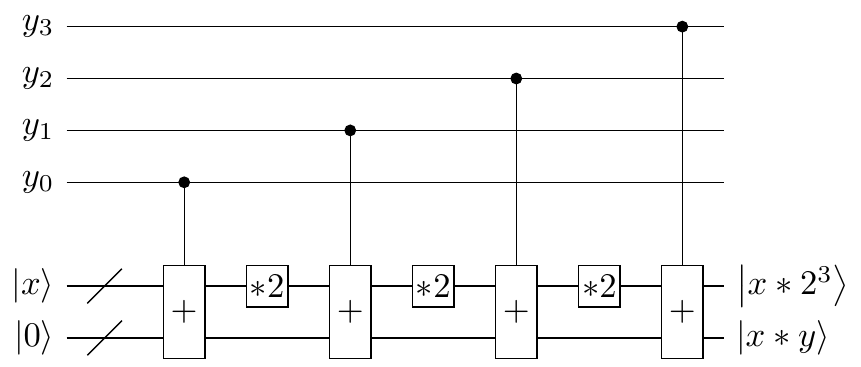}
\caption{A 4-bit quantum multiplier constructed from a sequence of controlled additions}
\label{fig:quantum:mult:mult}
\end{figure}

A full quantum-quantum multiplier (\fig{quantum:mult:mult}) computes the product \ket{xy}
from to quantum input values \ket{x} and \ket{y}. As in the case of the quantum-classical multiplier, 
the quantum-quantum multiplier consists of a sequence of controlled additions into an accumulation product register.
If we use modular adders in the quantum-quantum multipliers then the resulting product will be reduced
modulo $N$. Otherwise an additional reduction step will be required.

\begin{figure}
\centering
\inputtikz{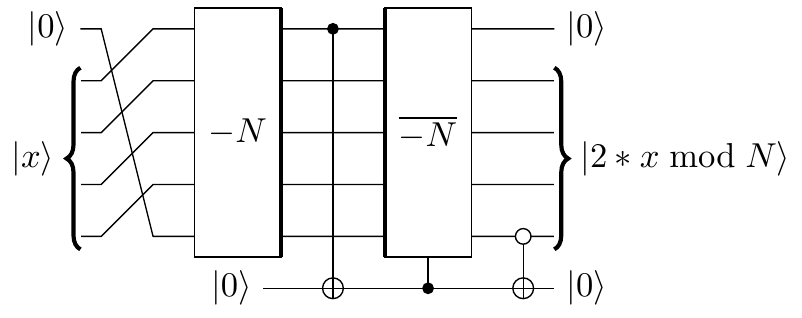}
\caption{Modular shift and reduce circuit. Shown here for a 4-qubit register.
The input value is shifted one position by relabeling the inputs and inserting a zero value at the LSB.
At most one reduction of $N$ is required, and after this reduction the most significant bit is cleared,
which replaces the input ancilla used at the LSB.
}
\label{fig:quantum:mult:mod_shift}
\end{figure}

As discussed previously, each reduced partial product is of the form $y_i(2^ix)\bmod N$. For the quantum-classical
modular multiplier we can calculate $2^iX\bmod N$ off-line; however, for the full
quantum-quantum multiplier $\ket{x}$ is a quantum value and therefore $2^ix\bmod N$ must be calculated with a quantum circuit. Each addition in the multiplier uses a value that is twice the previous value, therefore
we just need to shift the value by one position for each addition. A circuit that performs a shift and
reduce is shown in \fig{quantum:mult:mod_shift}. The circuit shifts the input forward by
one position by relabeling the input and inserting a zero at position 0.
Since this produces a value $<2N$ at most one reduction is required. We can perform this reduction by subtracting $N$ and 
checking to see if the result is negative. If it is, we reverse the subtraction, otherwise we complete
the in-place subtraction. We can clear the comparison bit by noting that since $N$ is always chosen
to be odd, and the pre-reduced shifted value is even, the output value is only odd when we have
done a reduction. Therefore, the least significant bit can be use to clear the comparison bit.

For the quantum-classical modular multiplier, we could classically pre-compute the modular inverse $X^{-1}\bmod N$ required for the reversed step.  In the quantum-quantum case, we instead require the inverse of a quantum value, requiring a reversible modular inversion routine.  Reversible inversion circuits employing the extended Euclidean algorithm have been demonstrated~\cite{Proos:2003}; however, their
cost is much higher than that of a single modular multiplication, and incorporating them is beyond the
scope of this paper. 
We will therefore describe and determine resources for only the out-of-place implementation of quantum-quantum modular multipliers.

%% file: division.tex

\section{Quantum Modular Multiplication with Division}
\label{sec:div}

Our first implementation of a quantum modular multiplier is the most straightforward: after an initial multiplication, we implement a reversible division operation comprising trial subtractions and controlled re-additions.  Standalone modular reduction being irreversible, we first define a quantum division operator,
\begin{equation}
  \eq{div:div}
  \ket[n+m]{t}  \lra[\QQ\DIV{N}]
  \ket[m]{t \bdiv N} \ket[n]{t \bmod N} =
  \ket[m]{q}         \ket[n]{t - qN},
\end{equation}
where \ket[n+m]{t} is the $(n+m)$-bit result of the initial multiplication, and $(\bdiv)$ indicates integer division such that $(t\bdiv N) = \flr{t/N} = q$ is the computed quotient.

Classically, modular reduction is constructed from a division operation by simply discarding the quotient and preserving the remainder.  In the reversible case, the quotient must instead be uncomputed, exacerbating the computational discrepancy between multiplication and division and sourcing the principal bottleneck in typical quantum implementations of modular multiplication.  Here, we utilize information present in both the input and output registers of the out-of-place modular multiplier in order to clear \ket[m]{q} while circumventing a full Bennett-style reverse computation.  The depth of the $\QQ\DIV{}$ operator is then poly-logarithmic in $n$, so that the out-of-place modular multiplication operation,
\begin{equation}
  \ket[n+m]{0}\ket[n]{x}  \lra
  \ket[n+m]{t}\ket[n]{x}  \lra[\QQ\DIV{N}]
  \ket[m]{q} \ket[n]{t \bmod N} \ket[n]{x}  \lra
  \ket[m]{0} \ket[n]{t \bmod N} \ket[n]{x}
\end{equation}
is asymptotically dominated by the initial computation of \ket{t}.

\subsection{Multiplication Stage}
\label{sec:div:mult}

We first must compute a state \ket{t} such that $t$ is congruent to the non-modular product $X*y$.  
For the purpose of modulo-$N$ multiplication, we can reduce partial products prior to their accumulation, replacing $(Xy)$ with,
\begin{equation}
  t \defeq  \sum_{k=0}^{n-1} y_k\qty(2^k X \bmod N),
\end{equation}
such that $t\equiv Xy\pmod{N}$ and is bound by $t<nN$.  We then require at most $n+m=\clog[2]{(Nn)}=n+\clog[2]{n}$ bits to hold \ket[n+m]{t}, so that the initial multiplication stage,
\begin{equation}
  \ket[n+m]{0}\ket[n]{y}  \lra[\QQ\MAC(X\mid N)]  \ket[n+m]{t}\ket[n]{y},
\end{equation}
consists of $n$ in-place, width-$(n+m)$ quantum adders, conditioned on the bits of \ket[n]{y}.

\subsection{Division Stage}
\label{sec:div:div}

Using \ket[n+m]{t} as the input to the quantum division operation, we require at most $m\defeq\clog[2]{n}$ bits for the quotient register \ket[m]{q}.  We compute the quotient bitwise, beginning with its MSB, $q_{m-1}$.
After unconditionally subtracting $2^{m-1}N$, the sign bit (MSB) of the register indicates whether the input $t<2^{m-1}N$.  
Using the sign to condition the in-place re-addition of $2^{m-1}N$ onto the remaining bits of the register, we return its modulo-$(2^{m-1}N)$ reduction.   We are left with the single sign bit indicating that the subtraction was undone, that is, that $q_{m-1}=0$.  After inverting the sign, we have therefore performed the operation,
\begin{equation}
  \ket[n+m]{t}  \lra
  \ket{q_{m-1}} \ket[n+m-1]{ t - q_{m-1}2^{m-1} N }  =
  \ket{q_{m-1}} \ket[n+m-1]{ t\bmod (2^{m-1}N) },
\end{equation}
where the resulting value in the accumulator is bound by $2^{m-1}N$, requiring only the $(n+m-1)$ LSBs of the register and allowing $q_{m-1}$ to be preserved in the MSB location.

\begin{figure}[ht]
    \begin{center}
    \includecircuit[8]{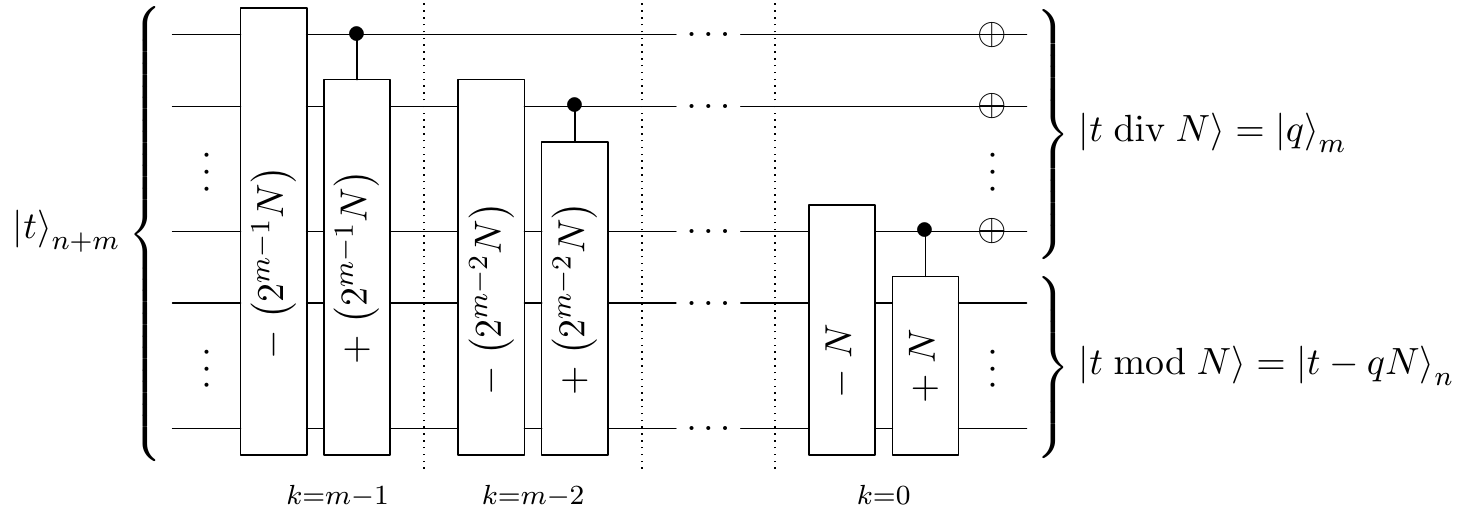}
    \end{center}
    \caption{Quantum division operation described in \sect{div:div}.  At each step $k$, we perform a trial subtraction and conditional re-addition of $2^{k}N$, computing one (inverted) bit of the quotient $q$ while reducing the input state modulo-$(2^kN)$.  The subtraction and re-addition of each stage can be merged into a single in-place quantum select-undo adder. (see \apx{adders:select-undo})}
    \label{fig:div:div}
\end{figure}

We iterate this process for $k=m-1,...,0$, each time reducing the register modulo-$2^{k}N$ and computing the bit $q_k$ of \ket[m]{q}.  After the final ($k=0$) reduction, we have constructed \ket[m]{q} over the $m$ MSBs of the input register, while leaving the remainder $\ket[n]{t\bmod N}$ in the remaining $n$ bits.  The complete \QQ\DIV{N} operation is shown as a quantum circuit in \fig{div:div}.

As described in \apx{adders:select-undo}, an in-place quantum adder generally comprises a pair of out-of-place adders.  In the select-undo quantum adder construction introduced in the appendix, control is efficiently added to the in-place adder through the second adder in the pair, which selectively undoes the addition performed by the primary adder.  In this structure, we require a control qubit only for the second adder.  We can use the select-undo adder to merge the trial subtractions and subsequent conditional re-additions of the division circuit, performing the trial subtractions out-of-place and using the sign (MSB) of the difference to conditionally undo the result.  As each subtraction by $2^kN$ affects only the bits more significant than $k$, the division operator comprises in total $m$ in-place, $n$-qubit select-undo adders.  Assuming logarithmic-depth prefix adders (\apx{circuits:prefix_adders}), the overall depth of this stage is $\ord{m\log n}=\ord{\log^2 n}$, with \ord{n\log n} total quantum gates.

Notably, the division stage constructed here assumes nothing of the initial multiplicands used to construct $t$, taking only the state \ket{t} and the classical modulus as inputs.  This stage is therefore independent of whether we are implementing a quantum-classical or quantum-quantum modular multiplier.

\subsection{Uncomputation Stage}
\label{sec:div:uncompute}

We now must uncompute the quotient register.  Unlike the division stage, this requires the incorporation of the individual multiplicands or partial products used in the calculation of $\ket{t}$.  However, we can avoid the complete reversal of the steps computing $\ket[m]{q}$ by utilizing information contained in both the input \ket[n]{y} and output \ket[n]{t\bmod N} registers.

Our strategy requires that we first multiply the quotient register by the modulus in-place:
\begin{equation}
  \ket[m]{q}    \lra[\QQ\MUL(N)]
  \ket[m]{qN},
\end{equation}
where modulo-$2^m$ reduction is implicit in the size of the $m$-bit register.
As described in \apx{appen:mult}, for odd $N$ we can reversibly multiply $q*N\pmod{2^m}$ in-place, with $m-1$ quantum adders of sizes $1,...,m-1$.  

We then add the $m$ LSBs of $\ket{t\bmod N}$ to the quotient register,
\begin{equation}
  \eq{div:uncompute:quotient}
  \ket[m]{qN}   \lra
  \ket[m]{qN + (t\bmod N)} = \ket[m]{qN + (t-qN)} = \ket[m]{t},
\end{equation}
leaving us with the result computed in the initial multiplication stage, truncated to its $m$ LSBs.  
We can clear $\ket{t}$ by a reverse of the multiplication stage, truncating all operations to their first $m$ bits.  

Though we now require only $m$-bit addition, our use of reduced partial products in computing $\ket{t}$ requires that we 
perform $n$ total additions each controlled by a bit of \ket[n]{y}.  Given logarithmic-depth quantum adders, the depth of this stage would then be \ord{n\log\log n}, dominating the \ord{\log^2n} depth of the \QQ\DIV{} operation.  
We therefore make use of the work bits necessary for the multiplication and division stage adders to parallelize the narrower adders required of the uncomputation stage.  

Dividing the \ord{n} work bits into $\ord{n/m}=\ord{n/\log n}$ separate accumulation registers, we can distribute and sum the $n$ addends in groups of $\ord{m}$.  The independent accumulators can then be combined with $\ord{\log(n/m)}=\ord{m}$ quantum-quantum adders in a binary tree structure in order to compute the complete sum $\ket[m]{t}$.  After using the result to clear the quotient register, we must uncompute the individual accumulators by reversing the parallelized adds.  The overall depth of this procedure is then $\ord{m\log m}=\ord{\log n\log\log n}$, no longer dominating the \ord{\log_2^2n}-depth division stage.

\begin{figure}[ht]
    \begin{center}
    \includecircuit[9]{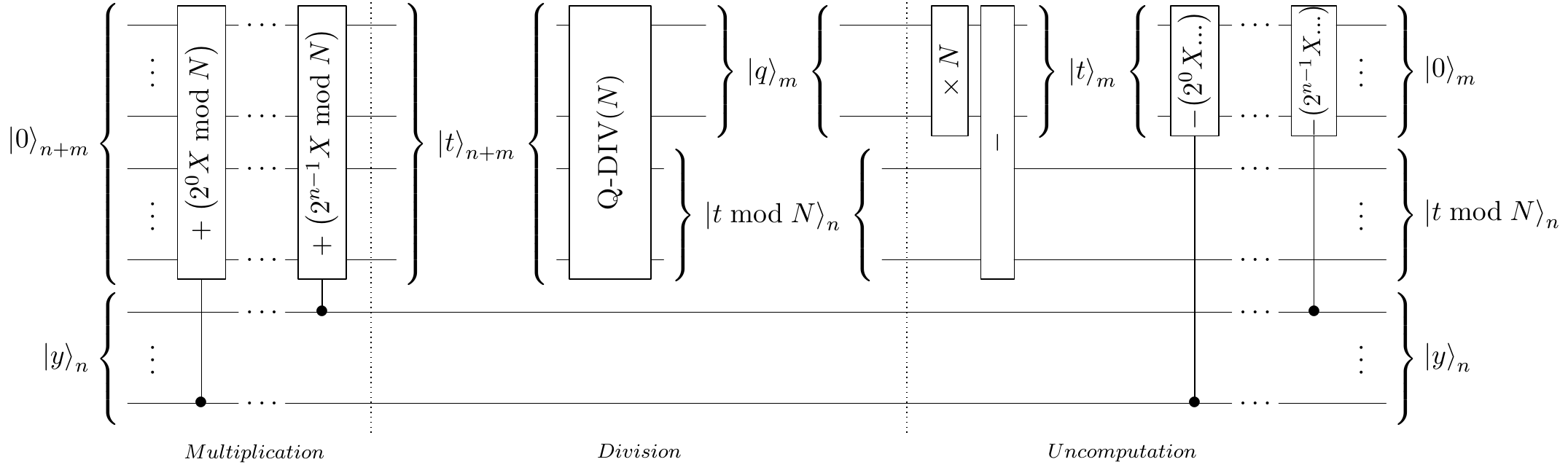}
    \end{center}
    \caption{Out-of-place modular multiplier constructed from the $\QQ\DIV{N}$ operation.  The final sequence of subtractions in the uncomputation stage is shown in series, but can be parallelized across work qubits in order to minimize the depth of this stage.}
    \label{fig:div:full}
\end{figure}

%% file: montgomery.tex

\section{Quantum Montgomery Multiplication}
\label{sec:montgomery}

\emph{Montgomery residue arithmetic}~\cite{Montgomery1985} is a technique for efficient multi-precision modular arithmetic, ubiquitous in applications such as hardware cryptography~\cite{Menezes1996}.  In the Montgomery framework, integers are mapped to a residue representation in which multiplication modulo $N$ requires calculations under a computationally friendly auxiliary radix $R$.  While the initial mapping of inputs to the Montgomery representation is not free, the constant overhead is quickly overcome by the efficiency of Montgomery multiplication when multiple calculations are required.

In our application, the advantage of moving to a power-of-two modulus is in flipping the reduction stage.  While the $\QQ\DIV{}$ procedure outlined in \sect{div} consists of quantum additions conditioned on the most significant bits of the product register, the corresponding Montgomery operator requires only the least significant bits.  This rearrangement has particularly profound advantages in the application of quantum Fourier-basis arithmetic.  As in \sect{div}, we are able to reduce the asymptotic complexity of quantum modular multiplication to that of a single non-modular multiplication.  
Remarkably, it turns out that the overhead incurred in mapping to and from the Montgomery representation can be relegated entirely to classical precomputation, making our construction immediately advantageous in terms of quantum overhead.

\subsection{\label{sec:montgomery:cl}Classical Montgomery Residue Arithmetic}

Montgomery residue arithmetic provides an efficient computational framework for extended modular arithmetic under a fixed modulus $N$.  Given an auxiliary radix $R=b^m$ such that $\gcd(b,N)=1$, we define the unique $N$-residue representation of an integer $x\in\mathbb{Z}_N$,
\begin{equation}
    x' \defeq xR \bmod N.
\end{equation}
The $N$-residues $\{ x' \mid 0\le x<N \}$ form a complete residue system such that $x$ represents the residue class containing $xR^{-1}\bmod N$.  Distributivity ensures the usual addition, subtraction, and negation operations behave normally in this representation:
\begin{equation}
    (x \pm y)'     \equiv    x' \pm y' \pmod N
\end{equation}
However, the $N$-residue representation of the product $xy$ is, 
\begin{equation}
  \qty(xy)'    \equiv \qty(xR)\qty(yR)R^{-1}    
               \equiv x'y'R^{-1} \pmod{N}.
\end{equation}
Multiplication in the $N$-residue representation therefore differs from the standard case, requiring the incorporation of the additional factor of $R^{-1}\pmod{N}$.  The new operation is known as the {Montgomery product}~\cite{Montgomery1985},
\begin{equation}
    \MonPro{x',y'}{N,R} \defeq  x'y'R^{-1}\bmod N.
\end{equation}

We can now perform modulo-$N$ arithmetic in the $N$-residue system.  After initially mapping input values to their respective residue representations, we proceed with calculations as in the standard case, but with each modulo-$N$ multiplication mapped to the corresponding Montgomery product.  The result of each operation acting on $N$-residues is then also an $N$-residue, and so extended arithmetic can be performed within the residue system without further conversion steps until we finally map results back to their standard $\mathbb{Z}_N$ representations.

\subsubsection{Montgomery Reduction}
\label{sec:montgomery:cl:redc}

Mirroring standard modular multiplication, the Montgomery product can be decomposed into a non-modular multiplication and a \emph{Montgomery reduction},
\begin{equation}
    \REDC{t}{N,b^m}   \defeq   t b^{-m}\bmod N,
\end{equation}
where $b^m=R$ is the chosen auxiliary radix.
Given $\gcd(N,b)=1$, \REDC{t}{N,b^m} provides a unique representation of $t\bmod N$.  It can be efficiently computed from $t<b^mN$ due to the identity,
\begin{equation}
    \eq{montgomery:cl:redc-equiv}
    t b^{-m}    \equiv    \qty(t-\U N)/b^m \pmod{N},  
\end{equation}
where,
\begin{equation}
    \eq{montgomery:cl:redc-U}
    \U    \defeq  t N^{-1}\bmod b^m,
\end{equation}
such that, for co-prime $N$ and $b$, $\U$ uniquely solves,
\begin{equation}
    \eq{montgomery:cl:redc-solves}
    t-\U N    \equiv  0 \pmod{b^m},
\end{equation}
ensuring that $(t-\U N)$ is divisible by $b^m$.\footnote{Note that in the above we have deviated slightly from the typical construction (for example, as presented in~\cite{Montgomery1985}), in which the estimate is bound by the range $[0,2N)$.  By rearranging signs slightly, we have shifted the bounds to the $(-N,N)$ range presented, which will serve to (very slightly) simplify the quantum construction we introduce below.}

The right hand side of \eq{montgomery:cl:redc-equiv} is bound by $-N<(t-\U N)/b^m<t/b^m$, such that its maximal value decreases toward zero for increasing $m$.  Taking $m \ge \clog[b]{(t/N)}$, this limit becomes,
\begin{equation}
    \eq{montgomery:cl:est-bound}
    -N \le (t-\U N)/b^m < N,
\end{equation} 
enabling the computation of Montgomery residue $t R^{-1}\equiv(t-\U N)/b^m\pmod{N}$ with a single comparison and corrective addition.  We refer to this term as the $m$-digit Montgomery estimate of $t$,
\begin{equation}
  \eq{montgomery:cl:est}
  \MonEst{t}{N,b^m} \defeq ( t - \U N )/b^m,
\end{equation}
and subdivide Montgomery reduction into independent \emph{estimation} and \emph{correction} stages (as in \alg{montgomery:cl:redc}).


Choosing $b=2$, the division by $b^m$ in \eq{montgomery:cl:est} becomes computationally trivial in binary architectures.  Due to \eq{montgomery:cl:redc-solves}, 
we can then compute the estimate while circumventing the explicit calculation of $\U$.  Re-expressing \eq{montgomery:cl:redc-solves} as a multiply-accumulate operation,
\begin{equation}
    \eq{montgomery:cl:redc-mac}
    t-\U N = t - \sum_{k=0}^{m-1}2^ku_kN \equiv 0 \pmod{2^m},
\end{equation}
we find that each subtraction of $2^ku_kN$ is the final operation affecting the $k$th bit of the accumulator, and therefore must clear that bit.  Beginning with the LSB $(k=0)$, we can therefore ignore $u_k$ and simply subtract $2^kN$ if bit $k$ of the accumulator needs to be cleared.  That is, each bit $u_k$ of $\U$ is equivalently the $k$th bit of the accumulator immediately prior to the corresponding conditional subtraction of $2^ku_kN$.

As each step ultimately clears a new bit of the accumulator, we can also right-shift the register by a single bit without loss of information.  As described in \alg{montgomery:cl:redc}, \alglines{montgomery:cl:redc}{est-start}{est-end}, after $m$ such iterations we have computed the Montgomery estimate $(t-\U N)/2^m$, requiring only the $m$ conditional subtractions necessary to compute $(-\U N)$ and $m$ computationally trivial right-shifts, and sidestepping the full-register comparison operations required of standard modular reduction.    Combined with a single modulo-$N$ correction (\alglines{montgomery:cl:redc}{mod-start}{mod-end}, in which we finally add $N$ if the estimate is negative), we have constructed the binary algorithm for Montgomery reduction presented in \alg{montgomery:cl:redc}.

\begin{algorithm}[h!]
\caption{Classical Montgomery reduction algorithm, \REDC{t}{N,2^m}}
\label{alg:montgomery:cl:redc}
\begin{algorithmic}[1]
    \Require{Modulus $N$, integers $t$, $m$ s.t. $t < N2^m$}
    \Ensure{$S = t2^{-m}\bmod N$, $\U= -t N^{-1}\bmod2^m$}
    \State {$S \gets t$}    
    \For   {$k\gets 0$ to $m-1$}             \label{alg-line:montgomery:cl:redc:est-start}
    \Comment {Estimation stage}
    \State {$u_k\gets S\bmod 2$}             \label{alg-line:montgomery:cl:redc:LSB}
    \State {$S \gets S - u_k\cdot N$}        \label{alg-line:montgomery:cl:redc:add}
    \State {$S \gets S / 2$}                 \label{alg-line:montgomery:cl:redc:shift}
    \EndFor                                  \label{alg-line:montgomery:cl:redc:est-end}
    \Statex{}
    \If    {$S < 0$}                       \label{alg-line:montgomery:cl:redc:mod-start}
    \Comment {Correction stage}
    \State {$S \gets S+N$}
    \EndIf                                   \label{alg-line:montgomery:cl:redc:mod-end}
    \Statex
    \State \Return {$S$}
\end{algorithmic}
\end{algorithm}


\newcommand{\yinvb}{\ensuremath{\bar{Y}{}}}

\subsection{Quantum Montgomery Reduction}
\label{sec:montgomery:qu:redc}

Given the initial construction of \ket[n+m]{t} outlined in \sect{div:mult}, we now introduce a reversible Montgomery reduction operator in imitation of the \QQ\DIV{} operator defined in \eq{div:div},
\begin{equation}
  \eq{montgomery:qu:redc}
  \ket{0} \ket[n+m]{t}    \lra[\QQ\REDC{N,2^m}{}] 
  \ket[n]{ t2^{-m} \bmod N } \ket[m+1]{ t N^{-1} \bmod 2^{m+1} },
\end{equation}
While the Montgomery reduction $(t2^{-m}\bmod N)$ is not unique for an unknown $t\ge N$, the \QQ\REDC{N,2^m}{} operation is bijective iff $t<2^{m+1}N$ and $\gcd(N,2) = 1$ by the Chinese remainder theorem.

As in the classical procedure, we split the quantum Montgomery reduction operation into distinct \emph{estimation} and \emph{correction} stages.  Mirroring \sect{div}, we will then couple the \QQ\REDC{}{} operator the standard initial multiplication stage (computing \ket{t}) and a final \emph{uncomputation} stage (clearing \ket{t M^{-1}\bmod2^{m+1}}) in order to construct a quantum Montgomery multiplication operator, \QQ\MonPro{}{}.

\subsubsection{Estimation Stage}
\label{sec:montgomery:qu:redc-est}

We first compute the Montgomery estimate, $(t-\U N)/2^m$.  Because this alone is not a unique representation of $t$, we preserve the $m$ bits of $\U$ that naturally fall out of the classical estimation procedure, arriving at the bijective mapping,
\begin{equation}
  \eq{montgomery:qu:redc-est}
  \ket{0}  \ket[n+m]{t}    
  {}\lra{}   
  \ket[n+1]{ (t - \U N)/2^m }  \ket[m]{  \U  }.
\end{equation}
By \eq{montgomery:cl:est-bound}, we require $n+1$ bits to represent the estimate state $\ket{(t - \U N)/2^m}_{n+1}$, necessitating a single ancillary bit in addition to the $n+m$ bits holding $t<nN$. 

We proceed as in \alg{montgomery:cl:redc}.  Prior to the $k$th iteration of the estimation stage, the LSB of the accumulation register is equivalently the $k$th bit of $\U$, or $u_k$ (\alg{montgomery:cl:redc}, \algline{montgomery:cl:redc}{LSB}).  Classically, $u_k$ is then used to condition the subtraction of $N$, so as to clear the LSB of the accumulator (\algline{montgomery:cl:redc}{add}).  This represents a two-to-one operation, necessitating the creation of a garbage bit in the reversible case.
However, after each iteration, we know also that the newly cleared LSB will be immediately shifted out (\algline{montgomery:cl:redc}{shift}); we can therefore consider only the effect of subtraction on the remaining bits of the register.   The subtraction occurs only when the accumulator is odd (and $N$ is odd by design), so no borrow will be generated by the first bit of the subtraction.  It is then equivalent to subtract $\flr{N/2}=(N-1)/2$ from the truncated register, conditioned on the LSB $\ket{s_0} = \ket{u_k}$:
\begin{equation}
  \eq{montgomery:qu:redc-est-step}
  \ket[w]{S}    
  =    \ket[w-1]{s_{w-1}...s_1} \ket{s_0}
  =    \ket[w-1]{\flr{S/2}} \ket{u_k}
  \lra \ket[w-1]{\flr{S/2} - u_k\cdot\flr{N/2}} \ket{ u_k }.
\end{equation}

\begin{figure}[H]
    \centering
    \includecircuit[8]{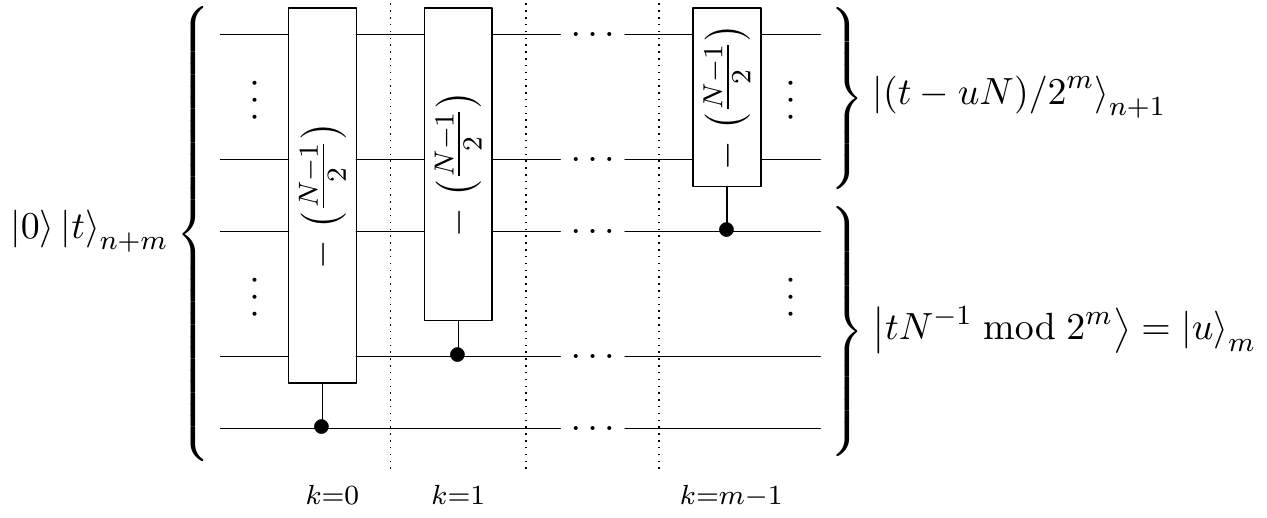}
    \caption{Estimation stage of the quantum Montgomery reduction algorithm (\sect{montgomery:qu:redc-est}), \QQ\MonEst{ N,2^m}{}.  Note the parallel to the division-based procedure (\fig{div:div}); where the latter computes the quotient $q$ with trial subtractions and re-additions conditioned on the MSBs of the accumulation register, Montgomery reduction allows for the computation of $\U$ from the LSBs of the register.}
    \label{fig:montgomery:qu:redc-est}
\end{figure}

In this way, we avoid accumulating new ancilla bits with each reduction step.  Iterating through $k=0,...,m-1$, we compute the Montgomery estimate $(t - \U N)/2^m$ with $n$ controlled subtractions.  The garbage information created in this sequence is simply the $m$ bits of $\U$, which are computed in the place of the $m$ least significant bits of the input state \ket[n+m]{t}.  As shown in \fig{montgomery:qu:redc-est}, the sequence of in-place subtractions mirrors that of the division case (\sect{div}, \fig{div:div}), but with adders controlled by the least significant bits of the product register.  Both reduction procedures have the same depth (in quantum adders)--while Montgomery reduction sidesteps the trial subtractions required in the $\QQ\DIV{}$ operation, the implementation of controlled, in-place quantum adders requires a pair of out-of-place adders identical to the paired adders of the division operation.

\subsubsection{Correction Stage}
\label{sec:montgomery:qu:redc-cor}

The second piece of the classical Montgomery reduction algorithm is a single modulo-$N$ correction (\alg{montgomery:cl:redc}, \alglines{montgomery:cl:redc}{mod-start}{mod-end}).  For $-N < (t-\U N)/2^m < N$, this correction requires a single controlled addition of $N$ conditioned on the sign of the estimate.  Labeling the sign bit \ket{s_\pm}, we perform,
\begin{equation}
    \ket[n+1]{(t-\U N)/2^m}    \lra    \ket{s_\pm} \ket[n]{ (t-\U N)/2^m + s_\pm \cdot N},
\end{equation}
where by \eq{montgomery:cl:redc-equiv} the final term is equivalently $\ket[n]{t2^{-m}\bmod N}$.

\begin{figure}[H]
    \centering
    \includecircuit[5]{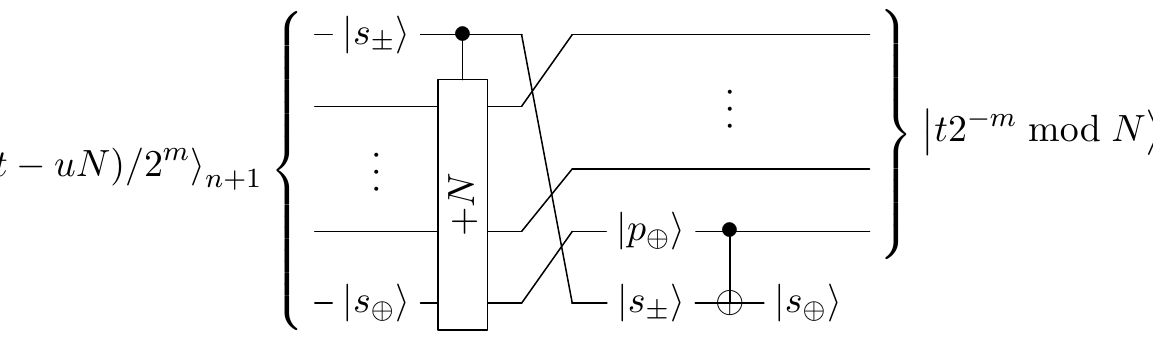}
    \caption{Quantum circuit demonstrating the correction stage of a quantum Montgomery reduction algorithm (\sect{montgomery:qu:redc-cor}), assuming $-N<(t-\U N)/2^m<N$.} 
    \label{fig:montgomery:qu:redc-cor}
\end{figure}

We are now left with the sign bit $\ket{s_\pm}$ as garbage.  For odd $N$, the conditional addition of $N$ must change the LSB of the register.  Defining the LSBs before and after the modular reduction,
\begin{align}
  p_\oplus    &\defeq     \qty( t2^{-m}\bmod N )   \bmod 2,\\
  s_\oplus    &\defeq     \qty( t - \U N )/2^m     \bmod 2,
\end{align}
we therefore find $s_\pm \oplus p_\oplus = s_\oplus$. By negating the sign bit $\ket{s_\pm}$, conditional on $\ket{p_\oplus}$, with a single \CNOT/ gate (as in \fig{montgomery:qu:redc-cor}), we return it to the pre-addition LSB $\ket{s_\oplus}$.  

We can then re-express,
\begin{equation}
  s_\oplus    \equiv    \qty(t - \U N) /{2^m}      
              \equiv    (\UU - \U) / 2^m \pmod{2}.
\end{equation}
where,
\begin{equation}
    \eq{montgomery:qu:redc-UU}
    \UU     \defeq   t N^{-1} \bmod 2^{m+1},
\end{equation}
descends from an $(m+1)$-bit Montgomery estimate $(t-\UU N)/2^{m+1}$.  In this form, \ket{s_\oplus} can be concatenated with \ket[m]{\U} and equivalently described,  
\begin{equation}
  \ket{s_\oplus} \ket[m]{\U}   =   
  \ket[m+1]{2^ms_\oplus +\U}   =   
  \ket[m+1]{ \UU }  = \ket[m+1]{t N^{-1}\bmod 2^m},
\end{equation}
completing the \QQ\REDC{N,2^m}{} operation introduced in \eq{montgomery:qu:redc}.

\subsection{Uncomputation}
\label{sec:montgomery:uncompute}

In order to construct a quantum Montgomery multiplier from the $\QQ\REDC{N,2^m}{}$ operator, we must finally uncompute the auxiliary output state \ket[m+1]{\UU}, mirroring the uncomputation of the quotient in \sect{div}.
Given the partial products composing $t$, we can express $\UU$,
\begin{equation}
    t N^{-1} \equiv  \sum_{k=0}^{n-1} y_k \qty( 2^k X \bmod N )N^{-1} \pmod{2^{m+1}}.
\end{equation}
We can therefore classically precompute the classical addends $((2^kX\bmod N)N^{-1}\bmod 2^{m+1})$ for $k=0,...,n-1$, and use $n$ controlled $(m+1)$-bit subtractions, conditioned on the bits of \ket[n]{y}, to clear the register.  Identically to \sect{div:uncompute}, the narrow register enables parallelization of the quantum adders to an overall depth of \ord{\log_2^2n}.

\begin{figure}[ht]
    \begin{center}
    \includecircuit[9]{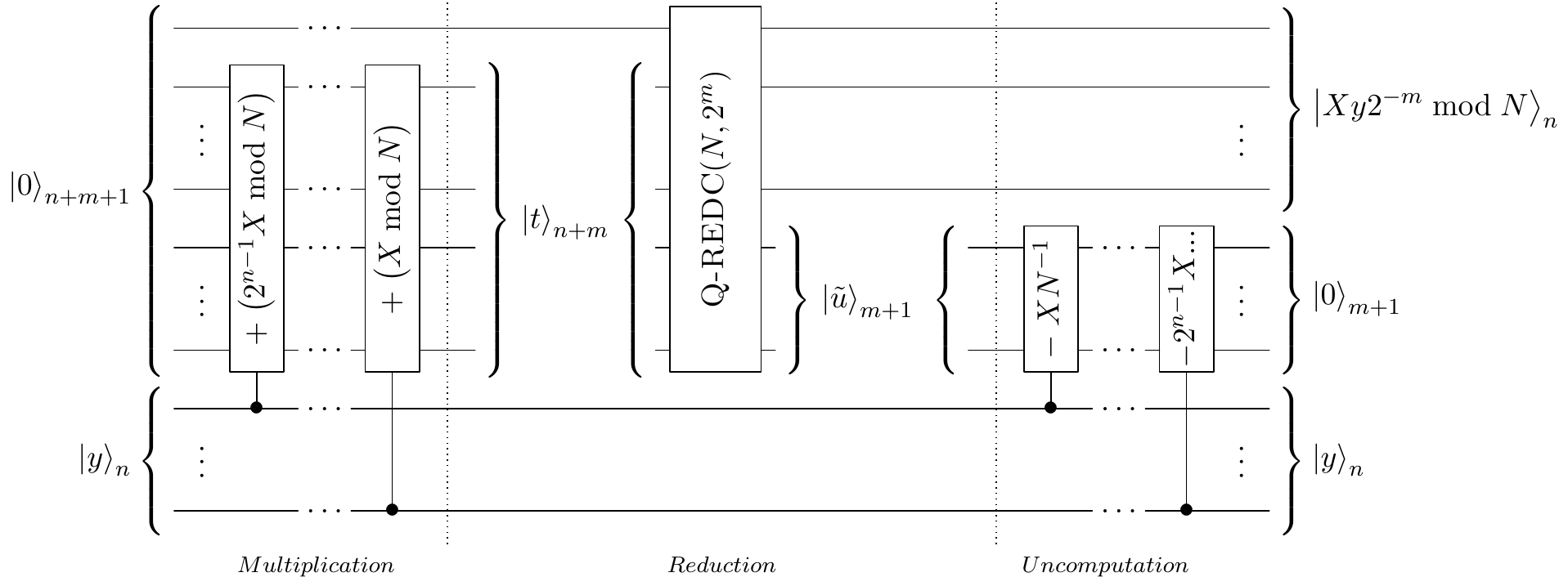}
    \end{center}
    \caption{Out-of-place quantum Montgomery multiplier \QQ\MonPro{t}{N,2^m}, comprising the initial computation of \ket[n+m]{t}, estimation ({\sect{montgomery:qu:redc-est}}) and correction (\sect{montgomery:qu:redc-cor}) stages of \QQ\REDC{N,2^m}{}, and a final uncomputation of \ket[n]{\UU}.}
    \label{fig:montgomery:qu:monpro}
\end{figure}

\subsection{Modular Multiplication via Quantum Montgomery Reduction}
\label{sec:montgomery:mult}

In the case of quantum-classical multiplication, we can use the quantum Montgomery multiplier developed here 
to compute a product where the quantum register remains in the standard representation.
Observing the Montgomery product of an $N$-residue $X'$ and a value $y$ in standard representation,
\begin{equation}
  \MonPro{X',y}{N,R}  =  (XR\bmod N)yR^{-1}\bmod N  =  Xy\bmod N,
\end{equation}
we find the modular product $Xy\bmod N$ returned in standard representation.  Classically, this would represent a disadvantage for most operations: we now need to remap the result to an $N$-residue representation for subsequent calculations.  In the quantum-classical case, however, we can precompute the residue of the classical multiplicand off-line, and use it to compute the residue,
\begin{equation}
  \eq{montgomery:mult:prod}
  t'  \defeq{}  \sum_{k=0}^{n-1} y_k (2^kX'\bmod N),
\end{equation}
from \ket{y} identically to the usual product \ket[n+m]{t}.  Applying \QQ\REDC{N,2^m}{} to \ket[n+m]{t'}, we compute the modular product \ket[n]{Xy\bmod N} in standard representation while relegating conversion overhead of Montgomery multiplication to classical precalculation.

\subsubsection{Montgomery Multiplication with Two Quantum Inputs}
\label{sec:montgomery:all-quantum}

We can also define a quantum Montgomery multiplier taking as input two $n$-bit quantum values.  This requires adapting the initial multiply so that it accepts two quantum inputs, from which we can compute \ket[n+m]{t} with $n$ quantum shift-and-reduce operations, as defined in \sect{mod-mult}.  
The \QQ\REDC{N,2^m}{} algorithm described in \sect{montgomery:qu:redc} acts only on \ket[n+m]{t} and is independent of the individual multiplicands, and therefore goes unchanged.  However, the final uncomputation of \ket[m+1]{\UU} requires some modification: we can no longer precompute the addends $\{(2^kx\bmod N)N^{-1}\bmod2^{m+1},k=0,...,n\}$.  Instead, noting \eq{montgomery:cl:redc-solves}, we perform an in-place multiplication of \ket[m+1]{\UU} by the classical value $N$ (requiring $m = \clog[2]{n}$ adders).  We are then left with the truncated product \ket[m+1]{t}, which we can clear by reversing the $n$ adders of the initial multiplication (but truncated to $m+1$ bits).  The reversed sequence will further undo the shift of \ket[n]{x} in the initial multiply:
\begin{alignat}{2}
    \ket[n+m+1]{0}\ket[n]{x}\ket[n]{y} 
      & \lraover[\QQ\REDC{N,2^m}{}]{\QQ\MAC}
        \ket[1]{0} \ket[n+m]{t}&& \ket[n]{2^nx\bmod N} \ket[n]{y} \nonumber\\
      & \lraover{\QQ\REDC{N,2^m}{}}
        \ket[n]{xy2^{-m}\bmod N}  \ket[m+1]{\UU}&& \ket[n]{2^nx\bmod N}            \ket[n]{y}  \nonumber\\
      & \lraover[\QQ\REDC{N,2^m}{}]{\QQ\MUL(N)}
        \ket[n]{xy2^{-m}\bmod N}  \ket[m+1]{t}&& \ket[n]{2^nx\bmod N} \ket[n]{y} \nonumber\\
      & \lraover[\QQ\REDC{N,2^m}{}]{\QQ\MAC^\dagger}
        \ket[n]{xy2^{-m}\bmod N}  \ket[m+1]{0}&& \ket[n]{2^nx\bmod N} \ket[n]{y}.
\end{alignat}

%% file: barrett.tex

\section{Quantum Barrett Multiplication}
\label{sec:barrett}

For classical modular arithmetic the Barrett reduction~\cite{Barrett1987} is beneficial because, for repeated
reductions using the same modulus, it reduces the number of divisions by that modulus and replaces them with
simpler multiply and shift operations. The standard method to reduce the number $t$ modulo $N$ would be to
calculate $q = \lfloor t/N\rfloor$ and then calculate the reduced value as: $t -qN$. The main idea behind the
Barrett reduction is to calculate a fixed-point fractional factor representing the division by the modulus
and then use this factor many times. The only divisions involving $t$ can be picked to be constant powers of
the machine word size and therefore can be implemented with shifts. Because of limited precision of the 
fixed-point factor, the Barrett reduction may not reduce $t$ completely to a value $<N$. However, we can
set the precision of the factor so that at most one additional subtraction by $N$ is required. Because of this
we can view the Barrett reduction as an approximate reduction technique.

With our quantum circuit we implement all operations on qubits using binary fixed-point logic.
Therefore, shift operations by a power of $2$ just redefine the fixed-point in our calculation. 
Additionally, the fixed-point fractional factor is a fixed classical value and can be 
pre-calculated offline. The quantum circuit to calculate $q$ is reduced to a quantum multiplication by
a classical value. As is the case for the classical Barrett reduction, the quantum circuit calculates an
approximate reduction factor, and we will want to carefully set the width of each of the individual operations
in the reduction to bound the error and reduce the total number of gates required. Completing the reduction 
is one of the challenges in constructing a reversible circuit for it, and is one of the main differences between
our implementation and that from~\cite{Zalka1998}. We show how to perform a complete reduction, 
whereas in~\cite{Zalka1998} they allow the case of a partial reduction and argue that doing so has an
insignificant impact on the fidelity of the circuit.

\subsection{The Barrett Reduction}
\label{sec:barrett:classical}

The Barrett reduction~\cite{Barrett1987} of an arbitrary number $t$ is defined as,
\begin{equation}
\REDC{t}{N} = t - \tilde{q}N\bmod N,
\end{equation}
where 
\begin{equation}
\label{eq:barrett:red:qhat}
\tilde{q} = \left\lfloor\left\lfloor\frac{t}{b^{k-1}}\right\rfloor \frac{\mu}{b^{k+1}}\right\rfloor
\quad\mu = \left\lfloor\frac{b^{2k}}{N}\right\rfloor
\end{equation}
and the $\bmod N$ involves at most one subtraction of $N$ because the parameters $b$ and $k$ can be picked
to ensure that $REDC(t) < 2N$. Typically $b$ is picked to be a small power of $2$ and $k$ must be picked
so that $b^{k-1} \leq N < b^{k}$. 
The only operation involving the value $t$ is a multiplication. If $b$ is picked to be a power of two then
the factors of $b^{k-1}$ and $b^{k+1}$ appearing in the denominators can be implemented as binary shifts, via
redefinition of the fixed-point binary number.

For our multiplier the value to be reduced is $t = Xy$. Further, we pick $k$ based on the value $N$, and we
pick $b=2$, therefore $k$ is just the bit-width of $N$, which we will denote as $n$. 
The value $Xy$ is calculated as the sum over $i$ as: $\sum{y_i (2^i X\bmod N})$. Because the shifted value 
is reduced before adding to the running sum, the total number of bits required for the product is $n+\log_2(n)$. 
In the calculation of
$\tilde{q}$ we only use the most significant bits of $Xy$ and $\mu$. To understand the impact of truncation it
it useful to look at the full-precision quotient $q$, defined as:
\begin{equation}
q = \left\lfloor\frac{X y}{2^{n-1}} \nu\right\rfloor
\quad\nu = \frac{1}{2^{n+1}} \left(\frac{2^{2n}}{N}\right),
\end{equation}
where $\nu$ corresponds to the shifted $\mu$ and because $2^{n-1} < N < 2^{n}$ we have $1/2 < \nu < 1$.
We can now write $q$ to separate the calculated values from the truncated ones,
\begin{equation}
q = \left\lfloor\frac{\widetilde{Xy}+(Xy)_t}{2^{n-1}} (\tilde{\nu}+\nu_t)\right\rfloor,
\end{equation}
where $\tilde{a}$ denotes the retained approximate value, and $a_t$ is the truncated portion of a value.
Separating the computed terms from the truncated ones we have:
\begin{equation}
q = \left\lfloor\frac{\widetilde{Xy}}{2^{n-1}} \tilde{\nu} + \frac{(Xy)_t \nu}{2^{n-1}} + \frac{\widetilde{Xy} \nu_t}{2^{n-1}}\right\rfloor =
\left\lfloor\tilde{q} + \frac{(Xy)_t \nu}{2^{n-1}} + \frac{\widetilde{Xy} \nu_t}{2^{n-1}}\right\rfloor.
\end{equation}
If we bound the truncated terms to be less than $2$ then only a single extra adjustment will be required.
We could use the upper bits of $Xy$ to calculate $\tilde{q}$,
however, it will be useful for our quantum implementation to have an independent $\widetilde{Xy}$ therefore we would like to 
minimize the width of this calculation. If we use the $n_k$ upper bits of each term in the sum then we can bound the first
truncated term as:
\begin{equation}
\label{eqn:barrett:classical:xyt}
\frac{(Xy)_t \nu}{2^{n-1}} < n \frac{2^{n_t}1}{2^{n-1}} = \frac{2^{\log_2(n_t)} 2^{n_t}}{2^{n-1}}.
\end{equation}
Where $n_t = n - n_k$ is the number of bits truncated from each term, and we have taken the upper bound of $\nu = 1$. 
If we pick $n_t$ such that: $\log_2(n)+n_t < n-1$ then the error term will be $<1$. This implies that each term must be
$n_k = \log_2(n) + 1$ bits, and the total approximate sum of $n$ values will require $2\log_2(n)+1$ bits.

For the second truncated term, if we use $n_v$ bits for $\nu$ we can bound this term as,
\begin{equation}
\frac{\widetilde{Xy} \nu_t}{2^{n-1}} < \frac{n 2^n 2^{-n_v}}{2^{n-1}} < \frac{2^{\log_2(n)+n-n_v}}{2^{n-1}}.
\end{equation}
We then need to pick $n_v > \log_2(n)+1$ to ensure that this term is less than $1$. The resulting bit widths for $\widetilde{Xy}$
and $\nu$ result in an $2\log_2(n)+2$ by $\log_2(n)+1$ bit multiplication to calculate $\tilde{q}$. 
We can truncate the $\widetilde{Xy}$ values used to calculate $\tilde{q}$ to $\log_2(n)+1$ bits and therefore we need a register of
length $2\log_2(n)+2$ to hold $\tilde{q}$.

\subsection{Quantum Modular Multiplier with Barrett Reduction}

In \alg{barrett:quantum:circuit} we describe the algorithm to compute the modular product of two numbers using the Barrett
reduction described in the previous section. A quantum circuit to calculate the out-of-place modular product of either a quantum register
with a classical constant or two quantum registers can be constructed directly from this algorithm. In the following discussion we
will just describe the case of a quantum product with a classical constant. We will return to the implications of a full quantum
multiply in \sect{barrett:quantum-quantum}.

The Barrett multiplier uses one input register, one output register, two work registers, and one single
qubit flag. The out-of-place multiplier performs the following operation:
\begin{equation}
\label{eqn:barrett:quantum:start}
  \ket[n]{y}
  \ket[n+m]{0}
  \ket[2m]{0}
  \ket[2m]{0}
  \ket[1]{0}
\lra{}\ket{y}\ket{yX\bmod N}\ket{0}\ket{0}\ket{0}.
\quad (m = \log_2(n)+1)
\end{equation}
In Steps~\ref{alg-line:barrett:quantum:xy} and~\ref{alg-line:barrett:quantum:xyapp} of the algorithm we calculate the full and approximate products and
produce the state
\begin{equation}
\ket{y}
\ket{Xy}
\ket{\smash{\widetilde{Xy}}}
\ket{0}
\ket{0}.
\end{equation}
The full product $(Xy)$ requires $\ord{n^2}$ basic gates and constitutes the majority of the operations in the entire circuit.
Calculating an approximate product eliminates the need to re-compute the full product later when we need to clear the
reduction factor $\tilde{q}$. 
In Steps~\ref{alg-line:barrett:quantum:qhat} and~\ref{alg-line:barrett:quantum:reduce} of the algorithm we calculate the approximate reduction factor
and use it to reduce the full product in-place producing the state:
\begin{equation}
\ket{y}
\ket{Xy-\tilde{q}N}
\ket{\smash{\widetilde{Xy}}}
\ket{\tilde{q}}
\ket{0}.
\end{equation}
As discussed above, the state reduced using $\tilde{q}$ may be greater than $N$ and therefore one more reduction by $N$ may be
required. We perform this reduction, however, doing so results in a one-bit flag indicating whether the additional
reduction was required. At this point we have the following state:
\begin{equation}
\ket{y}
\ket{Xy\bmod N}
\ket{\smash{\widetilde{Xy}}}
\ket{\tilde{q}}
\ket{adj}.
\end{equation}
and we have the reduced product, however, we have two registers with garbage and the one-bit adjustment flag that
need to be cleared. The two registers, containing $\widetilde{Xy}$ and $\tilde{q}$, can be cleared simply by reversing
steps~\ref{alg-line:barrett:quantum:xyapp} and~\ref{alg-line:barrett:quantum:qhat} of the algorithm, however the adjustment bit must be cleared
in some other way. If we add back $\tilde{q} N$ to $P$ then this register contains $\ket{Xy-adjN}$. If we subtract 
$\widetilde{Xy}$ from this register we obtain $\ket{(Xy)_t - adjN}$. But in \eqn{barrett:classical:xyt} we bounded the truncation
term $(Xy)_t < 2^{n-1} < N$ and therefore if $(Xy)_t - adjN < 0$ this indicates that an adjustment has occurred. This fact can be used to
clear the adjustment bit.
We also note that we only need to compute the high-order bits for the addition done in step~\ref{alg-line:barrett:quantum:addqn} of the algorithm.

\begin{algorithm}
\caption{\BarPro{X,y}{N}}
\label{alg:barrett:quantum:circuit}
\begin{algorithmic}[1]
\Require{Modulus $N$, integers $X,y < N$}
\Ensure{Integer $P = yX\bmod N$}
\State {$S \gets Xy$} 
\Comment{calculate the full product} \label{alg-line:barrett:quantum:xy}
\State {$\widetilde{Xy} \gets app(Xy)$} 
\Comment{calculate an approximate product} \label{alg-line:barrett:quantum:xyapp}
\State {$\tilde{q} \gets \widetilde{Xy} \tilde{\nu}$} 
\Comment{calculate the approximate reduction factor} \label{alg-line:barrett:quantum:qhat}
\State {$S \gets S - \tilde{q}N$} 
\Comment{reduce S s.t. $S<2N$} \label{alg-line:barrett:quantum:reduce}
\If {$S \ge N$}\label{alg-line:barrett:quantum:compare}
\State {$S \gets S - N$} 
\Comment{reduce by N if required}
\State {$adj \gets 1$}   
\Comment{reduction produces one bit of garbage}
\EndIf
\State{$S \gets S + \tilde{q}N$} 
\Comment{$S = Xy - adjN$} \label{alg-line:barrett:quantum:addqn}
\If {$S_{[n+\log_2(n):n-\log_2(n)]} - \widetilde{Xy} < 0$}\label{alg-line:barrett:quantum:appcompare}
\State {$adj \gets adj \oplus 1$} 
\Comment{clear adjustment flag}
\EndIf
\State {$S \gets S - \tilde{q}N$} 
\Comment{reset to modular product} \label{alg-line:barrett:quantum:subqn}
\State {$\tilde{q} \gets 0$} 
\Comment{reverse~\ref{alg-line:barrett:quantum:qhat} to clear $\tilde{q}$} \label{alg-line:barrett:quantum:qhat-rev}
\State {$\widetilde{Xy} \gets 0$} 
\Comment{reverse~\ref{alg-line:barrett:quantum:xyapp} to clear $\widetilde{Xy}$} \label{alg-line:barrett:quantum:xyapp-rev}
\end{algorithmic}
\end{algorithm}

From \eqn{barrett:quantum:start} we see that the out-of-place modular multiplication of a quantum
register by a constant requires $2n+5m+1 = 2n + 5\log_2(n)+6$ total qubits. This is compared to the $2n$ qubits
required by the standard method that utilizes modular adders. The additional overhead of $5\log_2(n)+6$ is 
small for realistic sized multipliers. 

In \tab{barrett:quantum:adders} we show the overhead in terms of the size and total number of
addition operations required per step in \alg{barrett:quantum:circuit}. For comparison the 
standard modular addition based adder would require $3n$ adders each of width $n$. The total number of gates is
linear in the number and width of the adders, therefore, the product of the adder-width times the number of
adders gives to first-order the number of gates required. For the Barrett multiplier this product is:
$n^2 + 14n\log_2(n) + n + 17(\log_2(n))^2 + \log_2(n)$, compared to the standard multiplier with $3nn = 3n^2$.
For realistic sized multipliers the $n^2$ term dominates for both adders and the Barrett multiplier provides
close to a factor of $3$ fewer gates than the standard method. The circuit depth of the multipliers will 
depend on the implementation of the adders used as well as how the individual steps of the multiplier overlap.

\begin{table}
\centering
\begin{tabular}{r|c|c}
\hline
\multicolumn{3}{c}{Width of Adder} \\
\cline{2-3}
Step & $n+\log_2(n)$ & $\log_2(n)$ \\
\hline
$1$ & $n$ & \\
$2$ &   & $2n$ \\
$3$ &   & $4\log_2(n)$ \\
$4$ & $3\log_2(n)$ & \\
$6$ & $1$ & \\
$9$ & $3\log_2(n)$ & \\
$13$ & $3\log_2(n)$ & \\
$14$ & & $4\log_2(n$) \\
$15$ & & $2n$ \\
\hline
& $n+9\log_2(n)+1$ & $4n$ + $8\log_2(n)$ \\
\end{tabular}
\caption{Full-width and log-width adders required by the steps of \alg{barrett:quantum:circuit}
The $n$ adders of width $n + \log_2(n)$ required to calculate the full product dominate the required resources}
\label{tab:barrett:quantum:adders}
\end{table}

\subsubsection{Quantum Barrett Multiplication with Two Quantum Inputs}
\label{sec:barrett:quantum-quantum}
The Barrett reduction method can be extended to a full-quantum multiplier, i.e., one where both inputs are
contained in quantum registers. For this adder we can either add the shifted terms $2^ix$ directly or
reduce them modulo $N$ before adding them. Since $x$ is a quantum value, reducing them would require the
shift and reduce circuit described in \sect{mod-mult:mod-shift}, but
the operation of the Barrett reduction is the same as the case when one input is classical.
If we assume that the accumulated shifts must 
be reversed at the end of the multiplier then $2n$ shifts are required per out-of-place multiplication. 
Therefore the full-quantum Barrett Multiplier requires $~3n$ additions compared to the $3n+2n=5n$ additions
that would be required by the corresponding standard quantum-quantum modular multiplier.
Adding the shifted terms directly would eliminate the $2n$ reduction steps, but would require a $2n-bit$
product register and would require higher precision in the Barrett reduction.
As in the case of the standard modular multiplier, constructing an in-place full-quantum multiplier would 
require calculating a multiplicative inverse, which would dominate the total cost of the multiplier.

%% file: fourier.tex

\section{Modular Multiplication with Quantum Fourier Arithmetic}
\label{sec:fourier}

Because of the elimination of division from the modular reduction step our 
Barrett and Montgomery modular multipliers constructions are uniquely amenable to arithmetic in the quantum Fourier number basis.  In the Fourier basis, number states are represented by the quantum Fourier transform (\QFT/) of their binary state,
\begin{equation}
  \eq{fourier:state}
  \fket[n]{y} \defeq \QFT/\ket[n]{y} = 
  \bigotimes_{k=0}^{n-1} \bigg\{ 
      \cos(\frac{y\pi}{2^k})\ket{0} 
    + \sin(\frac{y\pi}{2^k})\ket{1}
  \bigg\},
\end{equation}
where we have commuted and dropped Hadamard gates from the \QFT/ definition (see \apx{fourier:num-rep} for details).  Arithmetic in this representation  circumvents the ancillary 
work qubits and data dependencies required by the carry bits needed for binary-basis arithmetic, absorbing these bits into the continuous state of each qubit.  
Fourier-basis addition is decomposed into independent, commutable rotations acting on each qubit in the register independently, enabling large-scale parallelization of arithmetical operations with minimal ancilla and on a variety of computational topologies (e.g. a nearest-neighbor architecture)~\cite{Cleve2000,Moore2001,Pham2013}.  


The bottleneck of quantum Fourier-basis arithmetic is in the implementation of quantum control.  The continuous state \fket{y\pi/2^k} of the $k$th qubit of a Fourier-basis register \fket[n]{y} contains information about the bit $y_k$ as well each bit $y_{j<k}$ less significant than $y_k$.  In order to condition a quantum operation on $y_k$, we therefore require a $k$-bit \QFTd/ in order to extract the single bit as a $Z$-eigenstate.  Quantum Fourier arithmetic is generally characterized by these repeated transformations to and from the Fourier basis.

This limitation introduces a significant computation discrepancy between quantum Fourier-basis multiplication and division.  Each reduction step composing a typical division operation comprises a trial subtraction followed by a controlled re-addition conditioned on the sign bit (MSB) of the difference.  For a quantum register in Fourier representation, each such comparison requires a full-register \QFTd/ in order to extract the sign as a binary state, and subsequent \QFT/ to return the rest of the register to the Fourier basis prior to the re-addition.  In addition to the overhead of the \QFT/s themselves, each transform acts as a computational barrier point, inhibiting the parallelization of sequential adders.  A single modular adder constructed as in \sect{mod-mult:mod-add} then requires two \QFTd/-\QFT/ pairs (for the two embedded comparisons), totaling $4n$ full-register \QFT/-like operations for an out-of-place quantum modular multiplier constructed from Fourier modular adders~\cite{Draper00,Beauregard2003}.

\subsection{Quantum Fourier Division}
\label{sec:fourier:div}

The utility of Fourier-basis arithmetic for quantum modular multiplication can be improved by separating the component multiplication and division stages.  As described in \sect{div:mult}, the initial multiplication stage comprises only controlled additions to an accumulation register.  It can therefore be implemented with parallelized Fourier adders such that from a binary input state \ket[n]{y} we accumulate the Fourier-basis state \fket[n+m]{t},
\begin{equation}
  \ket [n+m]{0}\ket[n]{y} \lra[\QF\MAC(X\mid N)]
  \fket[n+m]{t}\ket[n]{y},
\end{equation}
with a total parallelized depth of $(n+m)$ controlled rotation gates.  

The quantum division operator (\QQ\DIV{}) defined in \sect{div} consists of $m=\clog[2]{n}$ reduction steps.  Using Fourier-basis arithmetic, each trial subtraction must be followed by $n$-bit \QFTd/ and \QFT/ operations in order to extract the resulting sign bit.  After $m$ steps, the remainder $(t\bmod N)$ is held in Fourier representation, while the quotient, computed by inverting the $m$ extracted sign qubits, is constructed in binary:
\begin{equation}
              \fket[n+m]{t}      \lra[\QF\DIV{N}]
  \ket [m]{q} \fket[n]{t\bmod N}.
\end{equation}
In order to construct the in-place modular multiplier described in \sect{mod-mult:in-place}, both outputs of the out-of-place operator must share the same representation.  We therefore apply one more $n$-bit \QFTd/ to the remainder \fket[n]{t\bmod N}, leaving both it and the input \ket[n]{y} in binary representation.

\begin{figure}[ht]
    \centering
    \includecircuit[7]{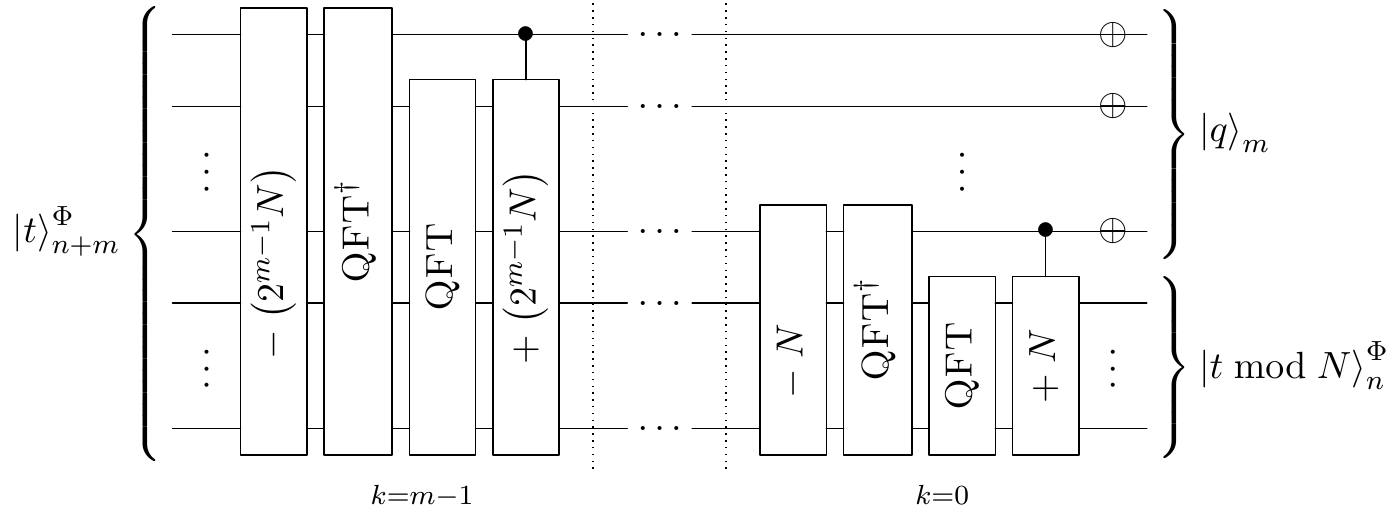}
    \caption{\label{fig:fourier:div}$\QF\DIV{N}$ circuit, incorporating the intermediate \QFTd/ and \QFT/ operators required to extract the sign after each trial subtraction.  The quotient, constructed by inverting the extracted sign bits, is computed in binary representation, while the remainder is output in the Fourier basis.}
\end{figure}

We finally uncompute the quotient register as in \sect{div:uncompute}:
\begin{equation}
  \ket [m]{ q} \lra[\QF\MUL(N)]
  \fket[m]{qN} \lra
  \fket{qN + (t\bmod N)} = \fket[m]{t} \lra
  \ket[m]{0}.
\end{equation}
Using the $\QF\MUL(N)$ operator defined in \apx{fourier:mult}, we first multiply the register by $N$ (modulo-$2^m$) in-place, while simultaneously transforming the result to its Fourier representation.  We then can add the remainder and uncompute the resulting $\fket[m]{t}$ register with a combined $(n+m)$ width-$m$ Fourier adders, controlled by the $m$ LSBs of the output register and all $n$ bits of \ket[n]{y}.  The gates composing the Fourier uncomputation stage can be overlapped with gates in either the \QF\DIV{N} operation or the final \QFTd/.

\subsubsection{Analysis}

The circuit dimensions of the out-of-place modular multiplier constructed from the \QQ\DIV{} operation are broken down in \tab{fourier:div:costs}.  The total two-qubit gate count of the multiplier is,
\begin{equation}
  \#(\text{gates}) = n^2m + 3n^2/2 + 3nm - n/2 + m^2/2 - m/2,
\end{equation}
parallelized to a depth of,
\begin{equation}
  4nm+2n+\ord{m}.  
\end{equation}
Though a significant speedup over Fourier modular multiplication via modular addition~\cite{Draper00,Beauregard2003}, the overall depth scales as \ord{n\log n}, offering no speedup over the binary-basis multiplier constructed with logarithmic-depth adders.  However, we require no work qubits for the Fourier adders, so that the total operator requires only the $(2n+m)$ qubits required to hold \fket[n+m]{t} alongside the input state \ket[n]{y}.
The in-place-multiplier doubles the gate count and depth requirements, and adding a quantum control add \ord{n} gates.

\begin{table}[H]
  \centering
  \caption{\label{tab:fourier:div:costs}%
    Total two-qubit gates and parallelized depth of an $n$-bit out-of-place Fourier modular multiplier constructed from the $\QF\DIV{N}$ operator (where $m=\clog[2]{n}$).  %
  }
  \vspace{6pt}
  \begin{tabular}{l l l l l l l}
  \textbf{Stage}   &       & \textbf{Adds}& \textbf{$\QFT/$s} & \textbf{Width} 
                           & \textbf{Gates} & \textbf{Depth} \\
  \hline\\[-9pt]
  \multicolumn{2}{l}{Multiplication} & $n$ & $0$  & $n+m$ & $n^2+nm$  & $n+m$ \\ 
  \multicolumn{2}{l}{Division (\QF\DIV{})}   & $m$ & $2m$ & $n$   & $n^2m+nm$ & $4nm$ \\
  \multicolumn{2}{l}{Output \QFTd/}         & $0$ & $1$  & $n$   & $(n^2-n)/2$ & $2n$ \\
  Uncompute: &$\ket[m]{q}\lra\fket[m]{qN}$
                 & $m-1$ & $0$ & $1,...,m-1$ & $(m^2-m)/2$ & $2m$ ${}^\ddagger$ \\
    & $\hphantom{\ket[m]{q}}\lra\fket[m]{t}$
                 & $m$ & $0$ & $m$ & $m^2$ & $m$ ${}^\ddagger$ \\
    & $\hphantom{\ket[m]{q}}\lra\ket[m]{0}$ 
                 & $n$ & $0$ & $m$ & $nm$ & $n$ ${}^\ddagger$ \\
  \hline\\[-9pt]
  \multicolumn{5}{l}{${}^\ddagger$Executed in parallel with \QF\DIV{} and output \QFTd/}

  \end{tabular}
\end{table}

The gate count of the \QF\DIV{}-multiplier can be further reduced if we bound the precision of the controlled rotations comprising each \QFT/.  By eliminating rotations by angles below a determined threshold, the number of gates required for an $n$-bit \QFT/ is decreased from $n^2/2$ to $\ord{n\log n}$, while overall fidelity, of a noisy circuit, is likely improved~\cite{Barenco1996}.  Applying this optimization to the \QFT/s embedded in the \QF\DIV{} operation, the total gate count of the multiplier becomes,
\begin{equation}
  \#(\text{gates, bound precision}) = n^2 + \ord{n\log^2n},
\end{equation}
where the magnitude of the second-order term is determined by the desired precision and fidelity of physical operations.
In this case, the asymptotic gate count of the modular multiplier is dominated by the $n^2$ gates of the initial multiplication.  Unfortunately, this optimization does not change the depth of the \QFT/ or the overall modular multiplier.

\subsection{Quantum Fourier Montgomery Multiplication}
\label{sec:fourier:redc}

Beginning with the Fourier state \fket[n+m+1]{t} (where we extend Fourier-basis multiply described in \sect{fourier:div} to incorporate the single ancilla necessary to hold the MSB of the quantum Montgomery estimate), we can reconstruct the quantum Montgomery reduction operator using Fourier-basis arithmetic. 
The bottleneck of the \QF\DIV{} circuit was in the extraction of the sign bit after each trial subtraction, necessitating $n$-bit \QFTd/ and \QFT/ operations due to the dependency of the MSB on the less significant bits of the Fourier register.  The Montgomery reduction algorithm sidesteps the requirement of trial subtraction, instead requiring additions controlled by the LSBs of the register.

\subsubsection{Estimation Stage}
\label{sec:fourier:redc-est}

As in the binary case, the Fourier Montgomery estimation stage is constructed from controlled Fourier subtractions and implicit right-shifts.  
For integral $t$, the LSB \fket{t_0} of the Fourier state \fket[n+m+1]{t} ($k=0$ term in \eq{fourier:state}) is equivalent (up to phase) to that of its binary representation.  We use this bit to condition a subtraction of $N$ from the register, ignoring the irreversible component of the subtraction affecting the control qubit.  In the continuous Fourier representation, this is equivalent to subtracting the half-integer $(N/2)$ from the truncated register,
\begin{equation}
  \fket[n+m+1]{t} = 
  \fket[n+m]{t/2}       \ket{t_0} \lra
  \fket[n+m]{(t/2)-t_0\cdot(N/2)}\ket{t_0},
\end{equation}
where \ket{t_0} is then equivalently the LSB $u_0$ of $\ket[m]{\U}$.

By design, the subtraction occurs only when $t$ is odd ($t_0=1$), so that the difference is always an integer.  The LSB of the truncated Fourier-basis state can therefore be used to condition the next subtraction of $(N/2)$ from the remaining bits of the register.  As shown in the first stage of \fig{fourier:redc}, after $m$ such iterations we are left with the Montgomery estimate in Fourier representation, while simultaneously extracting the $m$ bits of \ket[m]{\U} in binary:
\begin{equation}
  \fket[n+m+1]{t} \lra[\QF\MonEst{N,2^m}{}] \fket[n+1]{(t-\U N)/2^m} \ket[m]{\U}.
\end{equation}
Remarkably, we have thus far required no extra transformations.

\subsubsection{Correction Stage}
\label{sec:fourier:redc-cor}

\begin{figure}[H]
    \centering
    \includecircuit[8]{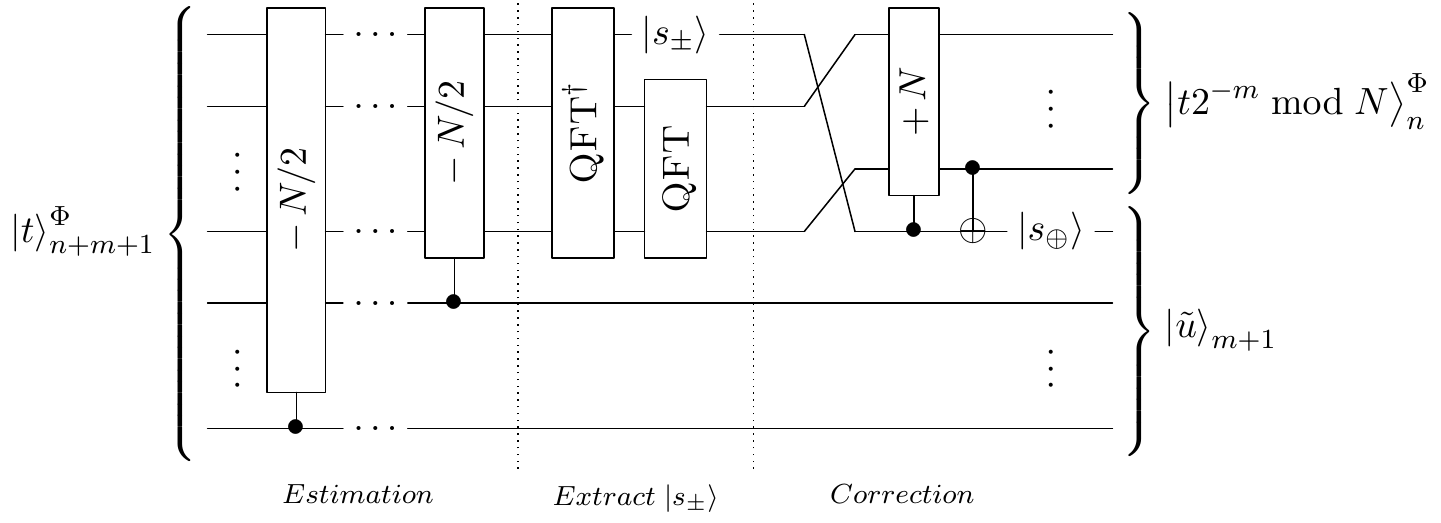}
    \caption{\label{fig:fourier:redc}\QF\REDC{N,2^m}{} circuit, with the requisite \QFTd/ and \QFT/ operators in order to extract the sign bit \ket{s_\pm}.  The estimation stage ($-N/2$) adders and \QFTd/ can then be sequenced like a single $\QFTd/$ over $n+m+1$ qubits.}
\end{figure}

Unfortunately, the quantum Montgomery reduction procedure does require a single modular reduction.  In the correction stage, we add $N$ to the estimate if it is negative, requiring a single $(n+1)$-bit \QFTd/ to extract the sign bit of the register, followed by an $n$-bit \QFT/ to return the remaining bits of to their Fourier representation.  As demonstrated in \fig{fourier:redc}, the binary sign bit is then used to control the addition of $N$, after which it is conditionally flipped by the LSB of the result and concatenated with \ket[m]{\U} to form \ket[m+1]{\UU} identically to the binary case.  Combined with the estimation stage, we have constructed the Fourier Montgomery reduction operator,
\begin{equation}
  \fket[n+m+1]{t} \lra[\QF\REDC{N,2^m}{}] \fket[n]{t2^{-m}\bmod N} \ket[m+1]{t N^{-1}\bmod 2^{m+1}},
\end{equation}
where the reduction is returned in Fourier representation, and the garbage state \ket[m+1]{\UU} in binary. 

We can then concatenate the \QFTd/ operation with the Fourier adders comprising the preceding estimation stage.  The resulting sequence of controlled rotations is identical in structure to that of a single \QFTd/ over all $n+m+1$ qubits, and can likewise be parallelized to a depth of $2(n+m)-1$.  Similarly, the controlled addition of $N$ can be concatenated with the preceding \QFT/ and parallelized as an $(n+1)$-qubit \QFT/ to depth $2n-1$ controlled rotation gates.

\subsubsection{Uncomputation Stage}

As in the case of the $\QF\DIV{N}$ procedure, in order to construct an in-place quantum Montgomery multiplier from the \QF\REDC{N,2^m}{} operator, we require a final $n$-bit \QFTd/ on the output so that it is returned in binary along with the input state \ket[n]{y}.   Simultaneously, we uncompute the \ket[m+1]{\UU} register.  After transforming the register to its Fourier representation with an $(m+1)$-bit \QFT/, we can replicate the sequence of subtractions in the binary procedure (\sect{montgomery:uncompute}) with $n$ controlled Fourier adders, each truncated to $(m+1)$ bits.  Being independent of the output state, the $n$-gate depth of this uncomputation is dominated by the $(2n-3)$-gate depth of the concurrent \QFTd/, fixing the total depth of this stage to the latter.

\subsubsection{Analysis}

Circuit characteristics of the out-of-place Fourier Montgomery multiplier are broken down by stage in \tab{fourier:montgomery:costs}.  In total, we require,
\begin{equation}
  \#(\text{gates}) = 5n^2/2 + 3nm - n/2 + m^2/2 - m/2,
\end{equation}
parallelized to a depth of,
\begin{equation}
  7n+\ord{m}.  
\end{equation}
Comparing the Montgomery circuit to the division-based circuit we see than the depth of the division circuit is higher by a factor of $m$.
This is a result of the extra \QFT/s required in the reduction portion of the division circuit.
As with the \QF\DIV{} operator, if we bound the precision of the rotation gates composing each \QFT/, the total gate count is reduced to $n^2 + \ord{nm}$, asymptotically equivalent to that of just the initial multiplication stage.

\begin{table}[H]
  \centering
  \caption{\label{tab:fourier:montgomery:costs}%
    Total two-qubit gates and parallelized depth of an $n$-bit out-of-place Fourier Montgomery multiplier (where $m=\clog[2]{n}$).  %
  }
  \vspace{6pt}
  \begin{tabular}{l l l l l l l l}
  \textbf{Stage}   &       & \textbf{Adds}& \textbf{$\QFT/$s} & \textbf{Width} 
                           & \textbf{Gates} & \textbf{Depth} \\
  \hline\\[-9pt]
  \multicolumn{2}{l}{Multiplication:}    
      & $n$ & $0$ & $n+m$         & $n^2+nm$  & $n+m$ \\ 
  \multicolumn{2}{l}{Estimation:}
      & $m$ & $0$ & $n+\{m,...,1\}$ & $nm+(m^2+m)/2$ & \multirow{2}{*}{$\Big\}\;2n+2m$} \\
  Correction
    & \QFTd/: & $0$ & $1$ & $n+1$ & $(n^2+n)/2$ & \\
    & \QFT/:  & $0$ & $1$ & $n$   & $(n^2-n)/2$ & \multirow{2}{*}{$\Big\}\;2n$} \\
    & Add $N$: & $1$ & $0$ & $n$   & $n$         & \\
  \multicolumn{2}{l}{Output transform:}
      & $0$ & $1$  & $n$   & $(n^2-n)/2$ & \multirow{3}{*}{$\bigg\}\;2n{}^\ddagger$} \\
  Uncompute 
    & \QFTd/:   & $0$ & $1$ & $m+1$ & $(m^2+m)/2$ & \\
    & $t\rightarrow0$: & $n$ & $0$ & $m$   & $nm$ & \\
  \hline\\[-9pt]
  \multicolumn{7}{l}{${}^\ddagger$Uncomputation steps executed in parallel with output \QFTd/}

  \end{tabular}
\end{table}

\subsection{Quantum Fourier Barrett reduction}

Like Montgomery reduction, the benefit of Barrett reduction in the classical case is in the replacement of division with integer multiplication (\alg{barrett:quantum:circuit}).  The quantum Barrett reduction procedure is therefore similarly well-suited for arithmetic in the quantum Fourier basis. 

As in the division and Montgomery reduction based multiplication procedures, we begin with Fourier calculation of \fket[n+m]{t} (\alg{barrett:quantum:circuit}, \step{barrett:quantum:xy}).  In the Barrett case, we also accumulate the approximate product $\xyapp$ (\step{barrett:quantum:xyapp}) with a simultaneous $\QF\MAC(\widetilde{X}\mid N)$ operation.  The approximate product is then used to compute $\qhat$ (\step{barrett:quantum:qhat}), requiring its transformation to binary representation prior its Fourier-basis multiplication by $\mu$.  Similarly, $\qhat$ is used as a quantum multiplicand in \step{barrett:quantum:reduce,barrett:quantum:addqn,barrett:quantum:subqn}, and so must be transformed to its binary representation prior to use in these steps.

The quantum Barrett reduction procedure requires two comparison operations (\alg{barrett:quantum:circuit}, \step{barrett:quantum:compare} and \step{barrett:quantum:appcompare}), requiring the usual \QFTd/ and \QFT/ operations in order to extract sign bits.  
However, while the first requires full-register transformations, the second comparison is limited to just the most significant bits of the accumulation register (the number of bits required is equal to the size of $\qhat$).   After a full-register \QFTd/ to extract the sign, we therefore only need to transform these MSBs of the register to the Fourier basis for the re-addition of $\xyapp$ and subtraction of $qN$ (\step{barrett:quantum:appcompare,barrett:quantum:subqn}).  We finally transform these bits back to their binary representation so that both the output and input states of the modular multiplier are binary.

The final uncomputation of $\qhat$ and $\xyapp$ (\step{barrett:quantum:qhat-rev,barrett:quantum:xyapp-rev}) requires the reversal of their initial computations and transformations.  As in the uncomputation stages of the \QF\DIV{} and \QF\REDC{}{} operations, the Fourier uncomputation of $\xyapp$ is requires subtractions controlled by each bit of the input state, and therefore maintains the initial multiplication's depth of $n$ controlled rotations.  Combined with the three full-register \QFT/s required for comparisons, the overall Fourier Barrett multiplication operation has a depth of $7n+\ord{m}$ controlled rotation gates, identical in leading order to the Montgomery reduction operator.  As with both \QF\DIV{} and \QF\REDC{}{}, if we bound the precision of the component \QFT/s, the total leading-order gate count is just that of the initial multiplier.

\begin{figure}[h!]
    \centering
    \includecircuit[14]{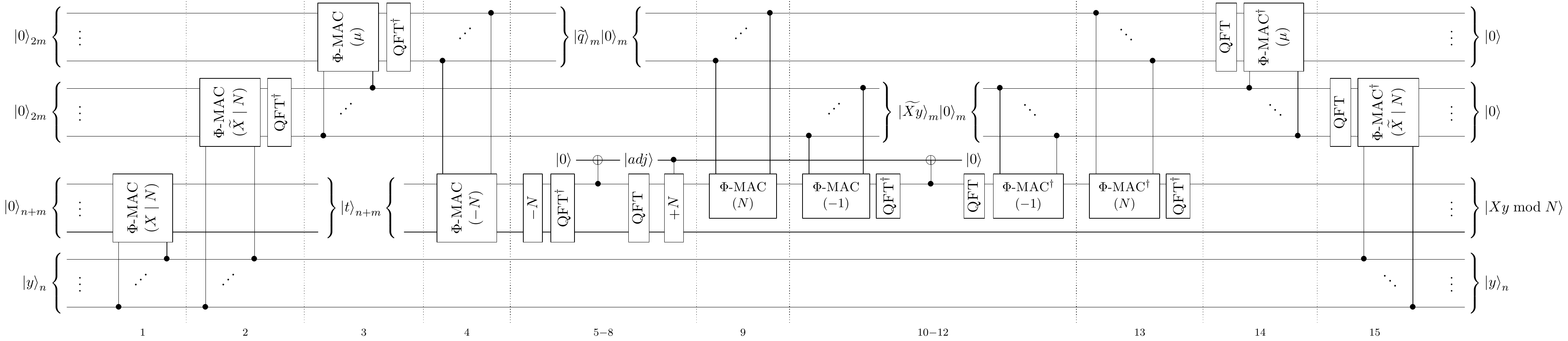}
    \caption{Barrett multiplication circuit using Fourier arithmetic. The numbers in the figure correspond to the steps of \alg{barrett:quantum:circuit}.}
    \label{fig:fourier:barrett}
\end{figure}

%% file: resources.tex

\section{Resource Evaluation}
\label{sec:resources}

In this section we present a detailed resource analysis of the modular multiplier designs introduced in the previous sections, alongside that of the standard modular-adder approach (\sect{mod-mult}).  We focus our analysis on the  quantum-classical modular multiplier (where one input is a classical parameter); as described in \sect{mod-mult:quantum-quantum}, in-place multiplication with two quantum inputs would require a circuit to calculate the multiplicative inverse of one of the inputs, and this inverse circuit would dominate the resources of the multiplier.  If the particular details of this analysis are not of interest, a summary important takeaways can be found in \sect{resources:summary}.

We explicitly generate circuits for each multiplier design with various sizes, different classical constants, and different classical moduli. 
We utilize the four adders discussed in the appendix:
the Fourier basis adder~\cite{Draper00}, the logarithmic-depth prefix (or carry-lookahead) adder~\cite{Draper2006}, the linear-depth majority ripple adder~\cite{Cuccaro2004}, 
and the rippled prefix adder defined in \apx{circuits:prefix_adders}. 
Resource analysis is performed assuming two different hardware models: the first treating all gates equally, and the second accounting for the relative complexity of each gate in the context of a fault-tolerant logical circuit using quantum error-correcting codes.  With each model, we determine an overall circuit \emph{size}, or combined cost of all gates, and \emph{depth}, or total latency of the circuit allowing for unbounded parallelization.  We do not include locality constraints in our hardware models. The impact of locality is primarily dependent on the particular architecture and the addition technique employed, with roughly the same impact on each multiplier design.

\subsection{Evaluation Methodology}
\label{sec:resources:methodology}

For each evaluation, we construct a full quantum circuit consisting of gates from a set natural to the adder that is being used. For example, the Fourier basis adders predominantly require \CRy/ controlled rotation gates, while all circuits utilize \X/ gates, with zero, one, or two controls. We use a C++ program to generate each circuit. The program can generate sequences of arbitrary size, to a practical limit.  The user specifies the classical modulus and multiplicand, which are folded into the circuit to produce an optimized circuit unique to those classical inputs.  Circuit depth for each hardware model is determined by then running the generated circuit through a scheduler, which places gates into parallel time slots based on the specified latency of each gate in that model.  The scheduler reorders gates according to standard commutation rules: gates that do not share any qubits can be trivially reordered, and two gates that do share qubits can be reordered if they both commute either with the Pauli \Z/ operator or the Pauli \X/ operator.  Finally, each circuit is verified for a variety of input values with a separate gate-level simulation program.

\begin{table}[ht]
\centering
\begin{tabular}{r|c|c}
\hline
&\multicolumn{2}{c}{Hardware Model} \\
\cline{2-3}
Gate & Equal-latency & Fault-tolerant \\
\hline
$\CX/$ & 1 & 1\\
$\H/$ & 1 & 1\\
$\T/$ & 1 & 10 \\
$\Toffoli/$ & 1 & 40 \\
$\GATE{Pauli}$ & 1 & 1\\
$\Ry/(\alpha)$ & 1 & $66\log_2(n) + 33$\\
$\CRy/(\alpha)$ & 1 & $\sim66\log_2(n) + 35$\\
\hline
\end{tabular}
\caption{Unit costs of gates for two hardware models. The first ``Equal latency'' model
assumes the same latency for all gates. The second, ``Fault-tolerant'' model enforces gate-dependent latencies, representing a rough cost in primitive gates available to an error-correcting code.
See the text for a description of the cost of the single-qubit $\Ry/(\alpha)$ and controlled $\CRy/(\alpha)$ rotation gates.
}
\label{tab:resources:hwmodel}
\end{table}

Our resource evaluations assume two different hardware models. In the first model (which we will label ``equal-latency''), all gates, including controlled arbitrary angled rotation gates, have the same latency.  This provides a useful characterization of our circuits, and comparison to other circuits in the literature.  However, it is not realistic for architectures that implement limited gate sets. The second model (``fault-tolerant'') is motivated by the gates that are typically available on logical qubits implemented using a quantum error correction code. For standard Calderbank-Steane-Shor (CSS) codes~\cite{Steane1996}, Clifford gates such as $\CX/$ and $\H/$ are easy to implement whereas gates such as $\T/$ and $\Toffoli/$ require the use of more complicated circuitry and potentially the use of magic-state distillation, which can be costly. For this reason we assign a cost of $10$ to the $\T/$ gate and a cost of $40$ to the $\Toffoli/$ gate.

\begin{figure}[!h]
  \centering
  \includecircuit[2.2]{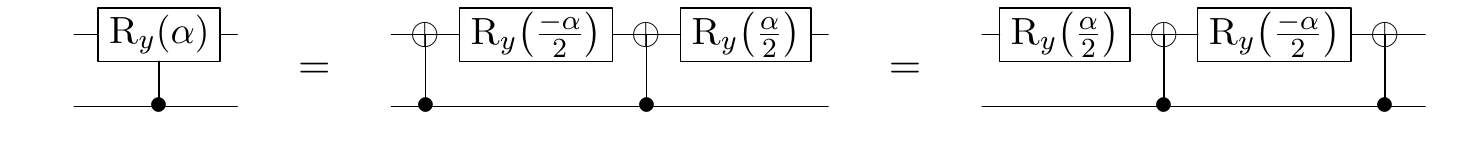}
  \caption{Decomposition of the $\CRy/(\alpha)$ controlled rotation gate into \CNOT/s and single-qubit rotations.  Because the control qubit is not active during the single-qubit rotations, this decomposition also allows for greater parallelization than a composite two-qubit gate of the same latency.}
  \label{fig:resources:cy-decomp}
\end{figure}

\begin{figure}[!h]
  \centering
  \includecircuit[6]{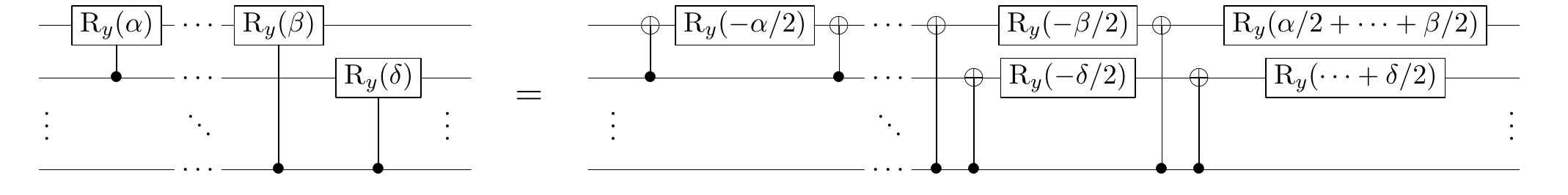}
  \caption{The outer rotation gate of the decomposed $\CRy/$ gates (\fig{resources:cy-decomp}) can be commuted through adjacent gates, and combined into a single rotation per qubit.  This freed control qubit enables further parallelization of adjacent gates, taking advantage of the low latency of the \CNOT/ gate relative to the rotation.}
  \label{fig:resources:cys-decomp}
\end{figure}

In order to construct controlled rotation gates fault-tolerantly, we decompose each into the sequence of two \CNOT/ gates and two unconditional rotations depicted in \fig{resources:cy-decomp}.  As shown, the unconditional $\Ry/(\alpha/2)$ rotation in the decomposed gate commutes with the remainder of the operation, as well as adjacent single-qubit and controlled rotations targeting that qubit.  We can therefore collect and merge these commuting gates into a single unconditional rotation per qubit, as shown in \fig{resources:cys-decomp}.  We then expect that, in the fault-tolerant model, each decomposed controlled rotation has an amortized cost of one rotation gate in addition to the two \CNOT/ gates.  
Also shown in \fig{resources:cys-decomp}, the control qubit of the decomposed gate is only involved in the \CNOT/ operations; given the low latency of \CNOT/s relative to rotation gates in this model, the latter can be be further parallelized between the \CNOT/s of adjacent gates sharing a control qubit.

Arbitrary single-qubit rotations cannot be implemented directly on CSS codes, and instead must be approximated with sequences from the finite set of available gates.  
A sequence of $\T/$ and $\H/$ gates approximating a single-qubit rotation to an accuracy $\epsilon$ can be constructed with $3\log_2(1/\epsilon)$ steps, where each step 
contains an $\H/$ gate and a $\T/$ gate raised to an integer power~\cite{Ross2014}.  
Noting that the complexity of implementing a $\T/^k$ gate on a CSS code should be no more than that of a single $\T/$ gate, we assume that the cost of each $(\H/\T/^k)$ step is the same as that of one $\H/$ gate and one $\T/$ gate.  

The necessary accuracy $\epsilon$ for fault-tolerance is a function of the total number of rotation gates in the multiplier.  
To first order, the Barrett and Montgomery multipliers require $2n^2$ \CRy/ rotations, each of which can be implemented using two $\CX/$ gates and one effective
single qubit rotation. We therefore choose $\epsilon = 1/(2n^2)$, requiring $3\log_2(2n^2) = 6\log_2(n) + 3$ steps to approximate the single-qubit rotation.
Incorporating the costs of $\T/$ and $\H/$ gates determined above, we estimate a total cost of $66\log_2(n) + 33$ per rotation, or $66\log_2(n)+35$ 
for the \CRy/ gate in \fig{resources:cy-decomp}.  The cost of each gate in the two hardware models is summarized in \tab{resources:hwmodel}.

Given the finite precision with which we are approximating rotation gates, it is also reasonable to explicitly limit the precision of the controlled rotation gates we can implement.  In particular, we can drop rotations with an angle $\theta^2 \approx\epsilon$, as these are sufficiently well approximated by the identity operation.  This simplification can dramatically reduce the overall gate count, while in the presence of decoherence possibly improving operator fidelity~\cite{Barenco1996}.  For example, the set of rotation gates required by the \QFT/ is described by $\{\theta=\pi/2^k\mid k=1,...n\}$, and so can be truncated after $k\sim\log{n}+2$ as described in~\cite{Barenco1996}.  We use this approximation in the analysis of all Fourier-basis circuits.

For each analysis, we construct sequences with increasing input register widths $8\le n\le2048$.  To avoid pathological cases, for a given $n$ we generate and average the behavior of eight unique quantum circuits, with (when applicable) classical moduli $2^{n-1}<N<2^n$ and multiplicands $0<X<N$ randomly selected from uniform distributions.  We expect a slight reduction in circuit size from the worst case due to this randomization; for example, each binary adder contains a number of gates conditioned on the bits of the classical addend, half of which can be expected to be eliminated for a randomly chosen addend (with negligible variance for large $n$).

\subsection{Adder Comparison}
\label{sec:resources:adders}

\begin{figure}[!htb]
\centering
\includeplot{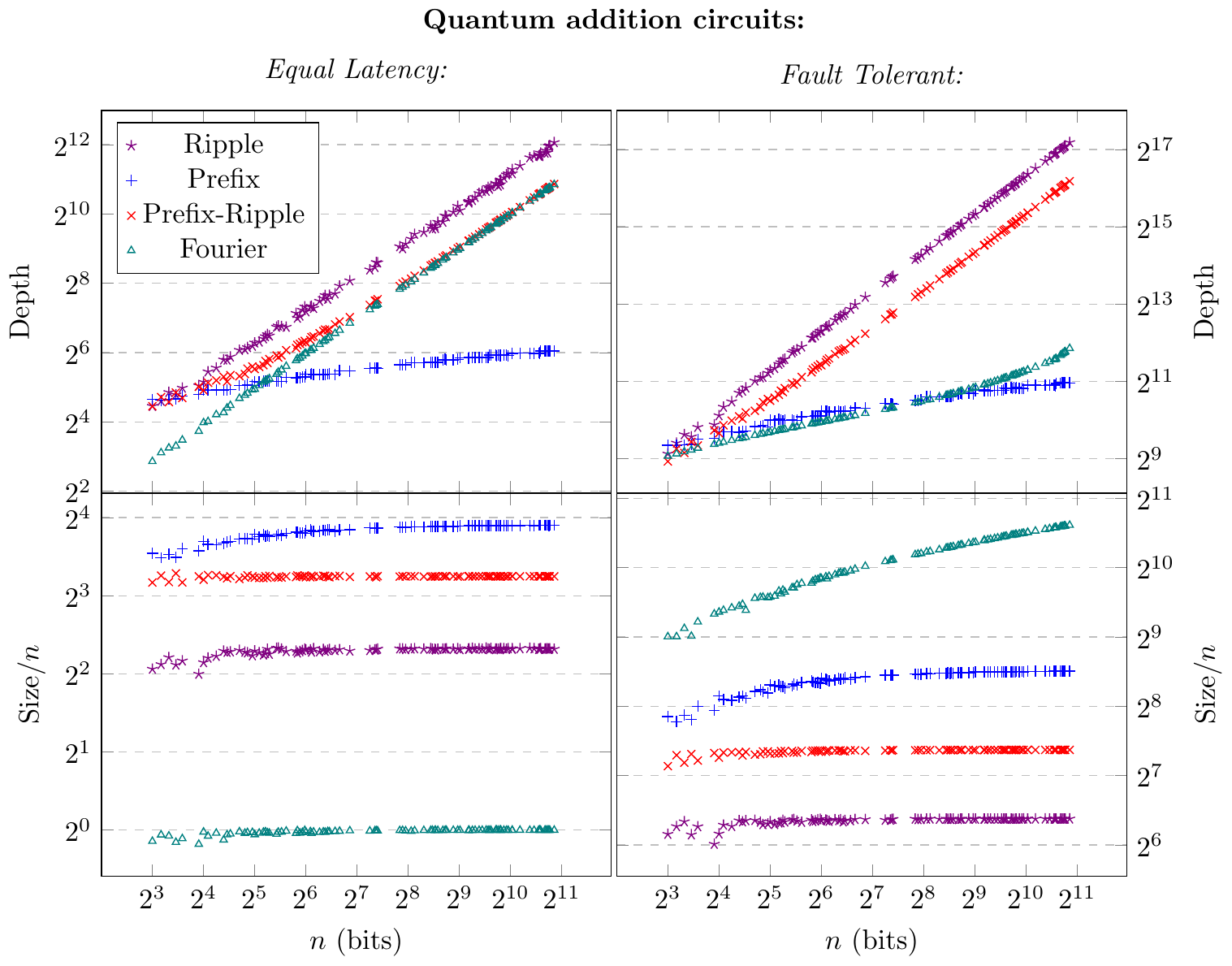}
\caption{Resources required for standalone quantum adder implementations. The details of the adders are described in the Appendix. The depth is the latency of the parallelized circuit and the size is the total number of gates in the circuit. The inflection point in the depth of the fault-tolerant Fourier-basis adder is where the $2n$ \CNOT/ gates from the adder's control qubit begin to dominate the logarithmic-depth fault-tolerant rotations.  Logarithmic depth could be preserved by fanning out the control, at the cost of an additional $\sim n$-qubit register, but this is unnecessary in any of our constructions.}
\label{fig:resources:plots:adders}
\end{figure}

We first evaluate the different quantum addition techniques we will use to benchmark our modular multipliers.  In \fig{resources:plots:adders}, we show the results of simulations of each of the adders using the two hardware models described above.  Each circuit is generated to perform in-place, controlled quantum-classical addition, with a quantum input register size $0<n\le2048$ randomly chosen from a logarithmic distribution.  In the equal-latency model (i.e. assuming all gates are equivalent), we see the expected logarithmic scaling of the prefix adder depth, and linear scaling of the ripple-carry and prefix-ripple constructions.  As in~\cite{Cuccaro2004}, asymptotic latency of the ripple-carry adder is $\sim2.0n$.  The prefix-ripple circuit improves upon the ripple-carry, approaching a depth of $\sim1.0n$ in our data set, but doubles the total number of gates required.  This observed depth matches what we would expect for a circuit dominated by one \Toffoli/ gate per two bits (where both the forward and reverse circuits required for the in-place operation contribute $n/2$). The Fourier-basis adder simply consists of $n$ controlled rotation gates in this model, and afford no parallelization due to the due to the shared control qubit.

The latencies of the three binary adders are dominated by \Toffoli/ gates, and so are increased proportionally in the fault-tolerant model.  However, the total gate count of the prefix adder, having a greater overall proportion of \Toffoli/ gates, is increased more significantly than the ripple and prefix-ripple circuits.  The decomposition of rotation gates required in the case of Fourier-basis addition in this model results in \ord{n\log n} total gates, dominating the linear sizes of the binary adders.  However, the latency of the fault-tolerant Fourier-basis adder is only increased by a factor of two at large $n$.  As described above, the logarithmic-depth single-qubit rotations can be parallelized such that the asymptotic latency is dominated by the $2n$ \CNOT/ gates sharing a control qubit.  Below $n<652$, the latency of two single qubit rotations is greater than $2n$ and so must dominate the circuit depth.  This transition is clearly visible in the Fault-tolerant latency plot of \fig{resources:plots:adders}.  The depth of the Fourier-basis adder is comparable to the logarithmic-depth prefix adder prior to the dominance of the $2n$ \CNOT/s, and consistently below that of the ripple-carry and prefix-ripple adders.

\subsection{Modular Multiplier Components}

\begin{figure}[!ht]
    \begin{center}
    \includecircuit[8]{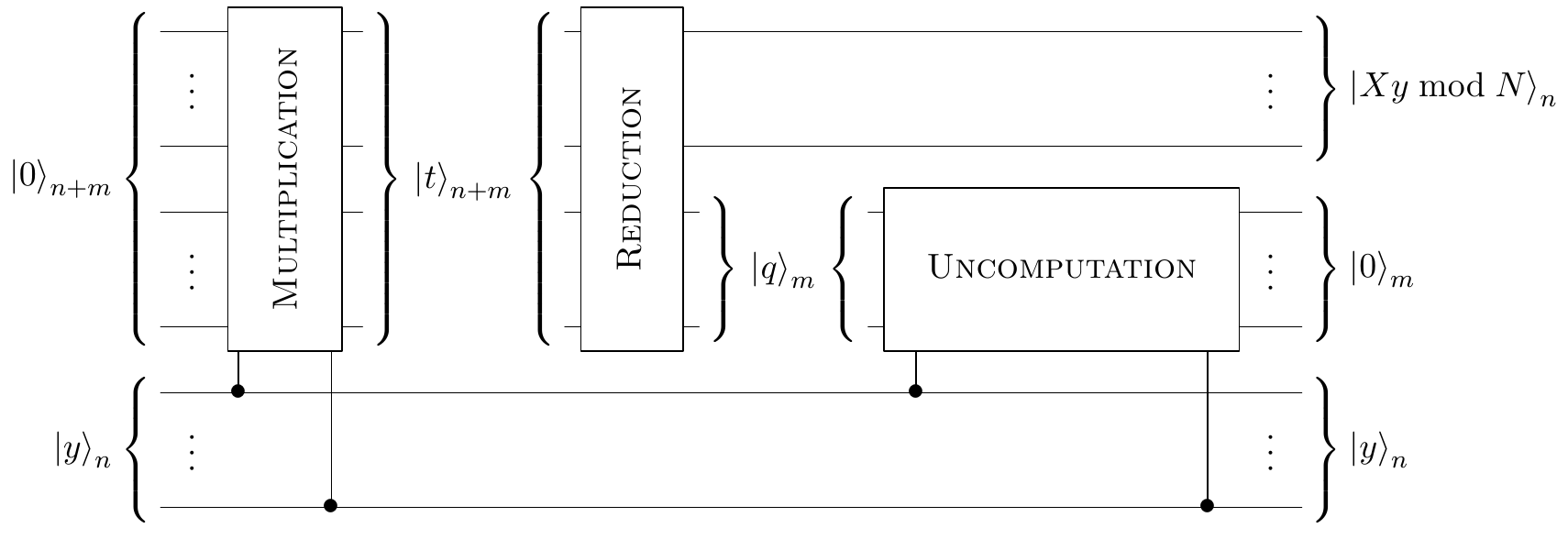}
    \end{center}
    \caption{Generalized three-stage structure of our three proposed out-of-place modular multipliers, where $m\approx\log_2n$.  The size of each circuit is asymptotically dominated by the initial $n$-addition \multiplication/ stage, with the \reduction/ stage requiring only \ord{\log n} adders, and the \uncomputation/ stage adders only spanning \ord{\log n} qubits.  As in \sect{mod-mult:in-place}, the in-place multiplier then requires the structure to be repeated in reverse.}
    \label{fig:resources:generic-mult}
\end{figure}

\begin{figure}[!htb]
\centering
\includeplot{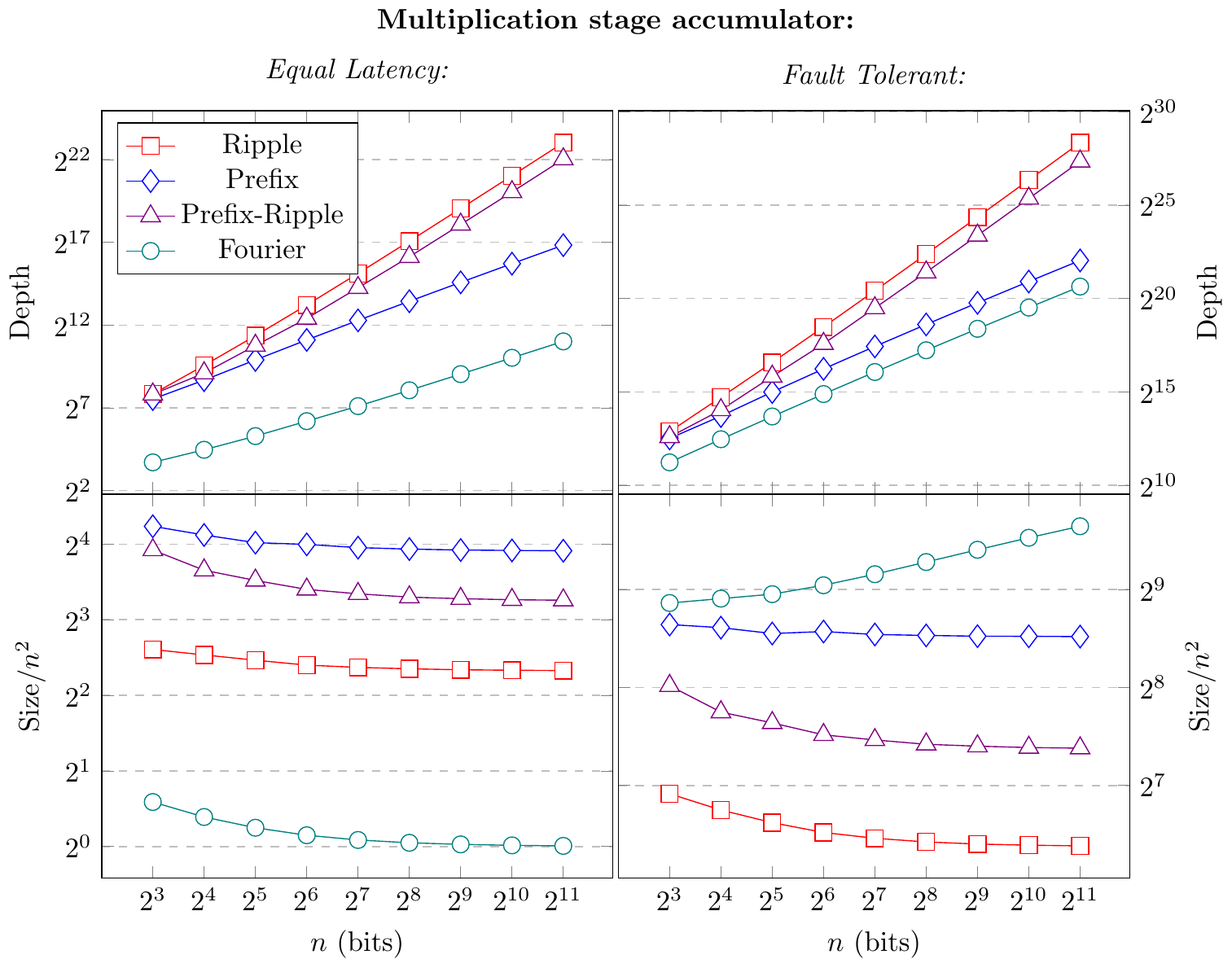}
\caption{Resource requirements of the initial \multiplication/ circuit common to the three modular multipliers, comprising $n$ sequential $(n+\log_2n)$-qubit quantum adders.  In both hardware models, we see the expected asymptotic speedup of the logarithmic-depth prefix adder over the linear-depth ripple and prefix-ripple circuits (upper plots), and the corresponding increase in circuit size (lower plots).  In the equal-latency model, the depth of the parallelized Fourier-basis multiplier is linear in $n$, while the prefix adder scales as \ord{n\log n} and the ripple and prefix ripple as \ord{n^2}.  In the Fault-tolerant model, the \ord{\log n} depth of controlled rotation gates results in the higher \ord{n^2\log n} asymptotic scaling of the Fourier-basis multiplication, while its parallelized depth remains the least of the four.}
\label{fig:resources:plots:multiply}
\end{figure}

The quantum division, Montgomery multiplication, and Barrett reduction circuits introduced in the preceding sections share a common high-level three-stage structure, diagrammed in \fig{resources:generic-mult}.  The initial \multiplication/ stage, identical in the three designs, consists of $n$ controlled adders arranged as an $(n+\log_2n)$-qubit quantum accumulator.  The goal of each design has been to reduce the overall complexity of quantum modular multiplication to that of this step.  We therefore begin by generating circuits for the \multiplication/ operation in isolation with each of the quantum addition techniques outlined above.
The sizes and depths of the generated circuits are displayed in \fig{resources:plots:multiply}.  Observing these results in comparison to \fig{resources:plots:adders}, in each case the total gate count predictably approaches that of $n$ isolated adders.   When constructed with any of three binary addition circuits, the overall circuit latency of the multiplier is increased by the same factor, indicating that the scheduler did not find significant opportunity to parallelize gates between adjacent adders.  The commutable rotations composing the Fourier-basis multiplier, however, enable parallelization across the entire accumulator.  In the equal latency model, this allows $n$ simultaneous controlled rotation gates in each time step, circumventing the control bottleneck of the single adder and reducing the total latency to $(n+\log_2n)$.  In the fault-tolerant model, the sequential adders further allow the commutation and combination of single-qubit rotations, as in \fig{resources:cys-decomp}.  Accordingly, the circuit latency observed in \fig{resources:plots:multiply} is close to $66n\log_2(n)$, or that of $n$ single-qubit rotations alone.  Even after this incurred cost in the Fault-Tolerant model, the Fourier-basis \multiplication/ circuit has the lowest depth among the adders.  Unfortunately, the total gate count in this model is \ord{n^2\log n}, asymptotically dominating the $\ord{n^2}$ gates required by the three binary circuits.

\begin{figure}[!htb]
\centering
\includeplot{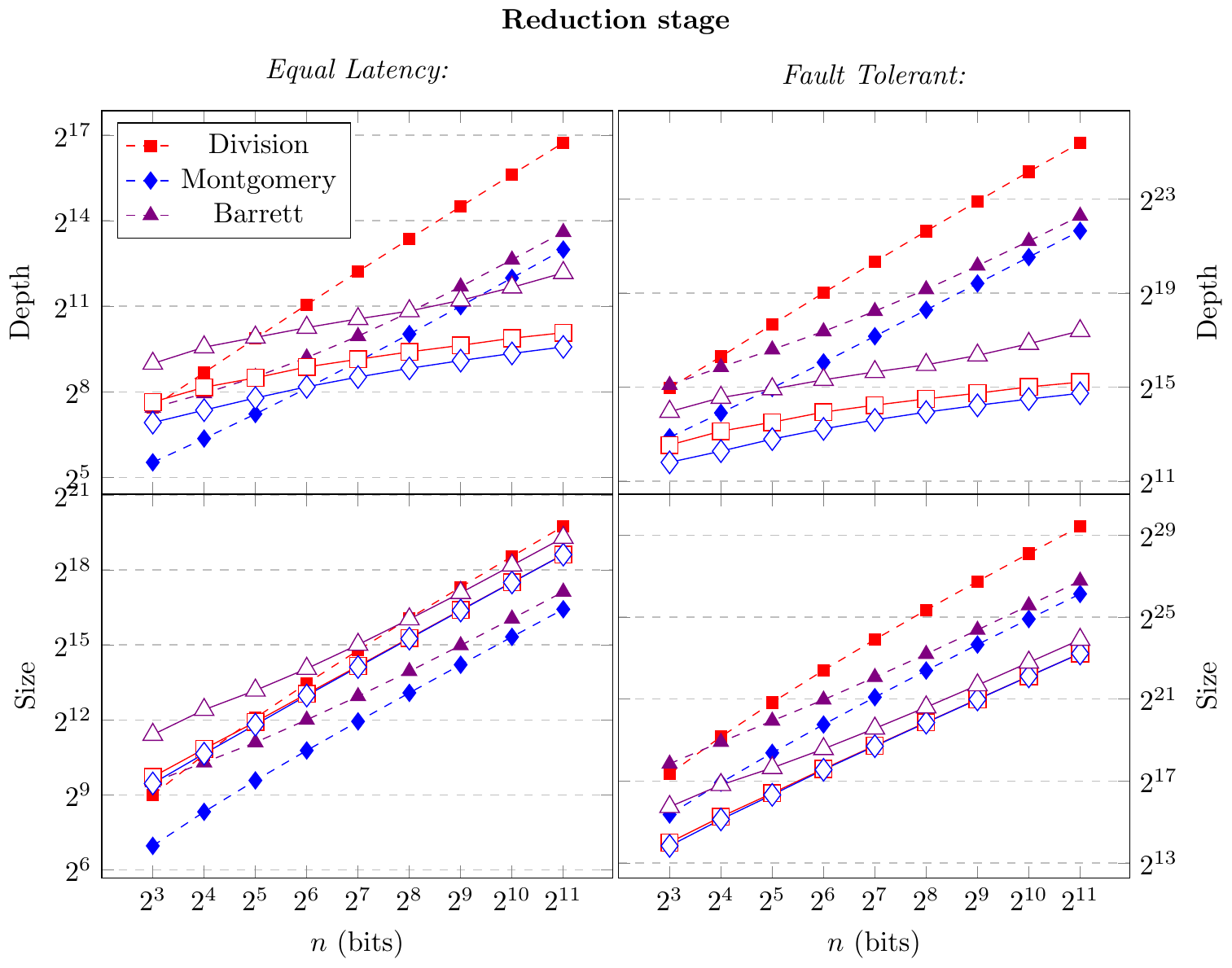}
\caption{\reduction/ stage of proposed modular multipliers, constructed with prefix (hollow marks) and Fourier-basis (solid marks) adders.  In the prefix case, the circuit size and latency of each circuit is comparable to the \uncomputation/ stage and asymptotically dominated by the initial \multiplication/.  The Fourier-basis \reduction/ stages also require asymptotically fewer gates than the corresponding \multiplication/ stages, but have greater depth due to the remaining of comparison operations.  Like the \multiplication/ stage, the Barrett and Montgomery circuits have linear depth in the equal-latency model, whereas the division circuit requires \ord{\log n} comparisons and therefore has \ord{n\log n} depth.}
\label{fig:resources:plots:reduce}
\end{figure}

The division, Montgomery, and Barrett circuits are differ in their respective \reduction/ stages, in which the modular reduction (or Montgomery reduction) is computed from the non-modular product, alongside an \ord{\log n}-qubit quotient-like term (which is necessarily produced by a generic information-preserving modular reduction).  The key to the performance of these circuits is in limiting the number of additions required in this stage to \ord{\log n}.  The resources required for the \reduction/ stage of each proposed multiplier, in both the Fourier-basis and binary (prefix) implementations, are compared in \fig{resources:plots:reduce}.  Due to its logarithmic dependency on register size, a small number of adds relative to the accumulator size makes the decreased latency of the prefix adder relative to the other binary circuits even more pronounced in this stage, allowing the reduction to be performed in poly-logarithmic time compared to the $\ord{n\log n}$ depth achieved with ripple and prefix-ripple adders.  For each binary adder, the circuit size and latency of this stage are overwhelmingly dominated by the initial \multiplication/ stage in \fig{resources:plots:multiply}.

Conversely, as described in \sect{fourier}, the need for at least one comparison operation within each multiplier's \reduction/ stage necessitates the invocation of intermediary \QFT/s in a Fourier-basis implementation, making this the highest-latency stage of the Fourier-basis modular multipliers.  The Barrett and Montgomery circuits, requiring a constant number of \QFT/s, have linear depth in the equal-latency model.  The \ord{\log n} comparisons in the \QF\DIV{} circuit increase this latency to \ord{n\log n}, asymptotically dominating the linear depth of the \multiplication/ stage.  However, assuming finite-precision rotations, the gate counts of the circuits are \ord{n\log n} and \ord{n\log^2 n}, respectively, all scaling comparably to the binary \reduction/ circuits and asymptotically smaller than the Fourier-basis \multiplication/ stage.

\begin{figure}[!htb]
\centering
\includeplot{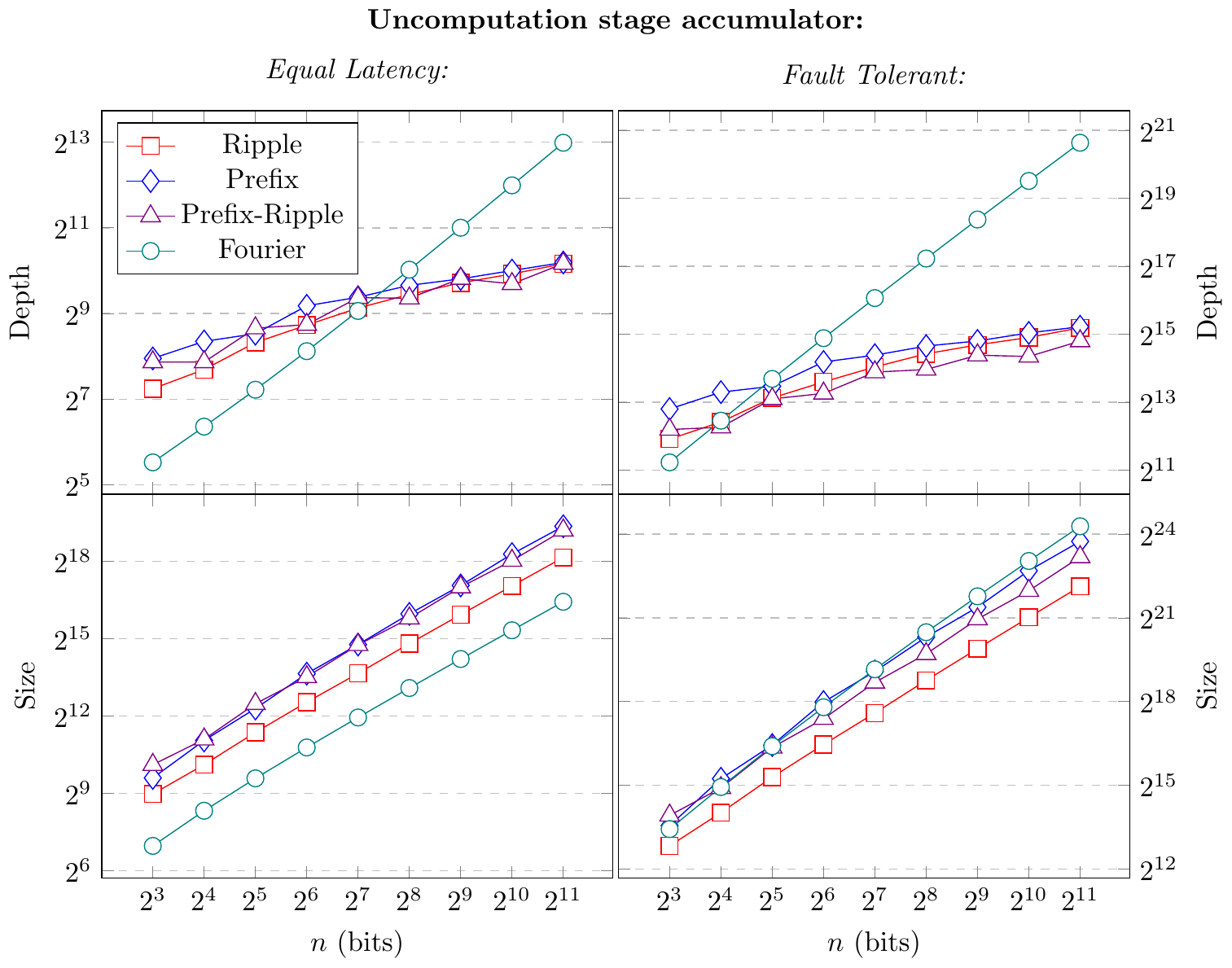}
\caption{\uncomputation/ stage accumulator: each of the proposed modular multipliers requires the clearing of a $(\log_2n)$-qubit quotient register, requiring additions conditioned on each bit of the $n$-qubit input register.  Using binary adders, these can either be executed sequentially or parallelized over the work qubits necessary for the preceding stages.  The Fourier-basis implementation does not afford this parallelization, resulting in its linear depth in the equal-latency model.}
\label{fig:resources:plots:uncompute}
\end{figure}

The final step of each modular multiplier is a $(\log n)$-bit quantum accumulator, consisting of $n$ adders controlled by the bits of the input state \ket[n]{x}.  The \uncomputation/ step in \fig{resources:generic-mult} serves to uncompute the $(\log_2n)$-qubit quotient-like term produced by the preceding \reduction/ stage.  As described in \sect{div:uncompute}, with the binary circuits we can parallelize adders over work qubits already necessary for the preceding stages, so as to match the poly-logarithmic circuit depth achieved with prefix adders in the \reduction/ stage.
The resource requirements for the parallelized \uncomputation/ stage constructed with each addition method are shown in \fig{resources:plots:uncompute}.  In the case of the binary adders, we immediately find that the size and depth of this step is then asymptotically insignificant in comparison to the initial \multiplication/ stage (\fig{resources:plots:multiply}).  In the equal-latency model, we find that the ripple-carry circuit has by a small margin the lowest latency for $n<2^6$, at which point it becomes comparable to the prefix-ripple circuit.  The latter performs marginally better in the fault-tolerant model.  In all cases, the prefix circuit has the greatest size and latency of the binary adders. Observing \fig{resources:plots:adders}, in the range of $n$ being analyzed, we expect the depth of a $(\log_2n)$-qubit adder to be comparable with any of the three circuits, while the additional qubits required by the prefix adder reduces the number of additions we can perform simultaneously in the parallelized accumulator.  We expect that for even larger multiplier sizes the prefix adder would again provide the fastest circuit.  In the Fourier-basis case, we cannot further parallelize adders, resulting in a linear depth in the equal-latency model and \ord{n\log n} depth in the fault tolerant model.  In both models, the gate count of the Fourier-basis \uncomputation/ remains asymptotically dominated by the \multiplication/ stage.

\subsection{Multiplier Comparison}

\begin{figure}[!htb]
\centering
\includeplot{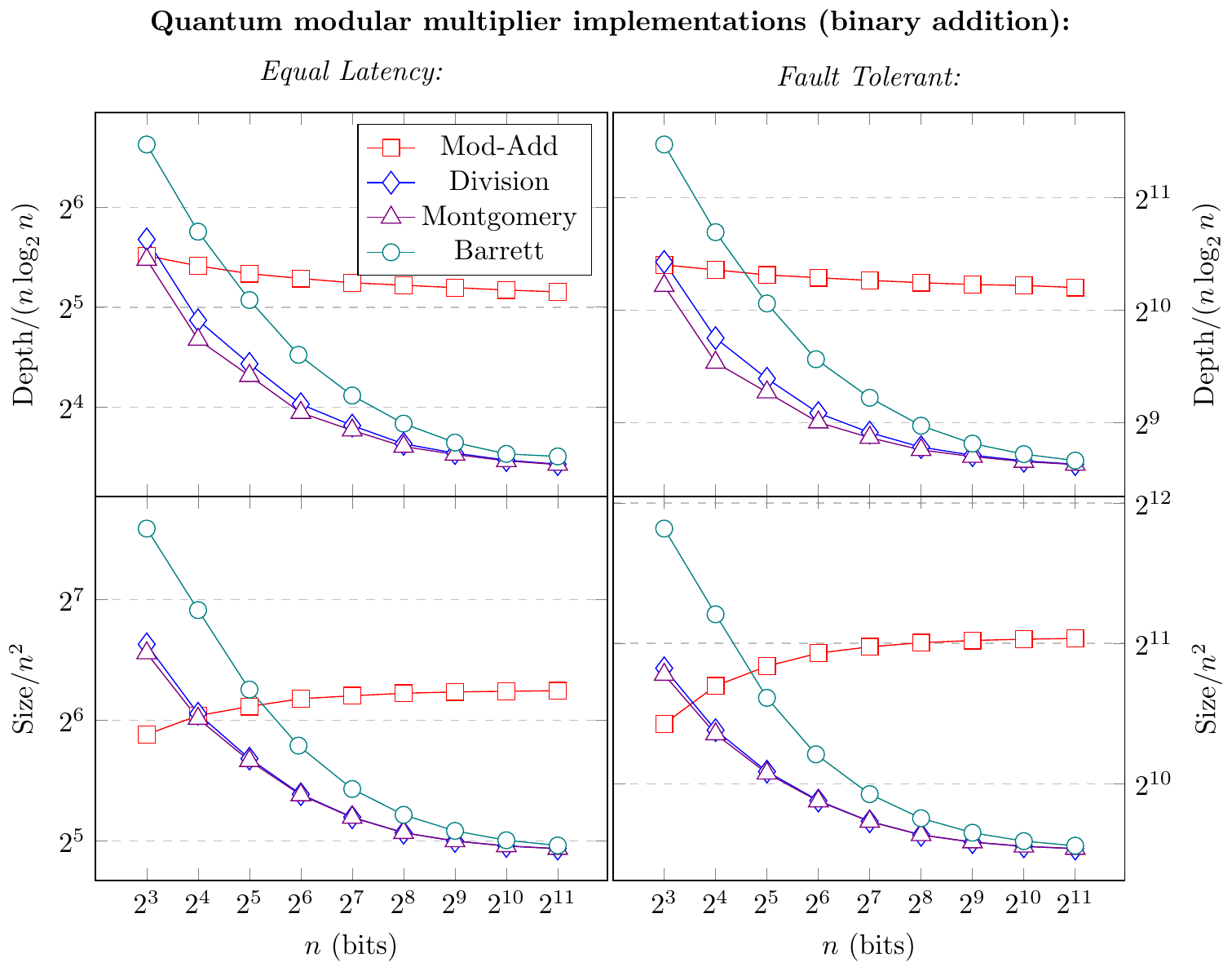}
\caption{\label{fig:resources:mults-binary}In-place, controlled modular multipliers constructed with binary quantum addition circuits.  We use the logarithmic-depth prefix circuit for the \ord{n}-qubit adders, but the prefix-ripple circuit for the $(\log_2n)$-qubit \uncomputation/-stage accumulator.  Both size and depth are normalized by their expected asymptotic scaling.}
\end{figure}

We now benchmark the optimized resource requirements of each of the quantum modular multipliers, constructed with both binary and Fourier-basis adders.  For each design, we construct circuits for in-place modular multiplication, incorporating quantum control with the third method outlined in \sect{mod-mult:control} (at the cost of $3n$ controlled-\SWAP/ operations, decomposed into \Toffoli/ and \CNOT/ gates).   We begin with the binary-basis multipliers.  Given the results of the previous section, we use the prefix adder for all \ord{n}-qubit quantum adders, and the prefix-ripple adder for the $(\log_2n)$-qubit \uncomputation/-stage accumulator.  The results for each multiplier, for both hardware models, are shown in \fig{resources:mults-binary}, where we have normalized by the expected asymptotic scaling ($\ord{n\log_2n}$ depth and $\ord{n^2}$ size).  As promised, we find that the asymptotic latency and gate count of each of the three modular multipliers proposed in this work is twice that of the initial \multiplication/ stage alone (where the factor of two results from the two out-of-place circuits required for in-place modular multiplication).  Circuits generated with standard modular addition approach, also shown in \fig{resources:mults-binary}, increase both the size and depth by another factor of three, corresponding to the three quantum additions required for each modular adder.

\begin{figure}[!bht]
  \centering
  \textbf{Qubits Required for Modular Multipliers}%
  \\\vspace{2ex}
  \begin{minipage}[c]{0.5\textwidth}
    \includeplot{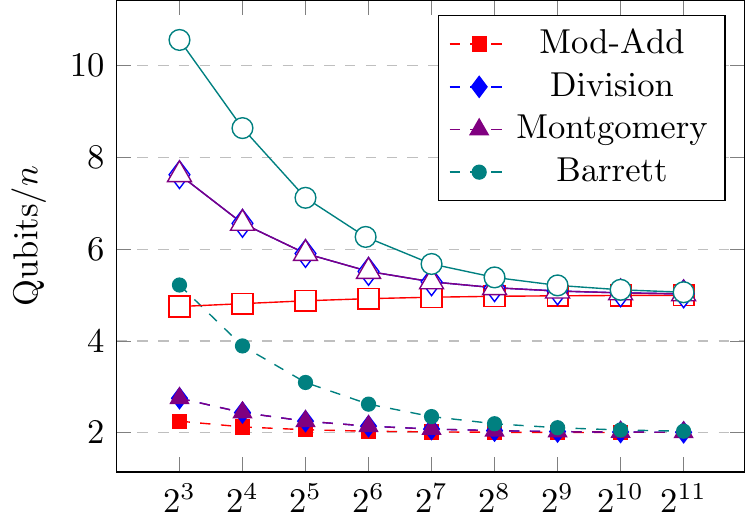}
  \end{minipage}\hfill
  \begin{minipage}[c]{0.5\textwidth}
    \caption{Qubit cost of modular multipliers with prefix (hollow marks) or Fourier-basis (solid marks) addition. Asymptotically, each proposed circuit requires $5n$ qubits with prefix adders, identically to the standard modular-addition approach.  The Fourier circuits require just $2n$ qubits asymptotically, matching the low-ancilla but much more costly circuit in~\cite{Beauregard2003}.  At small $n$, the additional $\ord{\log n}$ qubits required by our circuits become apparent.  The Barrett multiplier further requires an \ord{\log n}-qubit register for \ket{\xyapp}, causing its dominance at low $n$.\label{fig:resources:plots:mod-mult-qubits}}
  \end{minipage}
\end{figure}

\begin{figure}[!htb]
\centering 
\includeplot{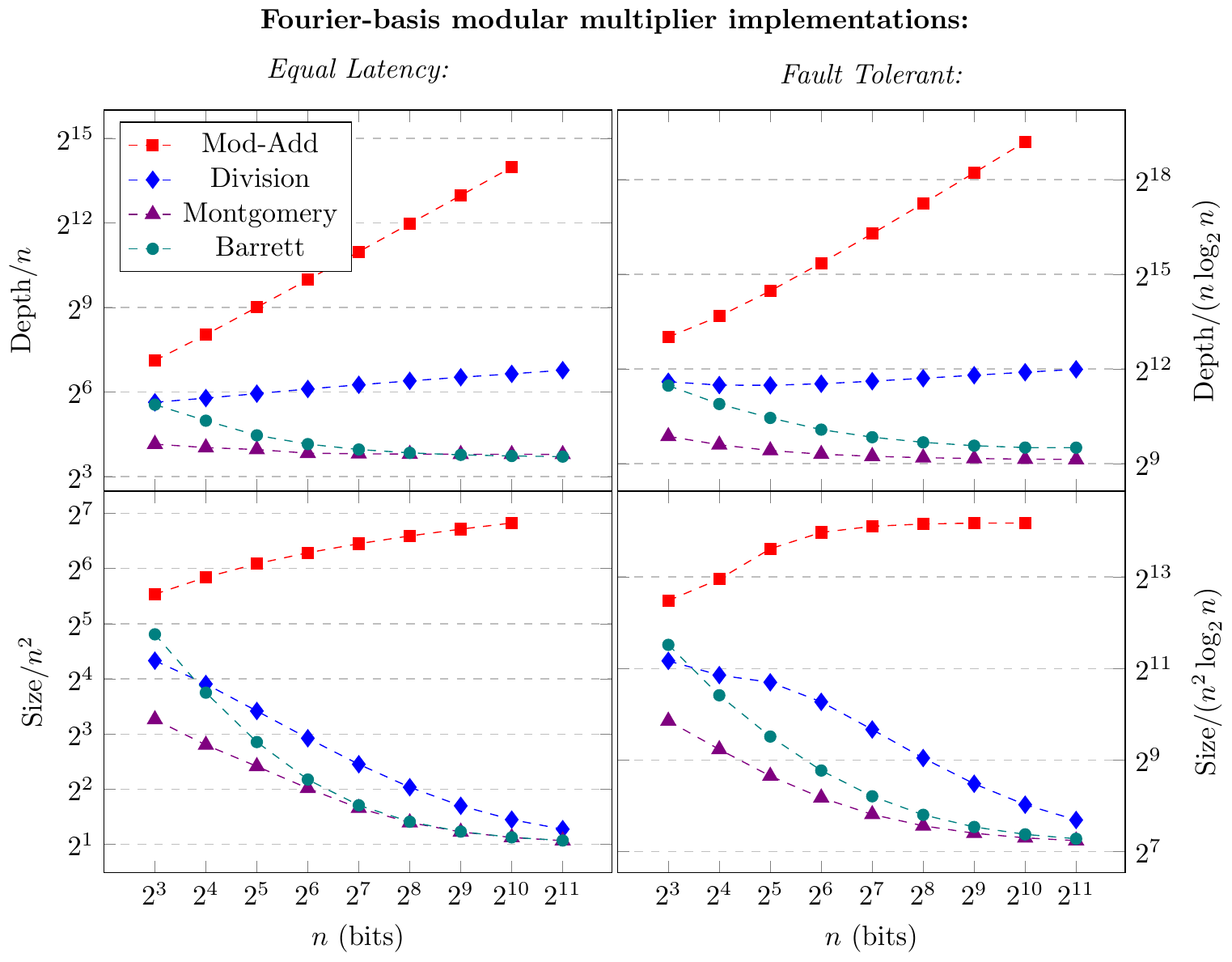}
\caption{In-place, controlled modular multipliers constructed with Fourier-basis adders. Both size and depth are normalized by the expected best-case asymptotic scaling.}
\label{fig:resources:plots:fourier}
\end{figure}

In \fig{resources:plots:fourier}, we show the resources consumed by the same multipliers constructed with Fourier basis operations. Here, the circuit depths of the different modular multipliers deviate significantly, driven by the intermediary \QFT/s necessary for comparison or modular reduction.  The standard modular-addition technique, as introduced in~\cite{Beauregard2003}, requires a total of $8n$ \QFT/s, resulting in the \ord{n^2} depth observed in the equal-latency model in \fig{resources:plots:fourier}.  The \reduction/ stage of the division-based circuit drives its \ord{n\log n} depth, while both the Barrett and Montgomery circuits display linear depth at large $n$, plateauing below the $14n$ worst-case latency determined in \sect{montgomery}.
At large $n$, the total number of gates required by all three proposed circuits coalesce to about $2n$.  This asymptotic size is twice that observed in \fig{resources:plots:multiply}, indicating that all three circuits are dominated in size by their non-modular multiplication operations.  As a result of the imposed finite precision of Fourier rotations, the number of gates required by the division algorithm's \reduction/ stage is eventually dominated by the $n^2$ gates in the initial \multiplication/ stage (however, as seen in \fig{resources:plots:fourier}, the size of the division-based circuit remains greater in the range of circuits observed).

In the fault-tolerant model, the total gate count of each circuit is increased by a factor of $\sim66\log_2(n)$ factor, corresponding to the cost of decomposing rotation gates.  However, after scheduling, we find that the total latency is only increased by $\sim40\log_2(n)$, demonstrating the additional parallelization enabled by the decomposition of controlled rotation gates described in \sect{resources:methodology}.

\begin{figure}[!htb]
\centering 
\includeplot{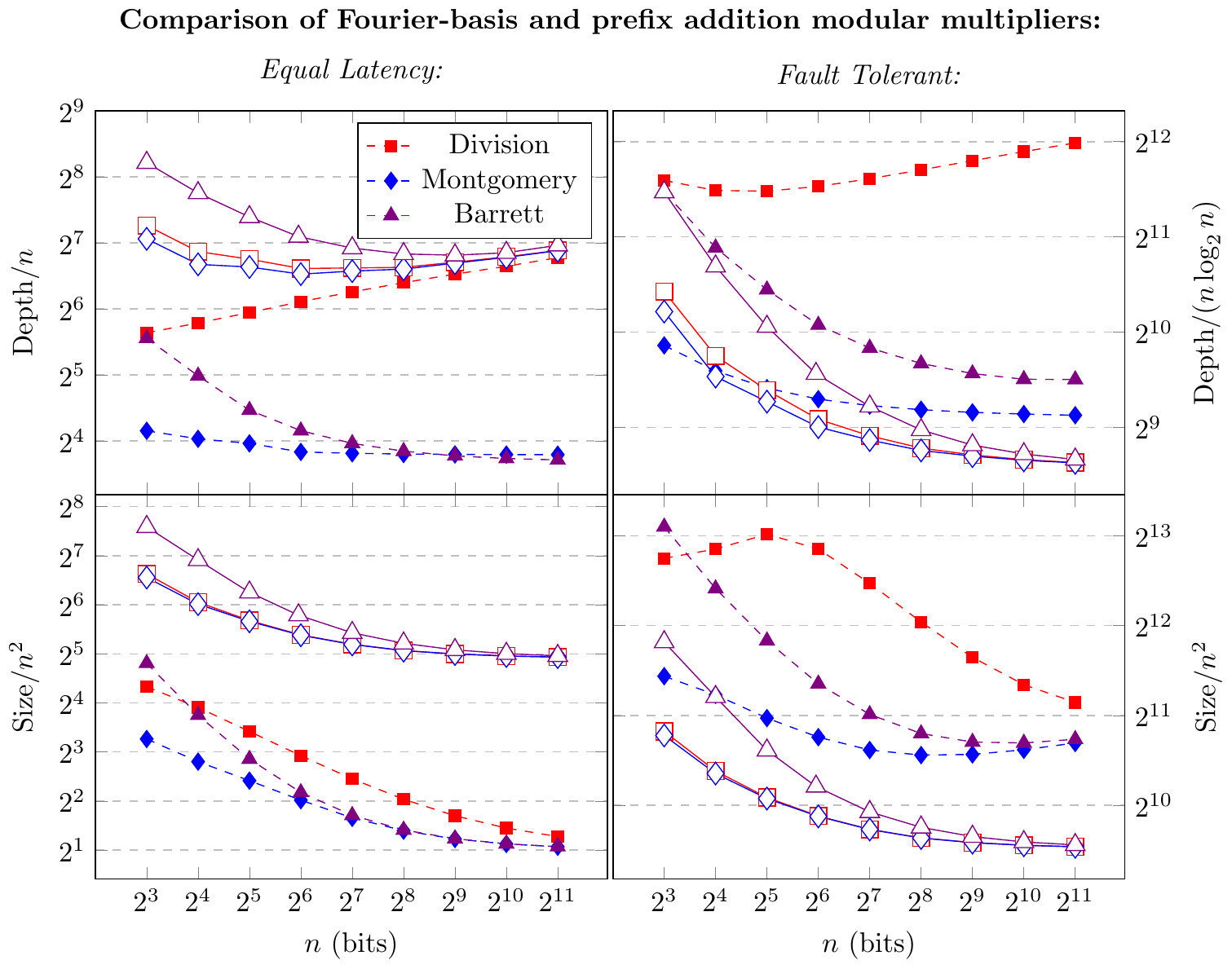}
  \caption{Proposed modular multipliers constructed with both Fourier-basis arithmetic (solid marks) and prefix addition (hollow marks). Both size and depth are normalized by the expected best-case asymptotic scaling.}
\label{fig:resources:plots:prefix-fourier}
\end{figure}

Finally, we plot the prefix and Fourier-basis implementations of the three proposed multipliers together in \fig{resources:plots:prefix-fourier}, and their corresponding qubit costs in \fig{resources:plots:mod-mult-qubits}.  A principal motivator for Fourier-basis quantum arithmetic is in reducing the number of required qubits.  Motivating its introduction in~\cite{Beauregard2003}, the Fourier-basis modular-addition-based circuit requires at most $2n+3$ qubits for any $n$.  The number of qubits required for Fourier-basis implementation of each of our proposed circuits also approaches $2n$ asymptotically, requiring only a logarithmic increase in ancilla.  Notably, prior speedups to Fourier-basis modular multiplication have come at the expense of additional \ord{n}-qubit registers~\cite{Kutin2006,Pavlidis2014}.  Comparatively, the ripple and prefix-ripple circuits each consume $\sim3n$ qubits, requiring an additional $n$-bit register to contain carry bits.  The prefix circuit, while having the lowest latency, of the binary adders requires $5n$ qubits.  In each case, the asymptotic ancilla cost for modular multiplication is that of $n$-qubit addition with the chosen adder.

While the $2n^2$ gate count of and linear depth of the Fourier-basis Barrett and Montgomery multiplication circuits in the equal latency model clearly outperform the $\sim2^5n^2$ gates and \ord{n\log n} latency observed of the multipliers constructed from prefix adders, this comparison is likely not valid in conjunction with quantum error correction.  In the more realistic fault-tolerant model, the Barrett and Montgomery circuits implemented with Fourier-basis addition and all three proposed multipliers constructed with prefix adders have \ord{n\log n} asymptotic latency.  In our simulations, the latencies of the Fourier-basis circuits were about $75\%$ greater than those from the prefix-adder circuits, and fall below the latencies of circuits constructed in the typical modular-addition approach with either adder.  The small latency increase indicates a potentially reasonable tradeoff for the qubit reduction enabled by the Fourier-basis circuits, given reasonable architectural assumptions. Comparatively, the latencies of the modular addition circuits are increased from \ord{n\log n} to \ord{n^2\log n} when implemented with fault-tolerant Fourier-basis arithmetic instead of prefix adders.  
Further, while the gate counts of all of the Fourier-basis circuits grow faster than the binary circuits by a factor of \ord{\log n}, the discrepancy remains within a factor of three in the large range of $n$ observed.

\subsection{Summary}
\label{sec:resources:summary}

We find a number important takeaways from the above experimental analysis of the quantum division, Barrett reduction, and Montgomery reduction circuits for quantum modular multiplication.  First, empirical analysis confirms the dominance of the non-modular multiplication process in the complexity of the three modular multiplication procedures introduced in this work.  Benchmarking the individual stages of the modular multipliers with a variety of binary quantum adders, we have found that the gate count and circuit latency of the modular reduction and uncomputation components become insignificant at large $n$ compared to the initial multiplication step of each circuit.  Additionally, the observation of the different addition circuits further enabled the characterization of multiplier stages in relation to their component adders, as well as corresponding optimizations.  For example, in the range of register sizes considered, the normally-fast prefix adder turns out to perform the worst of the adders in the case of the width-sensitive parallelized \uncomputation/ accumulator.  For the combined, in-place modular multiplier, the total asymptotic gate count and circuit latency observed of each design was twice that of a single $n$-bit quantum accumulator (where the factor of two results from the two out-of-place modular multipliers required for one in-place circuit).

In total, this represents a factor of three reduction in complexity from the typical modular addition approach to quantum modular multiplication, in which each $n$-bit addition required for product calculation is coupled with two more to perform a modular reduction step.  Accordingly, the number of ancilla qubits required for each modular multiplier is principally determined by the requirements of the $n$-bit addition circuit implemented, with all three of the proposed circuits requiring only a logarithmic number of ancilla beyond those required of for modular-addition-based circuit.  Further, in contrast to many previous proposals for fast modular multiplication~\cite{Zalka1998,Kutin2006,Pavlidis2014}, the proposed algorithms do not rely on inexact mathematical calculations or assumptions, and outperform the modular-adder design for input sizes as low as $n=2^5$.

Second, the modular multipliers introduced here, and the Barrett and Montgomery circuits in particular, present a unique amenability to implementation with quantum Fourier-basis arithmetic.  All three of the proposed modular multipliers can be implemented with quantum Fourier-basis adders with $2n+\ord{\log n}$ qubits, matching asymptotically the low-qubit modular-addition circuit proposed in~\cite{Beauregard2003}.  Assuming equal cost of each quantum gate, the total gate count of all three circuits approaches $2n^2$ at large $n$, again determined by the single multiplication procedure.  Further, the Barrett and Montgomery reduction circuits circumvent the costly Fourier-basis comparisons that dominate circuit latency of the modular addition circuit (and, to a lesser extent, the division-based circuit introduced here).  Experimentally, both circuits were demonstrated with latencies just below the analytically-determined $14n$-gate worst-case depth.  Comparatively, the Fourier-basis multiplier constructed from modular adders has \ord{n^2} latency and requires \ord{n^2\log n} gates, while the faster circuit introduced in~\cite{Pavlidis2014} requires $9n$ qubits and has a depth of $1000n$, and the inexact multiplier introduced in~\cite{Kutin2006} has the slightly smaller $12n$-gate depth but requires $3n$ total qubits.

Finally, our analysis demonstrates the competitiveness of Fourier-basis arithmetic for realistic (fault-tolerant) quantum modular multiplication.  The arbitrary rotations composing Fourier-basis operations can not be implemented directly on CSS codes, but instead must be decomposed into a sequence of available operations.  Given a reasonable set of architectural assumptions and the performance bounds for approximating arbitrary quantum gates presented in~\cite{Ross2014}, we nonetheless find that Fourier-basis implementations of the proposed Barrett and Montgomery multipliers can be demonstrated which perform comparably to the equivalent implementations with fault-tolerant logarithmic-depth binary adders.  After optimizing specifically for the decomposed Fourier rotation gates, with the assistance of a computerized scheduler, the Fourier-basis multipliers had less than twice the latency of the binary circuits in our model, in exchange for the $60\%$ reduction in required qubits.  Gate counts for fault-tolerant Fourier-basis circuits continued to dominate their binary counterparts by a logarithmic factor, but remained within a factor of three in the $8\le n\le2048$ range of input sizes modeled.

%% file: conclusion.tex

\section{Conclusions and Future Work}
\label{sec:conclusion}

We have presented three novel techniques for high-performance quantum modular multiplication, adapting fast classical frameworks for division, Montgomery residue arithmetic, and Barrett reduction to efficient reversible procedures.  Our techniques are independent of the lower-level implementation of binary quantum addition, and therefore can make use of any current or future quantum addition implementations. As an illustration of this
we have constructed and numerically analyzed quantum circuits resulting from three different binary adder implementations.  Each modular multiplication technique implements exact, out-of-place modular multiplication with the asymptotic depth and complexity of a single non-modular multiplication, representing a factor of three improvement over the standard quantum modular multiplication technique comprising repeated modular addition. 
The added gate count and ancilla requirements of our constructions is only \ord{\log n}.
The asymptotic depth and gate count of exact, in-place, controlled modular multiplication (comprising two out-of-place modular multipliers and $3n$ controlled-shift gates) is therefore that of $2n$ quantum adders, comparable to that previously achieved with inexact binary multipliers \cite{Zalka1998} and improving the $6n$ adders required of the typical modular-addition approach.

A unique advantage of the modular multipliers introduced in this work is their particular amenability to quantum Fourier-basis arithmetic.  All three proposed circuits require only $2n+\ord{\log n}$ qubits when implemented with Fourier-basis operations, asymptotically matching the low ancilla requirements of Beauregard's Fourier-basis modular-addition-based multiplier~\cite{Beauregard2003}.  Both the Barrett and Montgomery reduction techniques circumvent the need for repeated comparison operations, and therefore the corresponding \QFT/ and \QFTd/ circuits, which dominate the depth and complexity in the modular addition approach, are not required.  Taking advantage of the gate parallelization afforded by Fourier-basis arithmetic, both circuits can then be constructed with an asymptotic depth of $14n$ two-qubit gates.  This compares favorably with the $1000n$-gate latency of the fastest prior exact Fourier-basis modular multiplier~\cite{Pavlidis2014}, and is comparable to the $12n$-gate latency of the fastest inexact circuit~\cite{Kutin2006}.  Crucially, both prior circuits also expand the ancilla cost of Beauregard's circuit, asymptotically requiring $9n$ and $3n$ qubits, respectively.

Direct comparison between quantum Fourier-basis and binary arithmetic circuits is generally difficult for fault-tolerant systems, as the resource cost of arbitrarily-angled Fourier-basis rotations and $\Toffoli/$ gates depends highly on the underlying quantum computing hardware and error correction strategy employed.
It remains an open question as to whether the efficiency and speedup afforded by Fourier-basis circuits will be applicable to real quantum systems.  However, in \sect{resources} we have shown that with reasonable architectural assumptions, Fourier-basis modular multipliers can be constructed with performance comparable to the fastest binary adder. The space-time tradeoff between the two types of addition circuits is roughly equivalent, with the Fourier-basis adders requiring fewer qubits but more total gates.

In this work, we have primarily discussed circuits for \emph{quantum-classical} modular multiplication, where one of the two input multiplicands is a classical value known at ``compile time.''  Certain important quantum applications, such as the breaking of elliptic-curve cryptographic keys~\cite{Proos:2003}, instead require the implementation of in-place, \emph{quantum-quantum} modular multiplication.  
In the circuits we have presented, only the initial \GATE{Multiplication} and final \GATE{Uncomputation} accumulators depend on the individual input multiplicands, while the \GATE{Reduction} stage differentiating each circuit acts only on the computed product register.  An \emph{out-of-place} quantum-quantum modular multiplier is then easily derived by adapting these accumulators to a second quantum input, as described in \sect{mod-mult}.
However, in order to construct an in-place modular multiplier from this operator, we now require the multiplicative modular inverse of a quantum input state.  Reversible circuits implementing the extended Euclidean algorithm have been demonstrated but overwhelmingly dominate the complexity of the operation~\cite{Proos:2003}.   We have shown in the context of modular multiplication that with the adaptation of numerical or representational techniques we can mitigate the overhead of reversible reduction operations.  As Euclid's algorithm predominately comprising the sequential calculation of quotients and remainders, the techniques applied here present a similar and potentially significant opportunity for improving the implementation of this operation and corresponding class of problems.

Finally, the modular multipliers we have introduced are not limited to the arithmetic techniques discussed in this paper.  For example, the techniques for inexact computation~\cite{Zalka1998,Kutin2006}, fast large-number multiplication~\cite{Zalka1998}, or parallel arithmetic over a superlinear number of qubits~\cite{Gossett1998,VanMeter2005,Pham2013} could be applied independently to our proposed frameworks.  Similarly, the techniques could be extended to different domains; for example, implementations of both Barrett and Montgomery reduction over the Galois field $GF(2^m)$ (critical to elliptic curve cryptography) are well-known classically.

%% file: adders.tex

\section{Implementation of Quantum Arithmetic}
\label{sec:adders}

Integer quantum adders and multipliers are the base underlying circuits for all the circuit constructions
described in this paper. Because an integer multiplier can be constructed from repeated controlled
integer adders, the integer addition circuit can be considered the fundamental circuit of all the 
constructions. The basic modular multiplier constructed from modular adders requires only one type of
adder circuit, while the Barrett and Montgomery modular multipliers require additional adders in the 
reduction circuitry. These adders have different widths compared to the adders used in the main multiplier,
and therefore, it may be advantageous to utilize several different types of adders in these circuits.

Because the adder is such a fundamental circuit in all our circuit constructions the design of each adder
used will have a significant impact on the design and resources required by our modular multiplier circuits. 
Many quantum adders have been proposed, and we summarize the ones used in our circuit constructions in \tab{adders:desc}.
The adders fall into two main categories: adders based on reversible implementations
of classical adders, and adders that operate in a transformed Fourier basis. Each of these adders presents a 
different trade-off with respect to the total gates required, the circuit depth, and the number of ancilla used. 
These resource trade-offs translate directly to the multiplier designs, however, the form of the adders can also impact the 
resource requirements of the multipliers. For example, the Fourier adders allow gates from multiple adders to be overlapped,
which can reduce the overall depth of the multiplier.

\begin{table}[h!]
\caption{Quantum adder circuits used in multiplier constructions. The resource requirements are assuming the
in-place addition of a classical value onto a quantum register of width $n$, and are given to leading order only.
The resources for the Fourier transform basis adder assume the decomposition of the rotation gates required to
a specified accuracy ($\epsilon$) using a technique such as described in~\cite{Ross2014}.
}
\vspace{5pt}
\centering
\begin{tabular}{l|c|c|c}
\textbf{Adder type} & \textbf{\Toffoli//\T/ depth} & \textbf{\Toffoli//\T/ gates} & \textbf{Qubits required} \\
\hline
Majority ripple~\cite{Cuccaro2004} & $2n$ & $2n$ & $2n+1$ \\
Prefix-ripple [\sect{circuits:prefix_adders}] & $n$ & $3n$ & $2n+1$ \\
Carry look-ahead~\cite{Draper2006} & $4log_2(n)$ & $10n$ & $4n - log_2(n) -1$\\
Fourier transform basis~\cite{Draper2000} & $3log_2(1/\epsilon)$ & $3nlog_2(1/\epsilon)$ & $n$ \\
\hline
\end{tabular}
\label{tab:adders:desc}
\end{table}

\input{prefix_adder.tex}

\input{undo_adder.tex}

\input{multipliers.tex}

\input{fourier_adder.tex}

%% file: prefix_adder.tex
\subsection{Prefix Adders}
\label{sec:circuits:prefix_adders}

\begin{figure}
\centering
\includegraphics[scale=0.68]{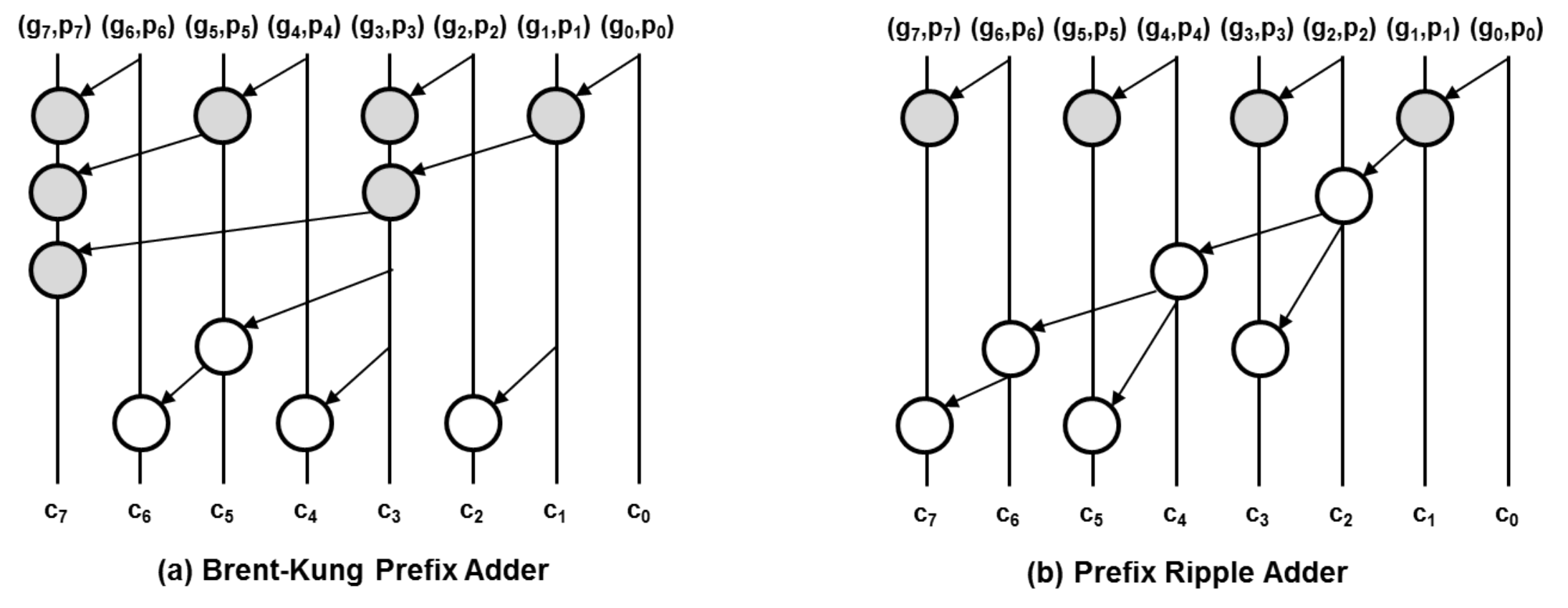}
\caption{Network structure for two different prefix adders. The shaded nodes produce both propagate (p) and
generate (g) bits from the inputs and the un-shaded nodes only produce the generate bits. 
The Brent-Kung adder has depth $2\log_2(n)-1$ for $n$ bits and the prefix-ripple adder has depth $n/2+1$.
}
\label{fig:adder:prefix:prefix_structure}
\end{figure}

The calculation of the carry bit-string for an adder can be thought of as a prefix operation, i.e., 
$C = c_{n-1}\circ\cdots\circ c_{2}\circ c_1\circ c_0$. Where each $c_i$ is represented by the tuple
$(p_i,g_i)$ and $p_i$/$g_i$ indicates that a carry is propagated/generated at position $i$.
The prefix composition function is defined as: $(p_{ij},g_{ij}) = (p_i\land p_j, g_j\lor(g_i\land p_j))$.
For a multi-bit adder with inputs $a_{[n-1:0]}$ and $b_{[n-1:0]}$ the single-bit inputs to the prefix network 
are calculated as: $p_i = a_i\oplus b_i$ and $g_i = a_i\land b_i$. 
The generate value from the first bit to bit $i$ (defined as $g_{0i}$) is the carry out of position $i$. 
For a multi-bit adder, a parallel network can be used to compute the prefix bits. 
For classical, non-reversible, adders many networks have been proposed and used to create adders.
Two example parallel-prefix networks are shown in \fig{adder:prefix:prefix_structure}.
For a description of these adders and others consult any textbook on computer arithmetic, for example~\cite{Ercegovac2004}.

Reversible adders that are suitable for quantum computing can be constructed from parallel prefix networks,
however, because of the constraints of reversible logic, which adders require the fewest resources
and have the lowest depth may be different than in the classical non-reversible case. For example
the adder described in~\cite{Draper2006} is based on the network structure shown 
in \fig{adder:prefix:prefix_structure}(a). The depth of this adder is logarithmic in the number of bits
and is well suited for a reversible implementation because it uses low fan-in/out nodes and requires 
fewer gates than many of the other proposed log-depth networks. 

\begin{figure}
\centering
\inputtikz{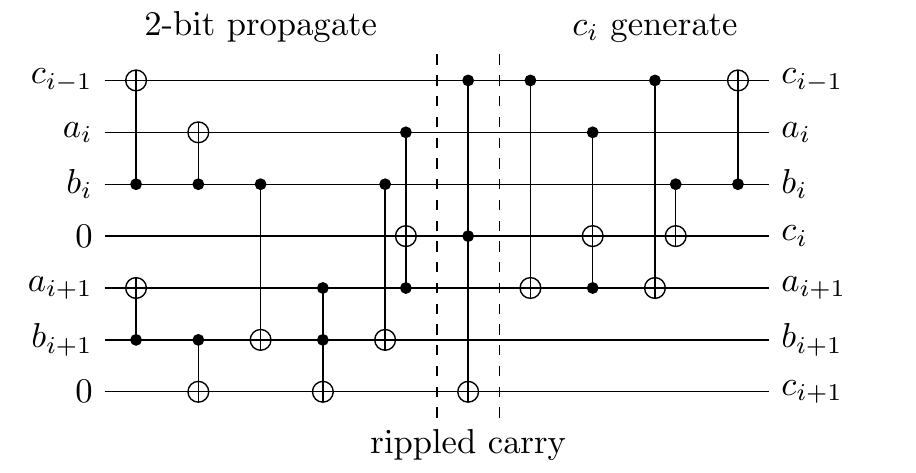}
\caption{Circuit to calculate two bits of the carry for the prefix-ripple adder. This adder requires $4$ \Toffoli/
gates when both inputs are quantum values. The section labeled \emph{2-bit propagate} can be executed in
constant time across all the bits in the adder. The \emph{rippled carry} step requires $n/2$ steps for
the entire adder and calculates the second carry bit in the pair. The first carry bit of the pair
is calculated after rippling the carry and can be done in constant time for all bits in the adder.
}
\label{fig:adder:prefix:half_prefix}
\end{figure}

Linear-depth parallel-prefix networks can also be defined, for example the network shown in 
\fig{adder:prefix:prefix_structure}(b) has depth $n/2+1$ for $n$ bits. This adder is similar
to a sequential ripple adder, but because we calculate two-bit propagate values across adjacent pairs of bits 
the carries can be rippled across two bits at each step. The odd-position carries are not in the critical 
path of the circuit and can be computed in a later step. We call this adder the prefix-ripple adder.
We have implemented a reversible circuit based on the network of \fig{adder:prefix:prefix_structure}(b),
and the repeating segment used to calculate pairs of carries is shown in \fig{adder:prefix:half_prefix}.
The first section of the circuit calculates the 2-bit propagate values and can be executed in parallel across
all bits in the adder. $n/2$ sequential \Toffoli/ gates are then used to ripple the carry to all even bit-positions 
of the adder. The last step is to calculate the odd-position carries. The carry at position $i$ can be calculated
in parallel with the calculation of an even-position carry in a later step, and therefore 
only adds a \Toffoli/ depth of one to the entire circuit. 
A full circuit built from the two-bit segments of \fig{adder:prefix:half_prefix} would require $4$ \Toffoli/ 
gates for every two bits in the full out-of-place addition of two quantum values. The circuit also requires
an additional $n$-bit ancilla register to hold the carries. Comparing the prefix-ripple adder to the ripple adder in~\cite{Cuccaro2004},
the new adder has half the \Toffoli/ depth but requires a factor of $2$ more \Toffoli/ gates, and an extra $n$-bit ancilla register.
If the prefix-ripple adder is used to add a classical value to a quantum one, then the first \Toffoli/ gate in the two-bit segment 
is reduced in degree and the cost becomes $1.5$ \Toffoli/ gates per bit. Additionally the total number of qubits for this
classical/quantum adder is $2n+1$, which is equivalent to that required by the adder in~\cite{Cuccaro2004} used in the
same way.

%% file: undo_adder.tex
\subsection{Select Undo Adder}
\label{sec:adders:select-undo}

The adders descried in this section thus far have assumed unconditional addition of two values. 
However, normally we require an adder that includes a global control. For example, this is the case when the adder is used
in a multiplier circuit. 
Typically a reversible in-place addition first calculates $\ket{x}\ket{y}\ket{0}\lra{}\ket{x}\ket{y}\ket{x+y}$ out-of-place and
then because the subtraction: $\ket{x}\ket{x+y}\ket{0}\lra{}\ket{x}\ket{x+y}\ket{y}$ produces the same set of outputs, we can run it in
reverse after the addition to produce our in-place addition: $\ket{x}\ket{y}\ket{0}\lra{}\ket{x}\ket{x+y}\ket{0}$. Because $\ket{y}$ 
and $\ket{x+y}$ are at positions $2$ and $3$ respectively after the addition, but the subtraction requires that they are in the
opposite order, we must swap the two registers. However, in most cases we can just relabel the two registers instead of swapping them.

\begin{figure}
\centering
\inputtikz{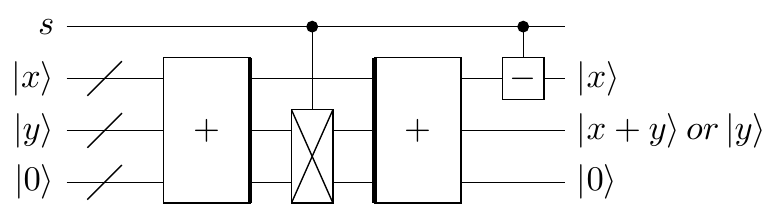}
\caption{Inplace select undo adder. When the select bit is $0$ the second reverse adder uncomputes the first
addition. When the select bit is $1$ the second addition acts as a reverse subtraction clearing the $\ket{y}$
input.}
\label{fig:quantum:adder:undo_adder}
\end{figure}

The similarity between addition and subtraction suggests a way to perform controlled in-place addition. 
A controlled adder should perform the in-place addition as described above when the control bit is set, but
when the bit is reset, it should undo the action of the addition. In the first case we are performing a reverse
subtraction, and in the later case a reverse addition. We can easily construct a circuit that selects between
these two cases based on the state of a control bit. A subtraction is just an addition where one of the inputs has
been negated. In two-complemented arithmetic this can be done by selectively flipping each bit with a \CNOT/ gate
and setting the input carry to one. The SWAP between the two out-of-place adders must now be controlled, requiring
two \CNOT/ gates and one \Toffoli/ per bit. This controlled adder, which we call the \emph{select undo} adder, is illustrated
in \fig{quantum:adder:undo_adder}. For this adder neither of the out-of-place adders is controlled and, depending on
the type of adder employed, this may lead to a reduction in the number of \Toffoli/ gates required. The main extra cost of the
select-undo adder are the $n$ \Toffoli/ gates required between the two out-of-place adders. However, the SWAP that they are used to
implement is between two existing registers and therefore no extra ancillas are required.

%% file: multipliers.tex

\subsection{Quantum Multiplication}
\label{sec:appen:mult}

Given an in-place quantum adder, we can construct the quantum multiply-accumulate circuit,
\begin{equation}
  \ket[w]{z}\ket[n]{y} \lra[\QQ\MAC(X)]  \ket[w]{Xy+z} \ket[n]{y}.
\end{equation}
Using the input \ket{z} as an initialized accumulator register, we add $2^kX$ in-place for each bit $y_k$ of \ket[n]{y}, requiring $n$ controlled, in-place quantum adders.
The \QQ\MAC{} operator can be generalized to any sum controlled by the bits of \ket{y}.  In particular, as is described in the text (\sect{div:mult}), we can accumulate the congruent value,
\begin{equation}
  t \defeq  \sum_{k=0}^{n-1} y_k\qty(2^k X \bmod N),
\end{equation}
such that $t\equiv Xy\pmod{N}$ and requires at most $w\le\clog[2]{nN}\le n+\clog{n}$ bits to represent.  As before, we require $n$ in-place controlled additions to the accumulation register \ket[w]{z}, now of the reduced partial products $2^kX\bmod N$.  If the accumulator is smaller than is required to hold $t$ (i.e. $w<\clog[2]{(nN)}$), these adders are truncated and the resulting product is computed modulo-$2^w$.

For classical $X$, we can similarly construct an in-place quantum multiplier,
\begin{equation}
  \ket[w]{0}\ket[n]{y} \lra[\QQ\MUL(X)]  \ket[n+w]{Xy},
\end{equation}
where the product is computed over the input register \ket{y} and is implicitly modulo-$2^{n+w}$.  For odd $X$, we can express the product,
\begin{equation}
  Xy  = y + y(X-1)
      = y + \sum_{k=0}^{n-1} y_k\cdot2^{k+1}\qty(\frac{X-1}{2}),
\end{equation}
where $(X-1)/2$ is an integer.  Each addition of $2^{k+1}(X-1)/2$ is then conditioned on the $k$th bit of $y$, and affects only the bits more significant than $k$ in the resulting sum.  Given a register initialized to \ket[n]{y}, we can therefore perform each of these additions in-place in descending order ($k=n-1,...,0$), so that each bit $y_k$ of \ket{y} controls an addition affecting only the more significant bits of the register before it is affected by any addition in the sequence.

For even $X$, the addend $(X-1)/2$ is not an integer.  Instead, we compute the equivalent product $(X/2^{\lambda})y$, where $\lambda=v_2(X)$ is the two-adic order of $X$.  The full product $Xy$ can then be produced by simply concatenating this result with $\lambda$ zero-initialized ancilla bits.
The in-place multiplier relies on the $k$ trailing zeros of each partial product $2^kX$, and so is not compatible with the partially-reduced multiply introduced above.  However, given the distribution of trailing zeros in the set of reduced partial products, it is likely that the result can be computed over about $\log_2(n)$ bits of the input state.

%% file: fourier_adder.tex

\subsection{Arithmetic in Fourier Transform Basis}
\label{sec:fourier:num-rep}

Central to quantum Fourier-basis arithmetic is the Fourier number state representation.  Defining the single-qubit state,
\begin{align}
    \fket[1]{\alpha}  &\defeq \cos(\alpha\pi/2)\ket{0} - i \sin(\alpha\pi/2)\ket{1}  \nonumber\\
                      &=      \Ry/(\alpha\pi)  \ket{0}, 
\end{align}
where $\Ry/(\alpha) = e^{-i\alpha \sigma_y/2}$ represents a single-qubit $y$-axis rotation, the Fourier representation of an $n$-bit number $y$ is equivalent to the product state,
\begin{equation}
    \eq{fourier:product-state}
    \fket[n]{y} \defeq  \bigotimes_{k=0}^{n-1}  \fKet[1]{ \frac{y}{2^{k}} }.
\end{equation}
Note that this differs from the typical definition~\cite{Nielsen2000}, as defined as $\fket[n]{y}^{'} \defeq \sum_{j=0}^{2^{n}-1} e^{-i jy\pi/2^{n}} \ket[n]{j}$.
The latter can be recovered as $(\S/\H/\S/^\dagger)^{\otimes n}\fket[n]{y}$, where \S/ and \H/ indicate single-qubit phase and Hadamard gates, respectively.

\subsubsection{Fourier-Basis Summation}
\label{sec:adders:fourier-add}

Uncontrolled, in-place addition of a classical parameter $X$ to a $n$-bit quantum Fourier register \fket[n]{y} requires $n$ unconditional $\Ry/$-rotations,
\begin{align}
  \eq{fourier:qft-add}
  \fket[n]{y+X}    
    &= \bigotimes_{k=0}^{n-1} \fKet[1]{ \frac{y+X}{2^{k}} }\nonumber\\
    &= \bigotimes_{k=0}^{n-1} \Ry/\qty( \frac{X\pi}{2^{k}} ) \fKet[1]{ \frac{ y}{2^k} } \nonumber\\
    &= \qty[ \prod_{k=0}^{n-1}  \Ry/^{(k)} \qty( \frac{X\pi}{2^{k}} )] \fket[n]{y},
\end{align}
where $\Ry/^{(k)}(\alpha)\ket[n]{y}$ indicates a $\Ry/(\alpha)$ rotation of the $k$th qubit of $\ket[n]{y}$.

Controlled Fourier addition of a classical parameter then requires conditioning each of these rotation gates on a single control qubit, inhibiting the parallelization of a standalone adder.  Instead, the commutability of the Fourier rotations (and lack of carries to propagate) admits large-scale parallelization for multiple additions.  In particular, as in \apx{appen:mult} we can construct an out-of-place multiply-accumulate operator,
\begin{equation}
  \fket{z}\ket[n]{y} \lra[\QF\MAC(X)]  \fket{Xy+z} \ket[n]{y},
\end{equation}
with each bit of a binary input register \ket[n]{y} controlling the Fourier-basis addition of a classical value to a Fourier-basis accumulation register.  The rotations required by this sum can be rearranged and executed in parallel, with a total depth of at most $\max(w,n)$ gates.

The addition of a computational-basis register \ket{y} to a Fourier-basis register \fket{x},
\begin{equation}
  \fket[n]{x  } \ket[n]{y} \lra
  \fket[n]{y+x} \ket[n]{y},
\end{equation}
requires constructing the $\Ry/^{(k)}(y\pi/2^k)$ rotations bitwise, performing the set of conditional rotations $\Ry/^{(k)}(2^ly_l/2^k)$ for each bit $y_l$ of $\ket[n]{y}$.  Equivalently, the quantum-quantum Fourier-basis adder is simply a special case of the Fourier multiply-accumulate operation, for which the multiplier is one.

Finally, observing \eq{fourier:qft-add}, the addition of a classical value $X\ll2^n$ will involve asymptotically small rotations on bits more significant than $k\sim\clog[2]{X}$.  As in~\cite{Barreiro2011a}, these operations can therefore be truncated to \ord{\log{nX}} gates with negligible loss of, or possibly improved, fidelity.

\subsubsection{In-place Fourier-Basis Multiplication}
\label{sec:fourier:mult}

Given a binary input state \ket[n]{y}, we can also perform an in-place multiply using Fourier-basis adders.  Observing the $k$th bit of the Fourier state $\fket{Xy}$ (again assuming odd $X$),
\begin{equation}
  \fkett[1]{ \frac{Xy}{2^k} } 
  = \fkett[1]{y_k + \sum_{j<k} \frac{y_j2^jX}{2^{k}}}
  = \pm\prod_{j<k}\Ry/\bigg(\frac{y_j2^jX}{2^{k}}\bigg)\ket{y_k},
\end{equation}
we find the binary input bit \ket{y_k}, rotated by the less significant qubits in the register.  Again beginning with the MSB, we perform in-place additions of $X/2$ controlled by each bit $y_k$ of the input \ket{y} and acting on the more significant bits of the register.  The resulting state is the Fourier representation of the product:
\begin{equation}
  \ket[n]{y} \lra[\QF\MUL(X)] \fket[n]{Xy}.
\end{equation}
The quantum Fourier transform, $\ket{y} \lra \fket{y}$, is then the special case $\QF\MUL(1)$.
Crucially, the $\QF\MUL(X)$ can be parallelized identically to the standalone \QFT/, with a depth of $2n+\ord{1}$ on a variety of computational topologies~\cite{Cleve2000, Moore2001, Pham2013}.

